\documentclass{article}
\usepackage[utf8]{inputenc}
\usepackage{bm,amsfonts,amsthm,amsmath,amssymb}
\usepackage{graphicx,color}
\usepackage{mathrsfs}
\usepackage{indentfirst}
\usepackage{bbm}
\usepackage{appendix}

\DeclareMathAlphabet{\mathpzc}{OT1}{pzc}{m}{it}

\def\eqdefa{\buildrel\hbox{\footnotesize def}\over =}
\newcommand{\ve}{\varepsilon}
\newcommand{\ud}{\mathrm{d}}

\newcommand{\uu}{\mathbf{u}}
\newcommand{\vv}{\mathbf{v}}

\newcommand{\aaa}{\mathbf{a}}
\newcommand{\xx}{\mathbf{x}}

\newcommand{\nn}{\mathbf{n}}
\newcommand{\ee}{\mathbf{e}}
\newcommand{\hh}{\mathbf{h}}
\newcommand{\mm}{\mathbf{m}}
\newcommand{\rr}{\mathbf{r}}

\newcommand{\sss}{\mathbf{s}}

\newcommand{\A}{\mathbf{A}}

\newcommand{\GG}{\mathbf{G}}

\newcommand{\NN}{\mathbf{N}}
\newcommand{\SSS}{\mathbf{S}}

\newcommand{\QQ}{\mathbf{Q}}

\newcommand{\FF}{\mathbf{F}}
\newcommand{\EE}{\mathbf{E}}

\newcommand{\CR}{\mathcal{R}}
\newcommand{\CB}{\mathcal{B}}
\newcommand{\CD}{\mathcal{D}}

\newcommand{\CS}{\mathcal{S}}

\newcommand{\CV}{\mathcal{V}}

\newcommand{\CN}{\mathcal{N}}
\newcommand{\CF}{\mathcal{F}}
\newcommand{\CG}{\mathcal{G}}

\newcommand{\CA}{\mathcal{A}}
\newcommand{\CP}{\mathcal{P}}
\newcommand{\CH}{\mathcal{H}}

\newcommand{\CM}{\mathcal{M}}
\newcommand{\CW}{\mathcal{W}}
\newcommand{\CK}{\mathcal{K}}
\newcommand{\CJ}{\mathcal{J}}

\newcommand{\ML}{\mathscr{L}}

\newcommand{\MP}{\mathscr{P}}

\newcommand{\MD}{\mathscr{D}}

\newcommand{\Fi}{\mathfrak{i}}
\newcommand{\Fp}{\mathfrak{p}}
\newcommand{\Fq}{\mathfrak{q}}
\newcommand{\Ft}{\mathfrak{t}}
\newcommand{\Fb}{\mathfrak{b}}

\newcommand{\BR}{{\mathbb{R}^3}}
\newcommand{\BA}{\mathbb{A}}
\newcommand{\BOm}{\mathbf{\Omega}}

\newtheorem{theorem}{Theorem}
\newtheorem{proposition}[theorem]{Proposition}
\newtheorem{lemma}[theorem]{Lemma}

\numberwithin{theorem}{section}
\numberwithin{equation}{section}

\allowdisplaybreaks[4]
\title{Frame hydrodynamics of biaxial nematics from molecular-theory-based tensor models}
\author{Sirui Li\footnote{ School of Mathematics and Statistics, Guizhou University, Guiyang 550025, China (srli@gzu.edu.cn) }, Jie Xu\footnote{LSEC and NCMIS, Institute of Computational Mathematics and Scientific/Engineering Computing (ICMSEC), Academy of Mathematics and Systems Science (AMSS), Chinese Academy of Sciences, Beijing, China (xujie@lsec.cc.ac.cn)}}

\date{}

\begin{document}
\maketitle
\begin{abstract}
    Starting from a dynamic tensor model about two second-order tensors, we derive the frame hydrodynamics for the biaxial nematic phase using the Hilbert expansion.
    The coefficients in the frame model are derived from those in the tensor model.
    The energy dissipation of the tensor model is maintained in the frame model.
    The model is reduced to the Ericksen--Leslie model if the biaxial bulk energy minimum of the tensor model is reduced to a uniaxial one.
\end{abstract}

\tableofcontents

\section{Introduction}

Liquid crystals are featured by local orientational order, typically originated from nonuniform orientational distribution of non-spherical rigid molecules.
One case that many of us are familiar with is the uniaxial nematic phase formed by rod-like molecules.
For the uniaxial nematic phase, the local orientational order can be described by a unit vector $\nn$.
The hydrodynamics of liquid crystals then involves dynamics of the vector $\nn$, for which the well-known Ericksen--Leslie theory is proposed \cite{E-61,Les}.
The Ericksen--Leslie theory, as well as its variants, has been studied extensively in both analysis \cite{Lin1,Lin-Liu,Lin2,Lin3,WZZ1,WW,Lin4} and simulation \cite{L-Walk,DGS,BFP,Walk,ZYLW}.
It has also been coupled with other systems \cite{YGCS,ZYSW}.
For a detailed survey on modeling, analysis and computation of liquid crystals, we refer to \cite{WZZ4}.

Constructed on the assumption of uniaxial local anisotropy, the Ericksen--Leslie theory  is opaque to the building blocks of liquid crystals.
Although the elastic constants can be related to experimental measurements, several other coefficients in the hydrodynamics are difficult to obtain.
This deficiency can be overcome by studying the relation of the Ericksen--Leslie theory to molecular models about the orientation density function \cite{KD,EZ}, or tensor models about a second-order tensor $Q$ \cite{HLWZZ}.
From molecular models or tensor models, one could derive the Ericksen--Leslie theory with its coefficients expressed by those in the molecular models or tensor models.
Such derivations are based on the fact that the minimum of the bulk energy must be uniaxial \cite{LZZ,FS}.
When the bulk energy dominates, the dynamics can be regarded as constrained in the states such that the bulk energy takes its minimum, so that it reduces to a dynamics of vector field.
The whole procedure is done through the Hilbert expansion that has been shown rigorously \cite{WZZ2,WZZ3,LWZ,LW}.

Local orientational orders other than the uniaxial type have also been considered, of which the biaxial nematics is discussed more \cite{S-J-P,MDN,APK,BVG}.
Its orientational elasticity is written down in various forms that turn out to be equivalent \cite{S-A,GV1,SV}.
Biaxial hydrodynamics are also proposed \cite{S-A,Liu-M,BP,S-W-M,GV2} in different forms.
Analysis has been carried out for a few simplified models \cite{LLWa}.
These works focus on the form of the model, in which many more coefficients are involved.
A couple of previous works attempt to relate the elastic constants to molecular parameters \cite{K-A,Xu2}, while other coefficients in the hydrodynamics are yet to be considered.

In this paper, we derive the hydrodynamics for the biaxial nematic phase from a tensor model.
The $Q$-tensor model for rod-like molecules can only possess isotropic and uniaxial bulk energy minima.
For this reason, it is necessary to start from a different tensor model that could exhibit other types of bulk energy minima.
We consider a dynamic tensor model for bent-core molecules derived from molecular theory \cite{XZ}, which has the biaxial nematic phase as an energy minimum.
Its free energy is constructed on molecular architecture by expanding the pairwise molecular interaction kernel \cite{XYZ}.
The interaction between the molecule and the fluid is also carefully derived from the molecular architecture.
As a result, the form of the dynamic tensor model is determined by molecular symmetry, with all the coefficients calculated from molecular parameters.

Rigorous analyses show that under certain coefficients, the stationary points of the bulk energy can only be isotropic, uniaxial or biaxial \cite{XZ1,XZ2,Xu3}.
Although further rigorous analysis is still not available, numerical studies indicate that we can indeed find some coefficients such that the biaxial nematic phase is the bulk energy minimum \cite{S-J-P,SVD,LGRA,MRV,MV,XZ1,XYZ,Xu3}.
Therefore, we assume that it holds and use the Hilbert expansion near this bulk energy minimum.
The free energy in the tensor model is rotationally invariant, which is an essential ingredient to be utilized in our derivation.
In particular, the rotational invariance of the bulk energy implies that its minimum, if not isotropic, can be freely rotated.
The biaxial nematic phase has its own symmetry other than axisymmetry.
When axisymmetry does not hold, the orientation of the bulk energy minimum shall generally be described by an orthonormal frame, or an element in $SO(3)$.
We would like to call it a `frame model' that gives the elasticity and dynamics of the field of orthonormal frame.

Two key ingredients are needed to be dealt with in the Hilbert expansion.
When the tensors are constrained at the biaxial minimum, it actually gives a three-dimensional manifold.
We shall constrain the equations of tensors on this manifold to obtain the evolution equation for the orthonormal frame field.
The tangent space of the manifold given by the bulk energy minimum gives a zero-eigenvalue subspace of the Hessian of the bulk energy.
This subspace is utilized to cancel the non-leading terms in the Hilbert expansion, thus closing the system of the leading order.
The free energy about tensors can then be reduced to the orientational elasticity for the biaxial nematic phase, with the elastic constants expressed as the coefficients in the tensor model, which is exactly the results in \cite{Xu2}.

Although the free energy can be reduced straightforwardly, we still need to handle several high-order tensors, which call for a closure approximation to express them as functions of the order parameter tensors.
Intuitively, these high-order tensors shall be consistent with the symmetry of the biaxial nematic phase, from which the form of high-order tensors can be written down.
This intuition can be made rigorously by the closure through minimization of the entropy term.
The entropy term can have two choices.
One is calculated from the density function of the maximum entropy state, which we call the original entropy.
The other is the quasi-entropy, an elementary function about tensors, which maintains essential properties and underlying physics of the original entropy \cite{Xu3}.
No matter we choose the original entropy or the quasi-entropy, their fine properties result in the particular form of high-order tensors consistent with the symmetry of the biaxial nematic phase.
From these symmetry arguments, we could further arrive at alternative expressions of these high-order tensors that are convenient for us to deduce the coefficients.

Using these properties, we could derive the frame model for the biaxial nematic phase.
Its form is actually determined by the symmetry of the biaxial nematic phase, which is consistent with early works \cite{GV2}.
The coefficients, on the other hand, are expressed as functions of the coefficients in the tensor model.
Again, since the coefficients in the tensor model are derived from physical parameters, the frame model we obtain is connected to rigid molecules with certain architecture.
We shall show that the energy dissipation of the tensor model is maintained in the frame model.
Furthermore, we will show that the biaxial hydrodynamics can be reduced to the Ericksen-Leslie theory when the bulk energy has a uniaxial minimum.
The corresponding coefficients are also derived, which turn out to be distinct from those derived from the $Q$-tensor hydrodynamics for rod-like molecules.

Below, we begin with introducing some notations for orthonormal frames and tensors in Section \ref{Preliminary}.
The tensor model is briefly described in Section \ref{tensor-models}.
Here, we also claim essential properties on the entropy term, bulk energy minima, and high-order tensors. The Hilbert expansion is carried out in Section \ref{frame-model}, from which we derive the biaxial hydrodynamics.
The biaxial hydrodynamics can be reduced to Ericksen--Leslie theory if the bulk energy minimum becomes uniaxial, which is shown in Section \ref{unaxial-dynamics}.
The relation of our model to the previous ones is discussed in Section \ref{comparison}.
Concluding remarks are given in Section \ref{concl}.
Detailed calculations and discussions on high-order tensors are presented in Appendices.

\section{Preliminary}\label{Preliminary}
Let us introduce some notations for orthonormal frames and tensors to be used subsequently.
For the rigid molecules forming liquid crystalline states, several essential quantities are defined through the orientational distribution.
The orientation of a single rigid molecule is described by an orthonormal, right-handed frame $(\mm_1,\mm_2,\mm_3)$ fixed on the molecule.
The axes of the frame are typically coincident with symmetry axes of the molecule.
Under a reference frame $(\ee_1,\ee_2,\ee_3)$, the coordinates of the molecular frame can be expressed by
\begin{align*}
\Fq_{ij}=\ee_i\cdot\mm_j,~i,j=1,2,3,
\end{align*}
which define a $3\times 3$ rotation matrix $\Fq\in SO(3)$.
The frame $\Fq$ can be parametrized by three Euler angles $(\varphi,\psi,\theta)$ as
\begin{align*}
\Fq=&\,(\mm_1,\mm_2,\mm_3)
=\left(
\begin{array}{ccc}
 m_{11}  &\,  m_{21} &\, m_{31} \\
 m_{12}  &\,  m_{22} &\, m_{32} \\
 m_{13}  &\,  m_{23} &\, m_{33}
\end{array}
\right)\\
=&\,\left(
\begin{array}{ccc}
 \cos\varphi  &\,  -\sin\varphi\cos\theta &\, \sin\varphi\sin\theta \\
 \sin\varphi\cos\psi  &\,  \cos\varphi\cos\psi\cos\theta-\sin\psi\sin\theta &\, -\cos\varphi\cos\psi\sin\theta-\sin\psi\cos\theta \\
 \sin\varphi\sin\psi  &\, \cos\varphi\cos\psi\cos\theta+\sin\psi\sin\theta  &\, -\cos\varphi\sin\psi\sin\theta+\cos\psi\cos\theta
\end{array}
\right),
\end{align*}
where $\varphi\in[0,\pi]$ and $\psi, \theta\in[0,2\pi)$. The corresponding invariant Haar measure on $SO(3)$ is denoted by
\begin{align*}
\ud\Fq=\frac{1}{8\pi^2}\sin\varphi\ud\varphi\ud\psi\ud\theta.
\end{align*}

In this paper, we also deal with fields of orthonormal frame.
To be distinguished from the molecular frame, we use the notation $\Fp=(\nn_1,\nn_2,\nn_3)$ for a frame field that is a function of the position $\xx$.
The notations for $\nn_i$ are similar to those for $\mm_i$ above.

Next, let us describe notations for tensors. An $n$th-order tensor $U$ in $\BR$ can be expressed as a linear combination of tensors generated by the axes of the reference frame $(\ee_1,\ee_2,\ee_3)$, written as
\begin{align}
U=U_{i_1\ldots i_n}\ee_{i_1}\otimes\cdots\otimes\ee_{i_n},~i_1,\ldots,i_{n}\in\{1,2,3\}, \label{tensor0}
\end{align}
where $U_{i_1\ldots i_n}$ are the coordinates of the tensor $U$.
Hereafter, we adopt the Einstein summation convention on repeated indices.
For any two $n$th-order tensors $U_1$ and $U_2$, the dot product $U_1\cdot U_2$ is defined as
\begin{align*}
    U_1\cdot U_2=(U_1)_{i_1\ldots i_n}(U_2)_{i_1\ldots i_n}.
\end{align*}

A tensor can be symmetrized by calculating its permutational average,
\begin{align*}
U_{\rm sym}=\frac{1}{n!}\sum_{\sigma}U_{i_{\sigma(1)}\ldots i_{\sigma(n)}}\ee_{i_1}\otimes\cdots\otimes\ee_{i_n},
\end{align*}
where the summation is taken over all the permutations $\sigma$ of $\{1,\ldots,n\}$.
If $U=U_{\rm sym}$, we say that the tensor $U$ is symmetric.
For an $n$th-order symmetric tensor, we define its trace as the contraction of two of its indices, giving an $(n-2)$th-order symmetric tensor,
\begin{align*}
({\rm tr}U)_{i_1\ldots i_{n-2}}=U_{i_1\ldots i_{n-2}kk}.
\end{align*}
If a symmetric tensor $U$ satisfies ${\rm tr}U=0$, then $U$ is called a symmetric traceless tensor.
For any symmetric tensor $U$ of the order $n$, there exists a unique symmetric traceless tensor $(U)_0$ of the form
\begin{equation}
    (U)_0=U-(\Fi\otimes W)_{\rm sym},\label{symtrlsnot}
\end{equation}
where $W$ is a tensor of the order $(n-2)$. We call $(U)_0$ the symmetric traceless tensor generated by $U$.

It could be convenient to express symmetric traceless tensors by polynomials.
The basic monomial notation is defined as
\begin{align}\label{mono-notation}
\mm^{k_1}_1\mm^{k_2}_2\mm^{k_3}_3\Fi^l=\Big(\underbrace{\mm_1\otimes\cdots\otimes\mm_1}_{k_1}\otimes\underbrace{\mm_2\otimes\cdots\otimes\mm_2}_{k_2}\otimes\underbrace{\mm_3\otimes\cdots\otimes\mm_3}_{k_3}\otimes\underbrace{\Fi\otimes\cdots\otimes\Fi}_{l}\Big)_{\rm sym},
\end{align}
where $\Fi$ is the second-order identity tensor that can be expressed as
\begin{align*}
\Fi=\mm^2_1+\mm^2_2+\mm^2_3.
\end{align*}
This equality holds independent of what frame $(\mm_1,\mm_2,\mm_3)$ is chosen.
As we have commented, the above definitions are also suitable for $\nn_i$.
When the symbol $\otimes$ is absent in a product, it means that the resulting tensor is symmetrized.
A few examples are given below,
\begin{align*}
\mm_1=&\,\mm_1,\\
\mm_1\mm_2=&\,\frac{1}{2}(\mm_1\otimes\mm_2+\mm_2\otimes\mm_1),\\
\mm^2_1=&\,\mm_1\otimes\mm_1,\\
\mm_1\mm_2\mm_3=&\,\frac{1}{6}\big(\mm_1\otimes\mm_2\otimes\mm_3+\mm_2\otimes\mm_3\otimes\mm_1+\mm_3\otimes\mm_1\otimes\mm_2\\
&\,+\mm_1\otimes\mm_3\otimes\mm_2+\mm_2\otimes\mm_1\otimes\mm_3+\mm_3\otimes\mm_2\otimes\mm_1\big),\\
\mm_1\mm^2_2=&\,\frac{1}{3}\big(\mm_1\otimes\mm_2\otimes\mm_2+\mm_2\otimes\mm_1\otimes\mm_2+\mm_2\otimes\mm_2\otimes\mm_1\big),
\end{align*}
where the permutation of the subscripts on the right-hand side implies that the tensor are symmetric.
Using the polynomial notations, the explicit expressions for different bases of symmetric traceless tensors have been identified \cite{Xu1}.
We shall introduce these explicit expressions when they are encountered afterwards.

The orientational distribution of rigid molecules is denoted by $\rho(\xx,\Fq)$.
However, it is significant to introduce some simple quantities to classify the local anisotropy given by the density function $\rho$.
Such quantities are defined through the moments of $\mm_i$,
\begin{align}\label{m-tensor-moment}
\big\langle\mm_{i_1}\otimes\cdots\otimes\mm_{i_n}\big\rangle=\int_{SO(3)}\mm_{i_1}(\Fq)\otimes\cdots\otimes\mm_{i_n}(\Fq)\rho(\xx,\Fq)\ud\Fq,\quad i_1,\ldots,i_n=1,2,3.
\end{align}
Hereafter, the notation $\langle\cdot\rangle$ is employed to represent the average of the distribution function $\rho(\xx,\Fq)$ on $SO(3)$.
The linearly independent components in these moments are further extracted, for which it is necessary to choose a few symmetric traceless tensors averaged by $\rho$ \cite{Xu1}.
These chosen averaged symmetric traceless tensors are the so-called order parameters.
For example, the local alignment state of rod-like molecules is described by the second-order symmetric traceless order tensor
\begin{align*}
\Big\langle\mm^2_1-\frac{\Fi}{3}\Big\rangle=\int_{SO(3)}\Big(\mm^2_1-\frac{\Fi}{3}\Big)\rho(\xx,\Fq)\ud\Fq.
\end{align*}

Two useful quantities that are employed throughout this paper are the Kroneker delta and the Levi-Civita symbol, which are defined respectively by
    \begin{align*}
    \delta_{ij}=\left \{
\begin{array}{ll}
1,      &\,i=j,\\
0,      &\,i\neq j,
\end{array}
\right.
\end{align*}
and
\begin{align*}
\epsilon^{ijk}=\left \{
\begin{array}{ll}
1,      &\, \text{if}~i,j~\text{and}~k~\text{are unequal and in cyclic order},\\
-1,     &\, \text{if}~i,j~\text{and}~k~\text{are unequal and in non-cyclic order,}\\
0,      &\, \text{if any two of}~i,j~\text{or}~k~\text{are equal}.
\end{array}
\right.
\end{align*}
The following two properties are useful:
\begin{align*}
\epsilon^{ijk}\epsilon^{ipq}=&\,\delta_{jp}\delta_{kq}-\delta_{jq}\delta_{kp},\\
\epsilon^{ijk}\epsilon^{pqr}=&\,\det\left(
  \begin{array}{ccc}
    \delta_{ip} &\, \delta_{iq} &\, \delta_{ir}\\
    \delta_{jp} &\, \delta_{jq} &\, \delta_{jr}\\
    \delta_{kp} &\, \delta_{kq} &\, \delta_{kr}\\
  \end{array}
\right).
\end{align*}

We will frequently encounter derivatives involving orthonormal frames.
Let us first define rotational differential operators.
For any frame $\Fp=(\nn_1,\nn_2,\nn_3)\in SO(3)$, its tangent space in $SO(3)$ is spanned by three matrices, given by $(0,\nn_3,-\nn_2)$, $(-\nn_3,0,\nn_1)$, $(\nn_2,-\nn_1,0)$.
Thus, we define three differential operators $\ML_j$ by taking the inner products of the above three matrices and $\partial/\partial\Fp=(\partial/\partial\nn_1,\partial/\partial\nn_2,\partial/\partial\nn_3)$, i.e.,
\begin{align}
\left\{
\begin{array}{l}
    \ML_1=\nn_3\cdot\frac{\partial}{\partial \nn_2}-\nn_2\cdot\frac{\partial}{\partial \nn_3},
    \vspace{1ex}\\
    \ML_2=\nn_1\cdot\frac{\partial}{\partial \nn_3}-\nn_3\cdot\frac{\partial}{\partial \nn_1},
    \vspace{1ex}\\
    \ML_3=\nn_2\cdot\frac{\partial}{\partial \nn_1}-\nn_1\cdot\frac{\partial}{\partial \nn_2}.
\end{array}
  \right.
\end{align}
The subscript indicates the differential operator is along the infinitesimal rotation about $\nn_j(j=1,2,3)$.
This can be verified by acting the differential operators on the axes of the frame, resulting in
\begin{align}
\ML_j\nn_k=\epsilon^{jkl}\nn_l. \label{ndev}
\end{align}

For a frame field $\Fp(\xx)$, its orientational elasticity is characterized by an elastic energy about the spatial derivatives of $\Fp(\xx)$.
Let us express these spatial derivatives under the local frame $\Fp$.
The derivative of $\nn_{\mu}$ along the direction $\nn_{\lambda}$ is given by $\nn_{\lambda}\cdot\nabla\nn_{\mu}$.
Its $\nu$-component in the frame $\Fp$ can be written as $n_{\lambda i}n_{\nu j}\partial_in_{\mu j}$. Using the equality $n_{\mu j}n_{\nu j}=\delta_{\mu\nu}$,
we obtain the following relation
\begin{align}\label{rel-nn}
n_{\lambda i}n_{\nu j}\partial_i {n}_{\mu j}
=-n_{\lambda i}n_{\mu j}\partial_i {n}_{\nu j}.
\end{align}
Consequently, the first-order derivatives of the frame $\Fp$ has nine degrees of freedom:
\begin{align}\label{Dij}
\left\{
\begin{array}{l}
  D_{11}=n_{1i}n_{2j}\partial_i n_{3j},\ D_{12}=n_{1i}n_{3j}\partial_i n_{1j},\ D_{13}=n_{1i}n_{1j}\partial_i n_{2j},
  \vspace{1ex}\\
  D_{21}=n_{2i}n_{2j}\partial_i n_{3j},\ D_{22}=n_{2i}n_{3j}\partial_i n_{1j},\ D_{23}=n_{2i}n_{1j}\partial_i n_{2j},
  \vspace{1ex}\\
  D_{31}=n_{3i}n_{2j}\partial_i n_{3j},\ D_{32}=n_{3i}n_{3j}\partial_i n_{1j},\ D_{33}=n_{3i}n_{1j}\partial_i n_{2j}.
  \end{array}
  \right.
\end{align}

\section{Tensor model}\label{tensor-models}

In tensor models, the local orientational order is described by one or several order parameter tensors.
From the structure of nonzero components in the tensors, local anisotropy could be divided into several classes.
Each class is recognized as a phase, and phase transitions between them can be described.
For examples, the transition between the isotropic and uniaxial nematic phases for rod-like molecules can be described by an energy about a second-order symmetric traceless tensor $Q$.
The dynamic tensor models could either be phenomenological, such as the Beris-Edwards model \cite{BE} and the Qian-Sheng model \cite{QS} based on the Landau-de Gennes theory, or be derived from the molecular theory \cite{HLWZZ}.
In the vicinity of a bulk energy minimum, the tensors possess the nonzero structure of a certain phase, so that the tensor model is reduced to a model with fewer variables.
For the uniaxial nematic phase of rod-like molecules, the models about a field of $Q$-tensor, which has five degrees of freedom, could be reduced to models about a field of unit vector, which has two degrees of freedom.

When rigid molecules of more complex architecture are taken into account, such as bent-core and star molecules, the corresponding molecular-theory-based tensor models have also been derived \cite{XYZ,XZ}.
The most notable feature of this model lies in the fact that its form and coefficients are  determined by molecular symmetry and molecular parameters, respectively.
Depending on the coefficients, the bulk energy may exhibit isotropic, uniaxial nematic or biaxial nematic phases. The modulated twist-bend nematic phase can also be described together with elastic energy.
Since the biaxial nematic phase is included in this tensor model, we choose this model as our starting point.

Compared with the original form in \cite{XZ}, we have made a couple of simplifications that are clarified below.
\begin{itemize}
\item The model in \cite{XZ} has three order parameter tensors, one first-order and two second-order. In the biaxial nematic phases, the first-order tensor takes zero. This is also maintained in the leading order of Hilbert expansion. As a result, keeping the first-order tensor makes no difference in our derivation. For this reason, we assume that the first-order tensor is zero to discard all the terms about it.
\item We ignore the spatial diffusion term. This is also adopted in the derivation from dynamic $Q$-tensor models to the Ericksen-Leslie model (see \cite{HLWZZ}).
\end{itemize}

The tensor model is then about two second-order symmetric traceless tensors, defined as
\begin{align*}
Q_1=\langle(\mm^2_1)_0\rangle=\Big\langle\mm^2_1-\frac{\Fi}{3}\Big\rangle,\quad Q_2=\langle(\mm^2_2)_0\rangle=\Big\langle\mm^2_2-\frac{\Fi}{3}\Big\rangle.
\end{align*}
Denote $\QQ=(Q_1,Q_2)^T$. Let us also define a projection on to symmetric traceless tensors,
\begin{align}
    (\MP R)_{ij}=\frac 12(R_{ij}+R_{ji})-\frac 13 R_{kk}\delta_{ij}.
\end{align}
The projection can also be imposed on an array of second-order tensors:
$$
\MP(R_1,\cdots,R_k)=(\MP R_1,\cdots,\MP R_k).
$$

\subsection{Free energy}

Assume that the concentration $c$ of rigid molecules is constant in space.
The free energy contains two parts, the bulk energy and elastic energy,
\begin{align}\label{free-energy}
\frac{\CF[\QQ,\nabla\QQ]}{k_BT}=\int\ud\xx\Big(\frac{1}{\ve}F_b(\QQ)+F_e(\nabla\QQ)\Big),
\end{align}
which is measured by the product of the Boltzmann constant $k_B$ and the absolute temperature $T$.
The bulk energy density, which can describe transitions between homogeneous phases, consists of an entropy term and pairwise interaction terms,
\begin{align}
    F_b=&\,cF_{entropy}+\frac{c^2}{2}\big(c_{02}|Q_1|^2+c_{03}|Q_2|^2+2c_{04}Q_1\cdot Q_2\big).\label{free-energy-bulk}
\end{align}
The elastic energy density penalizing spatial inhomogeneity contains a few quadratic terms of $\nabla\QQ$,
\begin{align}
    F_e=&\,\frac{c^2}{2}\Big(c_{22}|\nabla Q_1|^2+c_{23}|\nabla Q_2|^2 +2c_{24}\partial_iQ_{1jk}\partial_iQ_{2jk}\nonumber\\
    &\,\quad+c_{28}\partial_iQ_{1ik}\partial_jQ_{1jk}+c_{29}\partial_iQ_{2ik}\partial_jQ_{2jk} +2c_{2,10}\partial_iQ_{1ik}\partial_jQ_{2jk}\Big). \label{free-energy-elastic}
\end{align}
We have introduced a small parameter $\ve$ in the free energy \eqref{free-energy}.
It can be regarded as the squared relative scale $\tilde{L}$ between the rigid molecule and the domain of observation by a change of variable $\tilde{\xx}=\xx/\tilde{L}$.

The entropy term acts as a stabilizing term that guarantees the lower-boundedness of the bulk energy.
There can be different choices, but it is always independent of molecule architecture.
Moreover, the entropy term is related to expressing the tensors of higher order by $Q_1$ and $Q_2$.
For this reason, we shall specify the entropy term afterwards.

On the other hand, the coefficients $c_{ij}$ of the quadratic terms can be calculated as functions of molecular parameters.
For instance, if the hardcore molecular interaction is adopted, we are able to compute these coefficients from molecular shape parameters \cite{XYZ}.
This is also the case for the dynamic tensor model, which we introduce below.

\subsection{Dynamic model}
Based on the free energy functional \eqref{free-energy}, \eqref{free-energy-bulk}, and  \eqref{free-energy-elastic}, let us write down the molecular-theory-based dynamic tensor model derived in \cite{XZ}.
We define the variational derivative of (\ref{free-energy}) as
\begin{align}\label{vari-deriv-Q}
\mu_{\QQ}=&\,~\frac{1}{ck_BT}\frac{\delta \CF(\QQ,\nabla\QQ)}{\delta\QQ}=\frac{1}{ck_BT}\MP\Big(\frac{1}{\ve}\frac{\partial F_b(\QQ)}{\partial\QQ}-\partial_i\Big(\frac{\partial F_e(\nabla\QQ)}{\partial (\partial_i\QQ)}\Big)\Big)\nonumber\\
\eqdefa&\,~\frac{1}{\ve}\CJ(\QQ)+\CG(\QQ),
\end{align}
where $\mu_{\QQ}=(\mu_{Q_1}, \mu_{Q_2})^T$, $\CJ(\QQ)=\big(\CJ_1(\QQ),\CJ_2(\QQ)\big)^T$ and $\CG(\QQ)=\big(\CG_1(\QQ),\CG_2(\QQ)\big)^T$ are calculated as
\begin{align}
\mu_{Q_1}=&\,\frac{1}{\ve}\CJ_1(\QQ)+\CG_1(\QQ)\nonumber\\
=&\,\frac{1}{\ve}\left(\MP\frac{\partial F_{entropy}}{\partial Q_1}+cc_{02}Q_1+cc_{04}Q_2\right)\nonumber\\
&\,-cc_{22}\Delta Q_{1jk}-cc_{24}\Delta Q_{2jk}
-\MP(cc_{28}\partial_j\partial_iQ_{1ik}+cc_{2,10}\partial_j\partial_iQ_{2ik}),\label{mu-Q1}\\
\mu_{Q_2}=&\,\frac{1}{\ve}\CJ_2(\QQ)+\CG_2(\QQ)\nonumber\\
=&\,\frac{1}{\ve}\left(\MP\frac{\partial F_{entropy}}{\partial Q_2}+cc_{04}Q_1+cc_{03}Q_2\right)\nonumber\\
&\,-cc_{24}\Delta Q_{1jk}-cc_{23}\Delta Q_{2jk}
-\MP(cc_{2,10}\partial_j\partial_iQ_{1ik}+cc_{29}\partial_j\partial_iQ_{2ik}).\label{mu-Q2}
\end{align}

The dynamic tensor model is given by
\begin{align}
\frac{\partial\QQ}{\partial t}+\vv\cdot\nabla \QQ=&\,~\CK_{\QQ}+\CW_{\QQ},\label{MB-Q-tensor-1}\\
\rho_s\Big(\frac{\partial\vv}{\partial t}+\vv\cdot\nabla\vv\Big)=&\,-\nabla p+\nabla\cdot\sigma+\FF^e,\label{MB-Q-tensor-2}\\
\nabla\cdot\vv=&\,~0,\label{MB-Q-tensor-3}
\end{align}
where $\rho_s$ is the density of the fluid and $\vv$ the fluid velocity, and $p$ is the pressure to maintain the incompressibility.
Let us denote by $\kappa_{ij}=\partial_jv_i$ the velocity gradient.
The terms $\CK_{\QQ}=(\CK_{Q_1},\CK_{Q_2})$ and $\CW_{\QQ}=(\CW_{Q_1},\CW_{Q_2})$ on the right-hand side of (\ref{MB-Q-tensor-1}) characterize the rotational diffusions and rotational convections, respectively.
They are given by
\begin{align*}
-(\CK_{Q_1})_{kl}=&\,4\Gamma_2(\mu_{Q_1})_{ij}\big\langle\mm_1\mm_3\otimes\mm_1\mm_3\big\rangle_{ijkl}
+4\Gamma_3(\mu_{Q_1}-\mu_{Q_2})_{ij}\big\langle\mm_1\mm_2\otimes\mm_1\mm_2\big\rangle_{ijkl},\\
-(\CK_{Q_2})_{kl}=&\,4\Gamma_1(\mu_{Q_2})_{ij}\big\langle\mm_2\mm_3\otimes\mm_2\mm_3\big\rangle_{ijkl}
-4\Gamma_3(\mu_{Q_1}-\mu_{Q_2})_{ij}\big\langle\mm_1\mm_2\otimes\mm_1\mm_2\big\rangle_{ijkl},\\
(\CW_{Q_1})_{kl}=&\,2\kappa_{ij}\Big(\langle(\mm_1\otimes\mm_3)\otimes\mm_1\mm_3\rangle
+\frac{I_{22}}{I_{11}+I_{22}}\langle(\mm_1\otimes\mm_2)\otimes\mm_1\mm_2\rangle\\
&\,-\frac{I_{11}}{I_{11}+I_{22}}\langle(\mm_2\otimes\mm_1)\otimes\mm_1\mm_2\rangle\Big)_{ijkl},\\
(\CW_{Q_2})_{kl}=&\,2\kappa_{ij}\Big(\langle(\mm_2\otimes\mm_3)\otimes\mm_2\mm_3\rangle
-\frac{I_{22}}{I_{11}+I_{22}}\langle(\mm_1\otimes\mm_2)\otimes\mm_1\mm_2\rangle\\
&\,+\frac{I_{11}}{I_{11}+I_{22}}\langle(\mm_2\otimes\mm_1)\otimes\mm_1\mm_2\rangle\Big)_{ijkl},
\end{align*}
where $\Gamma_i=\frac{m_0}{\zeta I_{ii}}(i=1,2,3)$ are the diffusion coefficients, $m_0$ is the mass of a rigid molecule, $\zeta$ is the friction constant, and $I_{ii}(i=1,2,3)$ are diagonal elements of the moment of inertia for a molecule.

In \eqref{MB-Q-tensor-2}, the stress tensor $\sigma$ consists of the viscous stress $\sigma_v$ and the elastic stress $\sigma_e$. The viscous stress $\sigma_v$ includes the contribution of the fluid itself with a viscous coefficient $\eta$, and the fluid-molecule friction,
\begin{align}\label{visous-stress}
\sigma_v=\frac{1}{2}\eta(\kappa+\kappa^T)+\sigma_{vf}.
\end{align}
The second term $\sigma_{vf}$ is determined by the following equation,
\begin{align*}
(\sigma_{vf})_{ij}=c\zeta\kappa_{kl}\Big(I_{22}\langle\mm^4_1\rangle+I_{11}\langle\mm^4_2\rangle+\frac{4I_{11}I_{22}}{I_{11}+I_{22}}\langle\mm_1\mm_2\otimes\mm_1\mm_2\rangle\Big)_{ijkl}.
\end{align*}
The elastic stress $\sigma_e$ can be written as
\begin{align*}
(\sigma_e)_{kl}=&\,2ck_BT\Big[(\mu_{Q_2})_{ij}\big\langle\mm_2\mm_3\otimes(\mm_2\otimes\mm_3)\big\rangle_{ijkl}
+(\mu_{Q_1})_{ij}\big\langle\mm_1\mm_3\otimes(\mm_1\otimes\mm_3)\big\rangle_{ijkl}\\
&\,+\frac{1}{I_{11}+I_{22}}\Big((\mu_{Q_1}-\mu_{Q_2})_{ij}\big
\langle\mm_1\mm_2\otimes(I_{22}\mm_1\otimes\mm_2-I_{11}\mm_2\otimes\mm_1)\big\rangle_{ijkl}\Big)\Big].
\end{align*}
The external force $\FF^e$ is given by
\begin{align}
\FF^e_i=&\,ck_BT\mu_{\QQ}\cdot\partial_i\QQ\nonumber\\
\eqdefa&\,ck_BT(\mu_{Q_1}\cdot\partial_iQ_1+\mu_{Q_2}\cdot\partial_iQ_2).
\end{align}

The system (\ref{MB-Q-tensor-1})-(\ref{MB-Q-tensor-3}) obeys the following energy dissipation law (see \cite{XZ}):
\begin{align}\label{Q-energy-dis}
&\,\frac{\ud}{\ud t}\Big(\int\ud\xx\frac{\rho_s|\vv|^2}{2}+F(\QQ,\nabla\QQ)\Big)\nonumber\\
&\,\quad=\int\ud\xx\bigg\{-ck_BT\Big[\Gamma_1\big\langle(2\mu_{Q_2}\cdot\mm_2\mm_3)^2\big\rangle+\Gamma_2\big\langle(2\mu_{Q_1}\cdot\mm_1\mm_3)^2\big\rangle\nonumber\\
&\,\qquad+\Gamma_3\Big\langle\big(2(\mu_{Q_1}-\mu_{Q_2})\cdot\mm_1\mm_2\big)^2\Big\rangle\Big]-\eta\frac{\kappa+\kappa^T}{2}\cdot\frac{\kappa+\kappa^T}{2}\nonumber\\
&\,\qquad-c\zeta\Big[I_{22}\big\langle(\kappa\cdot\mm^2_1)^2\big\rangle+I_{11}\big\langle(\kappa\cdot\mm^2_2)^2\big\rangle+\frac{I_{11}I_{22}}{I_{11}+I_{22}}\big\langle(2\kappa\cdot\mm_1\mm_2)^2\big\rangle\Big]\bigg\}.
\end{align}

Note that several fourth-order tensors appear in the dynamic model.
In order to close the system, it is necessary to find certain way to express them by $Q_1$ and $Q_2$.
The closure approximation can be done by the entropy term, which will be introduced below.
Although there might be other ways of closure, one advantage of closure by the entropy term is that it guarantees the non-positiveness of the terms on the right-hand side of \eqref{Q-energy-dis}.

\subsection{Entropy and quasi-entropy}

We have mentioned that the entropy term plays a significant role in both free energy and closure approximation.
A general approach is to deduce the entropy term by minimizing $\int_{SO(3)} \rho\ln\rho\ud\Fp$ with the values of the tensors fixed, or finding the maximum entropy state.
When the two tensors $Q_1$ and $Q_2$ are involved, the maximum entropy state is given by
\begin{align}
    &\,\rho(\Fq)=\frac 1Z \exp(B_1\cdot \mm_1^2+B_2\cdot\mm_2^2), \label{maxentstate}
\end{align}
where the normalizing constant $Z$ and two second-order symmetric traceless tensors $B_1$ and $B_2$ are Lagrange multipliers for constraints,
\begin{align}
    &\,Z=\int_{SO(3)}\exp(B_1\cdot \mm_1^2+B_2\cdot\mm_2^2)\ud\Fq, \nonumber\\
    &\,Q_i=\frac 1 Z \int_{SO(3)}\left(\mm_i^2-\frac 13 \Fi\right)\exp(B_1\cdot \mm_1^2+B_2\cdot\mm_2^2)\ud\Fq. \label{maxent-cnstrt}
\end{align}
Taking \eqref{maxentstate} into $\int_{SO(3)} \rho\ln\rho\ud\Fp$, we obtain
\begin{align}
    &\,F_{\rm entropy}=B_1\cdot Q_1 +B_2\cdot Q_2-\ln Z. \label{orig-ent}
\end{align}
The maximum entropy state \eqref{maxentstate} is unique about $Q_1$ and $Q_2$ \cite{XYZ}.
Therefore, $F_{\rm entropy}$ can be viewed as a function about $Q_1$ and $Q_2$.
It is observed that $F_{\rm entropy}$ is invariant under rotations on $Q_1$ and $Q_2$.
Generally, we say a tensor $U$ is invariant under rotations, if the coordinates $U_{i_1\cdots i_n}$ in \eqref{tensor0} are kept but the basis $(\ee_i)$ is rotated to another frame $(\ee_i')$.
Specifically, a rotation on $Q_1$ and $Q_2$ is done by choosing certain $\Ft\in SO(3)$ and transforming $Q_i$ into $\Ft Q_i\Ft^{-1}$.
It is easy to verify that \eqref{orig-ent} is rotationally invariant under this transformation (see Appendix \ref{Apprendix-B1}).

The closure approximation can be done with the maximum entropy state, as the high-order tensors can be calculated using the density function \eqref{maxentstate}.
An equivalent viewpoint is that when $Q_1$ and $Q_2$ are given, the high-order tensors obtained in this way minimize $\int_{SO(3)} \rho\ln\rho\ud\Fp$.

The entropy term defined from maximum entropy state, which we call the original entropy, is given implicitly that involves integrals on $SO(3)$, which could bring difficulties in both analyses and numerical studies.
An alternative approach is proposed \cite{Xu3}, where the original entropy is substituted with the quasi-entropy.
The quasi-entropy is defined by log-determinant covariance matrix, which is an elementary function about the order parameter tensors.
To write down the expression of the quasi-entropy, it is necessary to specify the highest tensor order (that shall be even) to be involved.
When only second-order tensors are involved, the quasi-entropy for $Q_1$ and $Q_2$ is given by (see discussions in Appendix \ref{Apprendix-B2})
\begin{align}
    F_{\rm entropy}=&\,\Xi_2\big(Q_1,Q_2\big)=\nu\left(-\ln\det \Big(Q_1+\frac{\Fi}{3}\Big)-\ln\det\Big( Q_2+\frac{\Fi}{3}\Big)-\ln\det\Big(\frac{\Fi}{3}-Q_1-Q_2\Big)\right),\nonumber\\
    &\,\text{Domain: }Q_1+\frac{\Fi}{3},~ Q_2+\frac{\Fi}{3},~\frac{\Fi}{3}-Q_1-Q_2\text{ positive definite}. \label{qent-2nd}
\end{align}
A free parameter $\nu$ is introduced above.
It can be estimated as $\nu=5/9$ from special cases (see Section 6 in \cite{Xu3}), which we adopt in the current work.
Moreover, analyses show that the quasi-entropy possesses similar properties with the original entropy.
In particular, the results from the quasi-entropy \eqref{qent-2nd} are very similar to those from the original entropy \eqref{orig-ent}, provided that other terms in the free energy are identical.
These results have all been reported in \cite{Xu3}.

The quasi-entropy is also suitable for closure approximation.
To deduce high-order tensors in the dynamic model, we shall use the log-determinant covariance matrix up to fourth order, denoted by $\Xi_4$ which is provided in \eqref{qent-4th} of Appendix \ref{Apprendix-B2}.
The fourth-order tensors shall minimize $\Xi_4$ with the given values of $Q_1$ and $Q_2$.
Thus, we can see that the closure approximation by the original entropy and the quasi-entropy share the rationale, with the only difference lying in the function to be minimized.
In what follows, we shall see that these two approaches of closure approximation lead to high-order tensors of the same form due to the same symmetry arguments that will be shown in Appendix \ref{Apprendix-B}.

The properties of the quasi-entropy have been discussed previously \cite{Xu3}.
Here, let us state those to be utilized in this paper.
\begin{proposition}\label{qent-property}
The two functions $\Xi_2$ (see \eqref{qent-2nd}) and $\Xi_4$ (see \eqref{qent-4th}) have the following properties:
\begin{itemize}
    \item They are invariant under rotations on the tensors.
    \item They act as barrier functions in the following sense: $\Xi_2$ keeps covariance matrices up to the second order positive definite, while $\Xi_4$ keeps those up to the fourth order positive definite.
    \item They are strictly convex about the tensors.
\end{itemize}
\end{proposition}
As an example, it is straightfoward to see the rotational invariance for $\Xi_2$ by taking the rotation $Q_i\to\Ft Q_i\Ft^{-1}$ into \eqref{qent-2nd}.

The properties stated above are all crucial in our derivation below.
The rotational invariance is a foundation for the frame model to be established.
The positive-definiteness of covariance matrices is essential for energy dissipation to hold.
The strict convexity guarantees that the closure approximation by minimization results in a unique solution.

\subsection{Stationary points of bulk energy}

There are analyses on the stationary points of the bulk energy $F_b$, but they are far from well-understood.
We summarize the main results up to date in the following proposition \cite{XZ2,Xu3}.
To simplify the presentation, the conditions on the coefficients are stricter than they could be.

\begin{proposition}\label{bulk-minimum}
Assume that the matrix
\begin{align*}
\left(\begin{array}{cc}
    c_{02} &\, c_{04} \vspace{1ex}\\
    c_{04} &\, c_{03}
\end{array}\right)
\end{align*}
is not negative definite, or is negative but $c_{04}^2/c_{03}-c_{02}\le 2$.
No matter the entropy term takes \eqref{orig-ent} or \eqref{qent-2nd} (with $\nu=5/9$), at the stationary points $Q_1$ and $Q_2$ have a shared eigenframe.
\end{proposition}

When $Q_1$ and $Q_2$ has the same eigenframe, they can be written as
\begin{align}\label{Qi-biaxial}
Q_i=s_i\Big(\nn^2_1-\frac{\Fi}{3}\Big)+b_i(\nn^2_2-\nn^2_3),\quad i=1,2.
\end{align}
Numerical studies indicate that the global energy minimum could be either uniaxial (where $b_i=0$) or biaxial (where at least one $b_i\ne 0$).
Here, we assume that under certain coefficients $c_{02}$, $c_{03}$, and $c_{04}$, we have a biaxial global minimum $\QQ^{(0)}=(Q_1^{(0)},Q_2^{(0)})$ of the form \eqref{Qi-biaxial}.

It shall be noticed that the bulk energy $F_b$ is rotationally invariant, i.e. invariant of  $\Fp=(\nn_1,\nn_2,\nn_3)$.
This can be observed by combining Proposition \ref{bulk-minimum} and the fact that the three $c_{0i}$ terms are rotationally invariant.
Thus, a rotation of an energy minimum also results in an energy minimum.

At any energy minimum, we have $\CJ(\QQ^{(0)})=0$.
Let us fix $s_i$ and $b_i$ and let $\Fp=(\nn_1,\nn_2,\nn_3)$ vary, so that $\QQ^{(0)}=\QQ^{(0)}(\Fp)$ becomes a function of $\Fp$.
Since $\QQ^{(0)}$ is an energy minimum whatever $\Fp$ is, it implies that $\CJ\big(\QQ^{(0)}(\Fp)\big)=0$.
We then impose the operators $\ML_i$ on it.
By the chain rule, we obtain
\begin{align}
  \ML_{m}\CJ(\QQ^{(0)})_{ij}=\CJ'(\QQ^{(0)})_{ijkl}(\ML_{m}\QQ^{(0)})_{kl}=0,~m=1,2,3. \label{kernel}
\end{align}
This implies that the kernel of the Hessian $\CJ'(\QQ^{(0)})$ contains the space spanned by $\ML_m\QQ^{(0)}$.

With the form \eqref{Qi-biaxial}, the scalars $s_i$ and $b_i$ shall satisfy
\begin{align}
    \frac 23 s_i+\frac 13>0,\quad \frac 13-\frac 13 s_i\pm b_i>0,\qquad i=1,2,3, \label{range-2nd}
\end{align}
where we define $s_3=-s_1-s_2$ and $b_3=-b_1-b_2$.
This is exactly the range such that $F_{\rm entropy}$ is well-defined, no matter it takes the original entropy \eqref{orig-ent} or the quasi-entropy \eqref{qent-2nd}:
for the original entropy, the derivation can be found in \cite{XYZ};
for the quasi-entropy $\Xi_2$, the condition \eqref{range-2nd} is equivalent to the requirement that $Q_1+\Fi/3$, $Q_2+\Fi/3$ and $-Q_1-Q_2+\Fi/3$ are positive definite.

\subsection{High-order tensors and their symmetry}

We have mentioned that the high-order tensors in the dynamic model are determined from clousre approximation.
However, there are many linear relations between these high-order tensors.
It is necessary to specify their linearly independent components, which can be done with the help of symmetric traceless tensors.
The use of symmetric traceless tensors turn out to be crucial to figuring out symmetry arguments for these high-order tensors.

For the high-order tensors appearing in the dynamic tensor model, only the tensors below are involved other than $Q_1$ and $Q_2$,
\begin{align}
    \langle\mm_1\mm_2\mm_3\rangle,\ \langle(\mm_1^4)_0\rangle,\ \langle(\mm_2^4)_0\rangle,\ \langle(\mm_1^2\mm_2^2)_0\rangle. \label{symtrls-3rd-4th}
\end{align}
Here, we recall the notation $(U)_0$ in \eqref{symtrlsnot} for the symmetric traceless tensor generated by $U$.
The explicit expressions of these tensors are given in Appendix \ref{Apprendix-A1}.
Furthermore, in Appendix \ref{symtrls-decomp} we provide the explicit expressions of the fourth-order tensors in the dynamic model by the above four tensors together with $Q_1$ and $Q_2$.

Therefore, in closure approximation, our task is to determine the third-order and fourth-order tensors in \eqref{symtrls-3rd-4th}.
In particular, when $Q_1$ and $Q_2$ have the form \eqref{Qi-biaxial}, the tensors in \eqref{symtrls-3rd-4th} have the form indicated by the following theorem.

\begin{theorem}\label{Q-biaxial-theorem}
If $Q_1$ and $Q_2$ are biaxial of the form \eqref{Qi-biaxial}, then the third- and fourth-order symmetric traceless tensors, solved from closure approximation by the original entropy or the quasi-entropy, take the form
\begin{align}
  &\,\langle\mm_1\mm_2\mm_3\rangle=z\nn_1\nn_2\nn_3,\nonumber\\
  &\,\langle (\mm_1^4)_0\rangle=a_1(\nn_1^4)_0+a_2(\nn_2^4)_0+a_3(\nn_1^2\nn_2^2)_0,\nonumber\\
  &\,\langle (\mm_2^4)_0\rangle=\tilde{a}_1(\nn_1^4)_0+\tilde{a}_2(\nn_2^4)_0+\tilde{a}_3(\nn_1^2\nn_2^2)_0,\nonumber\\
  &\,\langle (\mm_1^2\mm_2^2)_0\rangle=\bar{a}_1(\nn_1^4)_0+\bar{a}_2(\nn_2^4)_0+\bar{a}_3(\nn_1^2\nn_2^2)_0. \label{biaxial_high}
\end{align}
The scalars $z$, $a_i$, $\tilde{a}_i$, $\bar{a}_i$ are solved as functions of $s_i$ and $b_i$.
Furthermore, if $s_i$ and $b_i$ satisfy \eqref{range-2nd}, these scalars can be uniquely solved.
\end{theorem}
The proof is left to Appendix \ref{Apprendix-B}.
This result actually determines the form of high-order tensors in the Hilbert expansion, which in turn makes a great difference in finding out the form of the frame hydrodynamics for the biaxial nematic phase.

\section{From tensor model to orthonormal frame model}\label{frame-model}

We make the Hilbert expansion (also called the Chapman--Enskog expansion) of solutions to the
with respect to the small parameter $\ve$.
The $O(1)$ system results in the orthonormal frame model for the biaxial nematic phase, with the energy dissipation maintained.
The coefficients in the frame model are derived from those in the tensor model.
Since the coefficients in the tensor model are derived from physical parameters, we finally build the relation between the frame model and the physical parameters.

For convenience,
we denote seven fourth-order tensor moments as follows:
\begin{align}\label{seven-tensors}
\left\{
\begin{array}{l}
\CR_1=\langle(\mm^2_1-\frac{\Fi}{3})\otimes(\mm^2_1-\frac{\Fi}{3})\rangle,\quad
\CR_2=\langle(\mm^2_2-\frac{\Fi}{3})\otimes(\mm^2_2-\frac{\Fi}{3})\rangle, \vspace{1ex} \\
\CR_3=4\langle\mm_1\mm_2\otimes\mm_1\mm_2\rangle,\quad
\CR_4=4\langle\mm_1\mm_3\otimes\mm_1\mm_3\rangle,\quad
\CR_5=4\langle\mm_2\mm_3\otimes\mm_2\mm_3\rangle, \vspace{1ex} \\
\CV_{Q_1}=2\Big(\langle\mm_1\mm_3\otimes(\mm_1\otimes\mm_3)\rangle+e_1\langle\mm_1\mm_2\otimes(\mm_1\otimes\mm_2)\rangle-e_2\langle\mm_1\mm_2\otimes(\mm_2\otimes\mm_1)\rangle\Big), \vspace{1ex} \\
\CV_{Q_2}=2\Big(\langle\mm_2\mm_3\otimes(\mm_2\otimes\mm_3)\rangle-e_1\langle\mm_1\mm_2\otimes(\mm_1\otimes\mm_2)\rangle+e_2\langle\mm_1\mm_2\otimes(\mm_2\otimes\mm_1)\rangle\Big),
\end{array}
\right.
\end{align}
where the coefficients $e_i(i=1,2)$ are expressed by
\begin{align*}
e_1=1-e_2=\frac{I_{22}}{I_{11}+I_{22}}.
\end{align*}

We frequently deal with contractions between fourth-order tensors and second-order tensors.
We could regard a fourth-order tensor as a matrix, and a second-order tensor as a vector, so that the contractions can be formulated as matrix-matrix and matrix-vector multiplications, as we explain below.
When a fourth-order tensor is contracted with a second-order tensor, we could write it in short as a matrix-vector product, say
\begin{align}
    (\CV_{Q_1})_{ijkl}\kappa_{kl}=(\CV_1\kappa)_{ij}. \label{4-2-product}
\end{align}
When using this short notation, we always assume that the second last index of the fourth-order tensor is contracted with the first of the second-order tensor, and the last of fourth-order tensor is contracted with the last of the second-order tensor.
By the convention \eqref{4-2-product}, we could define the transpose of a fourth-order tensor, such as
$$
(\CV_{Q_1}^T)_{ijkl}=(\CV_{Q_1})_{klij}.
$$

Let us define
\begin{align}
&\,\CM=\left(
  \begin{array}{cc}
    \CM_{11} &\, \CM_{12}\vspace{1ex} \\
    \CM_{12} &\, \CM_{22} \\
  \end{array}
\right)\eqdefa
\left(
  \begin{array}{cc}
    \Gamma_2\CR_4+\Gamma_3\CR_3 &\, -\Gamma_3\CR_3 \vspace{1ex}\\
    -\Gamma_3\CR_3 &\, \Gamma_1\CR_5+\Gamma_3\CR_3 \\
  \end{array}
\right), \label{diss-operator}\\
&\,\CV\eqdefa\left(\begin{array}{c}
    \CV_{Q_1}  \vspace{1ex}\\
    \CV_{Q_2}
\end{array}\right),
\quad
\CN\eqdefa(\CN_{Q_1},\CN_{Q_2})=(\CV^{T}_{Q_1}, \CV^T_{Q_2}),\label{CVQ-CNQ}\\
&\,\CP\eqdefa c\zeta\big(I_{22}\CR_{1}+I_{11}\CR_2+e_1I_{11}\CR_{3}\big). \label{CPQ}
\end{align}

The system (\ref{MB-Q-tensor-1})-(\ref{MB-Q-tensor-3}) can then be rewritten as
\begin{align}
\frac{\partial\QQ}{\partial t}+\vv\cdot\nabla\QQ=&\,-\CM\mu_{\QQ}+\CV\kappa,\label{Re-MB-Q-tensor-1}\\
\rho_s\Big(\frac{\partial\vv}{\partial t}+\vv\cdot\nabla\vv\Big)_i=&\,-\partial_ip+\eta\Delta v_i+\partial_j(\CP\kappa)_{ij}+ck_BT\partial_j(\CN\mu_{\QQ})_{ij}+ck_BT\mu_{\QQ}\cdot\partial_i\QQ,\label{Re-MB-Q-tensor-2}\\
\nabla\cdot\vv=&\,0, \label{Re-MB-Q-tensor-3}
\end{align}
where $\CM\mu_{\QQ}$ is carried out by matrix-vector multiplication,
\begin{align*}
    \CM\mu_{\QQ}=\left(
    \begin{array}{c}
        \CM_{11}\mu_{Q_1}+\CM_{12}\mu_{Q_2} \vspace{1ex}\\
        \CM_{12}\mu_{Q_1}+\CM_{22}\mu_{Q_2}
    \end{array}\right).
\end{align*}
Similarly are the terms involving $\CV$ and $\CN$ interpreted.
In the above, we have incorporated some simple calculations for the viscous stress, such as
\begin{align}\label{mm1-4-cdot-kappa}
    \langle\mm^4_1\rangle \kappa=&\,\big(\CR_1+Q_1\otimes\Fi+\Fi\otimes Q_1+\Fi\otimes\Fi\big)\kappa\nonumber\\
    =&\,\CR_1\kappa+(Q_1\cdot\kappa)\Fi,
\end{align}
because the incompressibility can also be written as $\Fi\cdot\kappa=0$.
Furthermore, the second term in (\ref{mm1-4-cdot-kappa}) can be merged into the pressure $p$, so that only the term $\CR_1\kappa$ remains in the operator $\CP$.

The fourth-order tensors $\CR_i(i=1,\cdots,5)$ are positive definite in the sense that for any second-order symmetric traceless tensor $Y$, we have $Y\cdot \CR_i Y\ge 0$ and the equality implies $Y=0$.
This result comes from the property of the entropy term, which we will show in Appendix \ref{Apprendix-B}.
Consequently, we deduce that for arbitrary second-order symmetric traceless tensors $Y_1$ and $Y_2$, it holds
\begin{align}
    Y_1\cdot \CP Y_1=&\,c\zeta \Big(I_{22} Y_1\cdot\CR_1 Y_1+I_{11}Y_1\cdot\CR_2 Y_1+e_1I_{11}Y_1\cdot\CR_3 Y_1\Big)\ge 0, \label{P-spd}\\
    (Y_1,Y_2)\CM\left(\begin{array}{c}
        Y_1 \\
        Y_2
    \end{array}\right)
    =&\,\Gamma_1 Y_2\cdot\CR_3 Y_2+\Gamma_2 Y_1\cdot\CR_4 Y_1 + \Gamma_1 (Y_1-Y_2)\cdot\CR_3 (Y_1-Y_2)\ge 0. \label{M-spd}
\end{align}
The equality in \eqref{M-spd} leads to $Y_1=Y_2=0$, so that $\CM (Y_1,Y_2)^T=0$ implies $Y_1=Y_2=0$.

\subsection{The Hilbert expansion}
Assume that $\big(\QQ(t,\xx),\vv(t,\xx)\big)$ is a solution to the molecular-theory-based $Q$-tensor system (\ref{Re-MB-Q-tensor-1})-(\ref{Re-MB-Q-tensor-3}).
We perform the following Hilbert expansion about $\ve$:
\begin{align}
\QQ(t,\xx)=&\,\QQ^{(0)}(t,\xx)+\ve\QQ^{(1)}(t,\xx)+\ve^2\QQ^{(2)}(t,\xx)+\cdots, \label{Q-expansion}\\
\vv(t,\xx)=&\,\vv^{(0)}(t,\xx)+\ve\vv^{(1)}(t,\xx)+\ve^2\vv^{(2)}(t,\xx)+\cdots, \label{v-expansion}
\end{align}
where $\QQ^{(i)}=(Q^{(i)}_1, Q^{(i)}_2)^T$, and $(\QQ^{(i)},\vv^{(i)})(i=0,1,2,\cdots)$ are independent of the small parameter $\ve$.

Based on the expansion \eqref{Q-expansion}--\eqref{v-expansion}, we could write down the expansion of other terms in (\ref{Re-MB-Q-tensor-1})-(\ref{Re-MB-Q-tensor-3}), frequently by Taylor expansion.
Since we focus on the $O(1)$ system, we only write down the terms up to $O(1)$.
In $\mu_{\QQ}$, the term $\CJ(\QQ)=\MP\frac{\partial f_b(\QQ)}{\partial \QQ}$ is multiplied by $\ve^{-1}$, so we need to expand it to $O(\epsilon)$,
\begin{align*}
\CJ(\QQ)=\CJ(\QQ^{(0)})+\ve\CJ'\big(\QQ^{(0)}\big)\QQ^{(1)}+O(\ve^2),
\end{align*}
where $\CJ'(\QQ^{(0)})\eqdefa \CH^{(0)}$ is a fourth-order tensor.
By (\ref{vari-deriv-Q}) we can deduce that
\begin{align}
\mu_{\QQ}=&\,\ve^{-1}\CJ(\QQ)+\CG(\QQ)\nonumber\\
=&\,\ve^{-1}\CJ(\QQ^{(0)})+\CH^{(0)}\QQ^{(1)}+\CG(\QQ^{(0)})+O(\ve).
\end{align}
Since the tensors in \eqref{seven-tensors} are solved from closure approximation, they are functions of $\QQ$.
Thus, $\CM$, $\CV$, $\CN$ and $\CP$ are functions of $\QQ$.
Let us use the notation $\CM^{(0)}$ for the $\CM$ when $\QQ$ takes $\QQ^{(0)}$.
Then we have
\begin{align*}
    \CM=&\,\CM^{(0)}+O(\ve), \quad \CV=\CV^{(0)}+O(\ve),\\\CN=&\,\CN^{(0)}+O(\ve)=\CV^{(0)T}+O(\ve), \quad \CP=\CP^{(0)}+O(\ve),
\end{align*}
where
\begin{align}
    \CM^{(0)}=&\,\left(
      \begin{array}{cc}
        \CM_{11}^{(0)} &\, \CM_{12}^{(0)} \vspace{1ex}\\
        \CM_{12}^{(0)} &\, \CM_{22}^{(0)} \\
     \end{array}
    \right)=\left(
      \begin{array}{cc}
        \Gamma_2\CR^{(0)}_4+\Gamma_3\CR^{(0)}_3 &\, -\Gamma_3\CR^{(0)}_3 \vspace{1ex}\\
        -\Gamma_3\CR^{(0)}_3 &\, \Gamma_1\CR^{(0)}_5+\Gamma_3\CR^{(0)}_3 \\
     \end{array}
    \right), \label{dissip-operator}\\
    \CV^{(0)}=&\,\left(\begin{array}{c}
        \CV^{(0)}_{Q_1} \vspace{1ex}\\
        \CV^{(0)}_{Q_2}
    \end{array}\right),\quad
    \CN^{(0)}=(\CN_{Q_1}^{(0)}, \CN_{Q_2}^{(0)})=(\CV^{(0)T}_{Q_1}, \CV^{(0)T}_{Q_2}),\label{V0+N0}\\
    \CP^{(0)}=&\,c\zeta\big(I_{22}\CR_{1}^{(0)}+I_{11}\CR^{(0)}_2+e_1I_{11}\CR_{3}^{(0)}\big). \label{P0}
\end{align}

Substituting the above expansion (\ref{Q-expansion}) and (\ref{v-expansion}) into the system (\ref{Re-MB-Q-tensor-1})-(\ref{Re-MB-Q-tensor-3}) and collecting the terms
with the same order of $\ve$, we can obtain a series of equations.
The $O(\ve^{-1})$ system requires that
\begin{align}\label{ve-1-system}
\CM^{(0)}\CJ(\QQ^{(0)})=0.
\end{align}
Since $\CM^{(0)}$ is positive definite, the above equation implies that
\begin{align*}
\CJ(\QQ^{(0)})=0.
\end{align*}
It means that $\QQ^{(0)}$ is just the critical point of $F_b(\QQ)$.
We shall consider the case that $\QQ^{(0)}$ is the biaxial global minimum, which takes the form \eqref{Qi-biaxial}.

The terms of the order $O(1)$ give
\begin{align}
\frac{\partial\QQ^{(0)}}{\partial t}+\vv^{(0)}\cdot\nabla \QQ^{(0)}
=&\,-\CM^{(0)}\big(\CH^{(0)}\QQ^{(1)}+\CG(\QQ^{(0)})\big)+\CV^{(0)}\kappa^{(0)},\label{ve1-QQ}\\
\rho_s\Big(\frac{\partial\vv^{(0)}}{\partial t}+\vv^{(0)}\cdot\nabla\vv^{(0)}\Big)_i
=&\,-\partial_ip^{(0)}+\eta\Delta v^{(0)}_i
+\partial_j(\CP^{(0)}\kappa^{(0)})_{ij}\nonumber\\
&\,+ck_BT\partial_j\Big(\CN^{(0)}\big(\CH^{(0)}\QQ^{(1)}+\CG(\QQ^{(0)})\big)_{ij}\Big)\nonumber\\
&\,+ck_BT\big(\CH^{(0)}\QQ^{(1)}+\CG(\QQ^{(0)})\big)\cdot\partial_i\QQ^{(0)},\label{ve2-v}\\
\nabla\cdot\vv^{(0)}=&\,~0.\label{ve3-impress}
\end{align}
In the $O(1)$ system \eqref{ve1-QQ}--\eqref{ve3-impress}, $\QQ^{(0)}$ is a function about $\Fp=(\nn_1,\nn_2,\nn_3)$.
The high-order tensors with the superscript $(0)$ are functions of $\QQ^{(0)}$, thus are functions of $\Fp$.
Therefore, if we could elminate $\QQ^{(1)}$ in the $O(1)$ system, we would arrive at a system about $\Fp$ and $\vv^{(0)}$.
Indeed, this can be done by examining  the kernel of $\CH^{(0)}$.

Our task becomes expressing terms with the superscript $(0)$ in terms of $\nn_1$, $\nn_2$, $\nn_3$.
It turns out that the form \eqref{Qi-biaxial} of $\QQ^{(0)}$ results in specific form of the following terms.
\begin{itemize}
    \item The derivatives of $\QQ^{(0)}$, which are related to the kernel of $\CH^{(0)}$.
    \item The variational derivative of the elastic energy, $\CG(\QQ^{(0)})$.
    \item The high-order tensors $\CM^{(0)}$, $\CV^{(0)}$, $\CN^{(0)}$ and $\CP^{(0)}$.
\end{itemize}

Up to now, all the equations are expressed by the components in the basis generated by the reference frame $(\ee_1,\ee_2,\ee_3)$.
In order to facilitate the specific form of the above terms, we shall first rewrite the $O(1)$ system in the basis generated by $\Fp=(\nn_1,\nn_2,\nn_3)$.

\subsection{Change to the local basis}\label{change-basis}
In what follows, we denote by $\A_0$ and $\BOm_0$ the
symmetric and skew-symmetric parts of the velocity gradient $\kappa^{(0)}_{ij}=\partial_jv^{(0)}_i$, respectively, i.e.,
\begin{align*}
\A_0=\frac{1}{2}(\kappa^{(0)}+\kappa^{(0)T}),\quad\BOm_0=\frac{1}{2}(\kappa^{(0)}-\kappa^{(0)T}).
\end{align*}

We consider the basis for second-order tensors given by $\Fi$, five symmetric traceless tensors,
\begin{align}\label{sss-five}
    \sss_1=\nn^2_1-\frac13\Fi,\quad \sss_2=\nn^2_2-\nn^2_3,\quad \sss_3=\nn_1\nn_2,\quad \sss_4=\nn_1\nn_3,\quad \sss_5=\nn_2\nn_3,
\end{align}
and three asymmetric traceless tensors,
\begin{align}\label{aaa-three}
    \aaa_1=\nn_1\otimes\nn_2-\nn_2\otimes\nn_1,\quad \aaa_2=\nn_3\otimes\nn_1-\nn_1\otimes\nn_3,\quad \aaa_3=\nn_2\otimes\nn_3-\nn_3\otimes\nn_2.
\end{align}

Let us look into the fourth-order tensors $\CM_{11}^{(0)},\,\CM_{12}^{(0)},\,\CM_{22}^{(0)}$. The first two components of $\CM_{11}^{(0)}$ are symmetric, and the contraction of the first two components gives a zero second-order tensor. So is the contraction of its last two components. Thus, it can be expressed as
\begin{align}
    \CM_{11}^{(0)}=(M_{11})_{ij}\sss_i\otimes\sss_j.
\end{align}
Similarly, $M_{12}$ and $M_{22}$ are also defined.

When the last two indices of $\CM_{11}^{(0)}$ are contracted with a second-order symmetric traceless tensor $Y=y_i\sss_i$, it gives
\begin{align}
    \CM_{11}^{(0)}Y=\big((M_{11})_{ij}y_k\big)(\sss_i\otimes\sss_j)\sss_k.
\end{align}
By the convention of the a fourth-order tensor times a second-order tensor \eqref{4-2-product}, the product $(\sss_i\otimes\sss_j)\sss_k$ gives a second-order tensor $(\sss_j\cdot\sss_k)\sss_i$.
Let us define a matrix $\Lambda$ by
\begin{align}
    \Lambda_{ij}=\sss_i\cdot\sss_j,
\end{align}
which equals
\begin{align}
    \Lambda=\mathrm{diag}\Big(\frac 23,2,\frac 12, \frac 12, \frac 12\Big). \label{Lambda}
\end{align}
So, $\CM_{11}^{(0)}Y$ is written as
\begin{align}
    \CM_{11}^{(0)}Y=\big((M_{11})_{ij}\Lambda_{jk}y_k\big)\sss_i=(M_{11}\Lambda y)_i\sss_i.
\end{align}
In other words, the coordinates of $\CM_{11}^{(0)}Y$ under the basis $\sss_i$ is given by $M_{11}\Lambda y$.

For a product involving $\CM^{(0)}$, we just combine the coordinates in a single vector.
For instance, for the term $\CM^{(0)}\CG(\QQ^{(0)})$, let us denote
\begin{align}
M=\left(\begin{array}{cc}
    M_{11} &\, M_{12} \vspace{1ex}\\
    M_{12} &\, M_{22}
\end{array}
\right), \quad
g=\left(\begin{array}{c}
    g_1 \vspace{0.5ex}\\
    g_2
\end{array}\right),
\quad \widetilde{\Lambda}=\left(\begin{array}{cc}
    \Lambda &\,  \vspace{0.5ex}\\
     &\, \Lambda
\end{array}\right), \label{Mcord}
\end{align}
where $g_1$ is the vector of the coordinates of $\CG_1(\QQ^{(0)})$, and $g_2$ that of $\CG_2(\QQ^{(0)})$, i.e. $\CG_1(\QQ^{(0)})=(g_1)_i\sss_i$, $\CG_2(\QQ^{(0)})=(g_2)_i\sss_i$.
Then, the term $\CM^{(0)}\CG(\QQ^{(0)})$ has the coordinates
\begin{align*}
M\widetilde{\Lambda}g=\left(\begin{array}{c}
    M_{11}\Lambda g_1+M_{12}\Lambda g_2 \vspace{0.5ex}\\
    M_{12}\Lambda g_1+M_{22}\Lambda g_2
\end{array}\right).
\end{align*}

We turn to the tensors $\CN_{Q_1}^{(0)},\,\CN_{Q_2}^{(0)}$.
For $\CN_{Q_1}^{(0)}$, its first two components are no longer symmetric, so that we can express it as
\begin{align}
    \CN_{Q_1}^{(0)}=&\,(N^u_1)_{ij}\sss_i\otimes\sss_j+(N^l_1)_{ij}\aaa_i\otimes\sss_j.
\end{align}
The matrices $N_2^u$ and $N_2^l$ are defined in the same way.
By $\CV^{(0)}=\CN^{(0)T}$, we can further write
\begin{align}
    \CV_{Q_1}^{(0)}=&\,(N_1^u)_{ij}\sss_j\otimes\sss_i+(N_1^l)_{ij}\sss_j\otimes\aaa_i.
\end{align}
Denote
\begin{align}\label{N-and-V}
    N=\left(\begin{array}{cc}
        N_1^u &\, N_2^u\vspace{1ex}\\
        N_1^l &\, N_2^l
    \end{array}
    \right), \quad
    V=N^T,
\end{align}
where the matrix $N$ has the size $8\times 10$, so that $V$ is $10\times 8$.
We have
\begin{align}
    \CV^{(0)}\kappa^{(0)}=\left(
    \begin{array}{c}
        \Big((N_1^u)_{ij}\sss_j\otimes\sss_i+(N_1^l)_{ij}\sss_j\otimes\aaa_i\Big)\kappa^{(0)} \vspace{1ex}\\
        \Big((N_2^u)_{ij}\sss_j\otimes\sss_i+(N_2^l)_{ij}\sss_j\otimes\aaa_i\Big)\kappa^{(0)}
    \end{array}
    \right).
\end{align}
We define an $8\times1$ vector $\omega$ by the contraction of $\kappa^{(0)}$ and the vector $\uu=(\sss_1,\cdots,\sss_5,\aaa_1,\aaa_2,\aaa_3)^T$ formed by eight tensors, which is given by
\begin{align*}
\omega=(\omega_s^T,\omega_a^T)^T,
\end{align*}
where $\omega_s$ and  $\omega_a$ are given by
\begin{align*}
\omega_s=(\A_0\cdot\sss_1,\cdots,\A_0\cdot\sss_5)^T,\quad\omega_a=(\BOm_0\cdot\aaa_1,\BOm_0\cdot\aaa_2,\BOm_0\cdot\aaa_3)^T.
\end{align*}
Then, the $10\times 1$ vector $V\omega$ gives the coordinates of $\CV^{(0)}\kappa^{(0)}$.

Similar to the vector $g$, we denote by $\bar{q}$ the coordinates of $\partial_t\QQ^{(0)}$, by $\tilde{q}_i$ the coordinates of $\partial_i\QQ^{(0)}$, and by $h$ the coordinates of $\CH^{(0)}\QQ^{(1)}$.
Then, the coordinates of the material derivative $\dot{\QQ}^{(0)}=\partial_t\QQ^{(0)}+\vv^{(0)}\cdot\nabla\QQ^{(0)}$ are given by $q=\bar{q}+v_1^{(0)}\tilde{q}_1+v_2^{(0)}\tilde{q}_2+v_3^{(0)}\tilde{q}_3$.
Therefore, we could write \eqref{ve1-QQ} in the coordinates,
\begin{align}
    q-V\omega=-M\widetilde{\Lambda}(h+g). \label{Mv1-QQ}
\end{align}

For \eqref{ve2-v}, the term $\CN^{(0)}\big(\CH^{(0)}\QQ^{(1)}+\CG(\QQ^{(0)})\big)$ can be expressed under the basis $\sss_i$ together with $\aaa_i$,
\begin{align}
    \sigma_e^{(0)}=&ck_BT\CN^{(0)}\big(\CH^{(0)}\QQ^{(1)}+\CG(\QQ^{(0)})\big)\nonumber\\
    =&ck_BT(\sss_1,\cdots,\sss_5,\aaa_1,\aaa_2,\aaa_3)N\widetilde{\Lambda}(h+g).
\end{align}
The term $\CP^{(0)}\kappa^{(0)}$ is symmetric traceless, so that it can be written as
\begin{align}
    \CP^{(0)}\kappa^{(0)}=(\sss_1,\cdots,\sss_5)P\omega_s.
\end{align}
The dot product $\big(\CH^{(0)}(\QQ^{(1)})+\CG(\QQ^{(0)})\big)\cdot\partial_i\QQ^{(0)}$ is given by $\tilde{q}_i^T\widetilde{\Lambda} (h+g)$.
Thus, \eqref{ve2-v} can be rewritten as
\begin{align}
\rho_s\Big(\frac{\partial\vv^{(0)}}{\partial t}+\vv^{(0)}\cdot\nabla\vv^{(0)}\Big)_i
=&\,-\partial_i p^{(0)}+\eta\Delta v^{(0)}_i
+\partial_j\Big((\sss_1,\cdots,\sss_5)P\omega_s\Big)_{ij}\nonumber\\
&\,+ck_BT\partial_j\Big((\sss_1,\cdots,\sss_5,\aaa_1,\aaa_2,\aaa_3)N\widetilde{\Lambda}(h+g)\Big)_{ij}\nonumber\\
&\,+ck_BT\tilde{q}_i^T\widetilde{\Lambda} (h+g).\label{Mv2-v}
\end{align}

\subsection{Expressions of matrices and vectors under the local basis}\label{matvec-local}
We begin to write down the matrices and vectors in \eqref{Mv1-QQ} and \eqref{Mv2-v}.
Let us first discuss the derivatives of $\QQ^{(0)}$, i.e. $\bar{q}$ and $\tilde{q}_i$.
Since $\QQ^{(0)}$ is a function of $\Fp$, any derivative of $\QQ^{(0)}$ can be expressed linearly by $\ML_i\QQ^{(0)}$.
For this reason, let us first look into the coordinates of $\ML_i\QQ^{(0)}$.

Using the relation \eqref{ndev}
and from (\ref{Qi-biaxial}) we
obtain
\begin{align*}
\ML_jQ^{(0)}_i=&\,s_i(\ML_j\nn_1\otimes\nn_1+\nn_1\otimes\ML_j\nn_1)\\
&\,+b_i\big(\ML_j\nn_2\otimes\nn_2+\nn_2\otimes\ML_j\nn_2-(\ML_j\nn_3\otimes\nn_3+\nn_3\otimes\ML_j\nn_3)\big),\quad i=1,2;~j=1,2,3,
\end{align*}
which implies that
\begin{align*}
\ML_1\QQ^{(0)}=&\,\left(\begin{array}{c}
    4b_1\nn_2\nn_3
    \vspace{0.5ex}\\
    4b_2\nn_2\nn_3
    \end{array}\right),\\
\ML_2\QQ^{(0)}=&\,-\left(\begin{array}{c}
    2(s_1+b_1)\nn_1\nn_3
    \vspace{0.5ex}\\
    2(s_2+b_2)\nn_1\nn_3
    \end{array}\right),\\
\ML_3\QQ^{(0)}=&\,\left(\begin{array}{c}
    2(s_1-b_1)\nn_1\nn_2
    \vspace{0.5ex}\\
    2(s_2-b_2)\nn_1\nn_2
    \end{array}\right).
\end{align*}
We denote by $L$ the coordinates of $\big(\ML_3 \QQ^{(0)},\ML_2 \QQ^{(0)},\ML_1 \QQ^{(0)}\big)$, which is a $10\times 3$ matrix.
The calculation above gives
\begin{align}
L=\left(
  \begin{array}{c}
    0_{2\times 3} \vspace{1ex}\\
    \mathrm{diag}\big(2(s_1-b_1),-2(s_1+b_1),4b_1\big)\vspace{1ex}\\
    0_{2\times 3} \vspace{1ex}\\
    \mathrm{diag}\big(2(s_2-b_2),-2(s_2+b_2),4b_2\big)
  \end{array}
\right).\label{L-matrix}
\end{align}

For any differential operator $\CD$, we have
\begin{align*}
    \CD \nn_1=(\CD \nn_1\cdot \nn_2)\nn_2+(\CD \nn_1\cdot \nn_3)\nn_3, \\
    \CD \nn_2=(\CD \nn_2\cdot \nn_1)\nn_1+(\CD \nn_2\cdot \nn_3)\nn_3, \\
    \CD \nn_3=(\CD \nn_3\cdot \nn_1)\nn_1+(\CD \nn_3\cdot \nn_2)\nn_2,
\end{align*}
and
\begin{align*}
    0=\CD(\nn_1\cdot\nn_2)=\CD \nn_1\cdot \nn_2+\CD \nn_2\cdot \nn_1.
\end{align*}
By the biaxial form (\ref{Qi-biaxial}) of $\QQ^{(0)}$, we have
\begin{align*}
    \CD Q^{(0)}_1=&\,(s_1+b_1)(\CD\nn_1\otimes\nn_1+\nn_1\otimes\CD\nn_1)
    +2b_1(\CD\nn_2\otimes\nn_2+\nn_2\otimes\CD\nn_2)\nonumber\\
    =&\,2(s_1+b_1)\Big((\CD\nn_1\cdot\nn_2)\nn_1\nn_2+(\CD\nn_1\cdot\nn_3)\nn_1\nn_3\Big)+4b_1\Big((\CD\nn_2\cdot\nn_1)\nn_1\nn_2+(\CD\nn_2\cdot\nn_3)\nn_2\nn_3\Big)\nonumber\\
    =&\,2(s_1+b_1)\Big((\CD\nn_1\cdot\nn_2)\sss_3+(\CD\nn_1\cdot\nn_3)\sss_4\Big)
    +4b_1\Big((\CD\nn_2\cdot\nn_1)\sss_3+(\CD\nn_2\cdot\nn_3)\sss_5\Big)\nonumber\\
    =&\,2(s_1-b_1)(\CD\nn_1\cdot\nn_2)\sss_3-2(s_1+b_1)(\CD\nn_3\cdot\nn_1)\sss_4
    +4b_1(\CD\nn_2\cdot\nn_3)\sss_5.
\end{align*}
Similarly is $\CD Q^{(0)}_2$ calculated.
Choose $\CD$ as $\partial_t$, $\partial_i$ and the material derivative $\partial_t+v^{(0)}_i\partial_i$, respectively.
Their coordinates are given by
\begin{align}
    &\,\bar{q}=L\left(\begin{array}{c}
        \partial_t{\nn}_1\cdot\nn_2 \vspace{0.5ex}\\
        \partial_t{\nn}_3\cdot\nn_1 \vspace{0.5ex}\\
        \partial_t{\nn}_2\cdot\nn_3
    \end{array}\right),\quad
    \tilde{q}_i=L\left(\begin{array}{c}
        \partial_i{\nn}_1\cdot\nn_2 \vspace{0.5ex}\\
        \partial_i{\nn}_3\cdot\nn_1 \vspace{0.5ex}\\
        \partial_i{\nn}_2\cdot\nn_3
    \end{array}\right),\quad
    q=L\left(\begin{array}{c}
        \dot{\nn}_1\cdot\nn_2 \vspace{0.5ex}\\
        \dot{\nn}_3\cdot\nn_1 \vspace{0.5ex}\\
        \dot{\nn}_2\cdot\nn_3
    \end{array}\right). \label{coordQdev}
\end{align}

Another important thing to be noticed is that \eqref{kernel} leads to $\CH^{(0)}\QQ^{(1)}\cdot\ML_i(\QQ^{(0)})=0$.
Recall that the coordinates of $\CH^{(0)}\QQ^{(1)}$ is $h$.
Thus, when writing this equation by the coordinates, we deduce that
\begin{equation}
    L^T\widetilde{\Lambda}h=0. \label{kernel-coord}
\end{equation}

The calculations of the matrices $M, V, N$ and $P$ involve high-order tensors, which are discussed in Appendices.
Here, we only present the result.
To express these matrices, we introduce six constant $5\times 5$ matrices $X_i(i=1,\cdots,6)$,
\begin{align}
X_1=&\left(
    \begin{array}{ccccc}
        -9 &\, 0 &\,  &\,  &\,
        \vspace{1ex}\\
        0 &\, -3 &\,  &\,  &\,
        \vspace{1ex}\\
        &\, &\, -12 &\, &\,
        \vspace{1ex}\\
        &\, &\, &\, -12 &\,
        \vspace{1ex}\\
        &\, &\, &\, &\, -12
    \end{array}
    \right),
\quad
X_2=\left(
    \begin{array}{ccccc}
        -\frac{3}{2} &\, 0 &\,  &\,  &\,
        \vspace{1ex}\\
        0 &\, \frac{1}{2} &\,  &\,  &\,
        \vspace{1ex}\\
        &\, &\, -1 &\, &\,
        \vspace{1ex}\\
        &\, &\, &\, -1 &\,
        \vspace{1ex}\\
        &\, &\, &\, &\, 2
    \end{array}
    \right),\label{X1-X2}\\
X_3=&\left(
    \begin{array}{ccccc}
        0 &\, \frac{3}{2} &\,  &\,  &\,
        \vspace{1ex}\\
        \frac{3}{2} &\, 0 &\,  &\,  &\,
        \vspace{1ex}\\
        &\, &\, -3 &\, &\,
        \vspace{1ex}\\
        &\, &\, &\, 3 &\,
        \vspace{1ex}\\
        &\, &\, &\, &\, 0
    \end{array}
    \right),\quad
X_4=\left(
    \begin{array}{ccccc}
        \frac{18}{35} &\, 0 &\,  &\,  &\,
        \vspace{1ex}\\
        0 &\, \frac{1}{35} &\,  &\,  &\,
        \vspace{1ex}\\
        &\, &\, -\frac{16}{35} &\, &\,
        \vspace{1ex}\\
        &\, &\, &\, -\frac{16}{35} &\,
        \vspace{1ex}\\
        &\, &\, &\, &\, \frac{4}{35}
    \end{array}
    \right),\label{X3-X4}\\
X_5=&\left(
    \begin{array}{ccccc}
        \frac{27}{140} &\, -\frac{3}{28} &\,  &\,  &\,
        \vspace{1ex}\\
        -\frac{3}{28} &\, \frac{19}{140} &\,  &\,  &\,
        \vspace{1ex}\\
        &\, &\, -\frac{16}{35} &\, &\,
        \vspace{1ex}\\
        &\, &\, &\, \frac{4}{35} &\,
        \vspace{1ex}\\
        &\, &\, &\, &\, -\frac{16}{35}
    \end{array}
    \right),\quad
X_6=\left(
    \begin{array}{ccccc}
        -\frac{9}{35} &\, \frac{3}{28} &\,  &\,  &\,
        \vspace{1ex}\\
        \frac{3}{28} &\, -\frac{1}{70} &\,  &\,  &\,
        \vspace{1ex}\\
        &\, &\, \frac{18}{35} &\, &\,
        \vspace{1ex}\\
        &\, &\, &\, -\frac{2}{35} &\,
        \vspace{1ex}\\
        &\, &\, &\, &\, -\frac{2}{35}
    \end{array}
    \right).\label{X5-X6}
\end{align}
They are associated with six tensors given in (\ref{six-tensors}).
The detailed calculations of (\ref{X1-X2})-(\ref{X5-X6}) can be found in Appendices \ref{Apprendix-A3} and \ref{Apprendix-C}.

By \eqref{R3}, \eqref{R4}, \eqref{R5} and \eqref{M-components}, the blocks of the matrix $M$ are expressed as
\begin{align}
M_{11}=&-\frac{1}{15}(\Gamma_2+\Gamma_3)X_1+\frac{4}{7}\Big(\big(\Gamma_2s_2-\Gamma_3(s_1+s_2)\big)X_2+\big(\Gamma_2b_2-\Gamma_3(b_1+b_2)\big)X_3\Big)\nonumber\\
&-4\Big(\big(\Gamma_2(a_1+\bar{a}_1)-\Gamma_3\bar{a}_1\big)X_4+\big(\Gamma_2(a_2+\bar{a}_2)-\Gamma_3\bar{a}_2\big)X_5\nonumber\\
&+\big(\Gamma_2(a_3+\bar{a}_3)-\Gamma_3\bar{a}_3\big)X_6\Big)\nonumber\\
\eqdefa&\,\left(
    \begin{array}{ccccc}
        \alpha_{11} &\, \alpha_{12} &\,  &\,  &\,
        \vspace{1ex}\\
        \alpha_{12} &\, \alpha_{22} &\,  &\,  &\,
        \vspace{1ex}\\
        &\, &\, \alpha_{33} &\, &\,
        \vspace{1ex}\\
        &\, &\, &\, \alpha_{44} &\,
        \vspace{1ex}\\
        &\, &\, &\, &\, \alpha_{55}
    \end{array}
    \right),\label{M11block}\\
M_{12}=&\frac{1}{15}\Gamma_3X_1+\frac{4}{7}\Gamma_3\Big((s_1+s_2)X_2+(b_1+b_2)X_3\Big)-4\Gamma_3\big(\bar{a}_1X_4+\bar{a}_2X_5+\bar{a}_3X_6\big)\nonumber\\
\eqdefa&\,\left(
    \begin{array}{ccccc}
        \beta_{11} &\, \beta_{12} &\,  &\,  &\,
        \vspace{1ex}\\
        \beta_{12} &\, \beta_{22} &\,  &\,  &\,
        \vspace{1ex}\\
        &\, &\, \beta_{33} &\, &\,
        \vspace{1ex}\\
        &\, &\, &\, \beta_{44} &\,
        \vspace{1ex}\\
        &\, &\, &\, &\, \beta_{55}
    \end{array}
    \right), \label{M12block}\\
M_{22}=&-\frac{1}{15}(\Gamma_1+\Gamma_3)X_1+\frac{4}{7}\Big(\big(\Gamma_1s_1-\Gamma_3(s_1+s_2)\big)X_2+\big(\Gamma_1b_1-\Gamma_3(b_1+b_2)\big)X_3\Big)\nonumber\\
&-4\Big(\big(\Gamma_1(\tilde{a}_1+\bar{a}_1)-\Gamma_3\bar{a}_1\big)X_4+\big(\Gamma_1(\tilde{a}_2+\bar{a}_2)-\Gamma_3\bar{a}_2\big)X_5\nonumber\\
&+\big(\Gamma_1(\tilde{a}_3+\bar{a}_3)-\Gamma_3\bar{a}_3\big)X_6\Big)\nonumber\\
\eqdefa&\,\left(
    \begin{array}{ccccc}
        \gamma_{11} &\, \gamma_{12} &\,  &\,  &\,
        \vspace{1ex}\\
        \gamma_{12} &\, \gamma_{22} &\,  &\,  &\,
        \vspace{1ex}\\
        &\, &\, \gamma_{33} &\, &\,
        \vspace{1ex}\\
        &\, &\, &\, \gamma_{44} &\,
        \vspace{1ex}\\
        &\, &\, &\, &\, \gamma_{55}
    \end{array}
    \right), \label{M22block}
\end{align}
where $a_i$, $\tilde{a}_i$, $\bar{a}_i(i=1,2,3)$ are those in \eqref{biaxial_high}, which we shall keep in mind are functions of $s_i, b_i(i=1,2)$.

By (\ref{CN-1}) and (\ref{CN-2}), it follows that
\begin{align}
    N_1^u=&-\frac{1}{15}e_1X_1-\frac{2}{7}\Big(\big((e_1-e_2)s_1-2e_2s_2\big)X_2+\big((e_1-e_2)b_1-2e_2b_2\big)X_3\Big)\nonumber\\
    &-2\Big((a_1+2e_2\bar{a}_1)X_4+(a_2+2e_2\bar{a}_2)X_5+(a_3+2e_2\bar{a}_3)X_6\Big)\nonumber\\
    \eqdefa&\,\left(\begin{array}{ccccc}
         \mu_{11} &\, \mu_{12}&\, &\, &\,
         \vspace{1ex}\\
         \mu_{12} &\, \mu_{22} &\, &\, &\,
         \vspace{1ex}\\
         &\,  &\,\mu_{33} &\, &\, \vspace{1ex}\\
         &\, &\, &\, \mu_{44} &\,
         \vspace{1ex}\\
         &\, &\, &\, &\,\mu_{55}
     \end{array}
     \right),\label{N1-u}\\
     N_2^u=&-\frac{1}{15}e_2X_1+\frac{2}{7}\Big(\big(2e_1s_1-(e_2-e_1)s_2\big)X_2+\big(2e_1b_1-(e_2-e_1)b_2\big)X_3\Big)\nonumber\\
     &-2\Big(\big(\tilde{a}_1+2e_1\bar{a}_1\big)X_4+\big(\tilde{a}_2+2e_1\bar{a}_2\big)X_5+\big(\tilde{a}_3+2e_1\bar{a}_3\big)X_6\Big)\nonumber\\
     \eqdefa&\,\left(\begin{array}{ccccc}
        \nu_{11} &\, \nu_{12} &\, &\, &\,
        \vspace{1ex}\\
        \nu_{12} &\, \nu_{22} &\, &\, &\,
        \vspace{1ex}\\
        &\,  &\,\nu_{33} &\, &\, \\
        &\, &\, &\, \nu_{44}&\,
        \vspace{1ex}\\
        &\, &\, &\, &\,\nu_{55}
     \end{array}
     \right), \label{N2-u}
\end{align}
and 
\begin{align}
    &\,V=\left(\begin{array}{c|c}
        N_1^u &   \\
        N_2^u & \raisebox{2.5ex}{$\frac 12 L$}
\end{array}
\right),\label{V-matrix}
\end{align}
where $L$ is given by (\ref{L-matrix}).

By \eqref{R1}, \eqref{R2}, \eqref{R3} and \eqref{P0}, we obtain
\begin{align}\label{P-matrix}
    P=&\,c\zeta\Big[-\frac{1}{45}\big(I_{22}+I_{11}(1+3e_1)\big)X_1-\frac{4}{21}\Big(\big((I_{22}+3I_{11}e_1)s_1+I_{11}(1+3e_1)s_2\big)X_2\nonumber\\
    &+\big((I_{22}+3I_{11}e_1)b_1+I_{11}(1+3e_1)b_2\big)X_3\Big)+\big(I_{22}a_1+I_{11}\tilde{a}_{1}+4I_{11}e_1\bar{a}_1\big)X_4\nonumber\\
    &+\big(I_{22}a_2+I_{11}\tilde{a}_{2}+4I_{11}e_1\bar{a}_2\big)X_5+\big(I_{22}a_3+I_{11}\tilde{a}_{3}+4I_{11}e_1\bar{a}_3\big)X_6\Big]\nonumber\\
\eqdefa&\,\left(\begin{array}{ccccc}
        \vartheta_{11} &\, \vartheta_{12} &\, &\, &\,
        \vspace{1ex}\\
        \vartheta_{12} &\, \vartheta_{22} &\, &\, &\,
        \vspace{1ex}\\
        &\,  &\,\vartheta_{33} &\, &\, \\
        &\, &\, &\, \vartheta_{44}&\,
        \vspace{1ex}\\
        &\, &\, &\, &\,\vartheta_{55}
     \end{array}
     \right).
\end{align}

In the above, we intentionally introduce notations for the components of $M$, $V$ and $P$, to emphasize that these matrices have specific forms.
These specific forms are significant in the forthcoming derivations.
We have claimed that $\CM$ and $\CP$ are positive definite in \eqref{M-spd} and \eqref{P-spd}.
As a result, the corresponding coefficient matrices $M$ and $P$ are also positive definite.
We do not consider the expressions of the vectors $h$ and $g$, because the terms involving them will be expressed by variational derivatives of the elastic energy.

\subsection{Orthonormal frame model}

We are now ready to derive the frame hydrodynamics for the biaxial nematic phase from the $O(1)$ system (\ref{ve1-QQ})-(\ref{ve3-impress}).

To begin with, we write down the elastic energy for the biaxial nematic phase.
In the tensor model, the elastic energy is a functional of $\QQ$.
When $\QQ$ takes $\QQ^{(0)}$ that is a function of $\Fp$, the corresponding elastic energy becomes a functional of the frame $\Fp$, which we denote by $\CF_{Bi}$.
Generally, the biaxial elastic energy consists of twelve bulk terms \cite{GV1,Xu2}, written as
\begin{align}
\frac{\CF_{Bi}[\Fp]}{ck_BT}=&\,\int\ud\bm{x}~\frac{1}{2}\Big(K_{1111}D_{11}^2+ K_{2222}D_{22}^2+K_{3333}D_{33}^2 \nonumber\\
&\,+K_{1212}D_{12}^2+ K_{2121}D_{21}^2+K_{2323}D_{23}^2
+K_{3232}D_{32}^2+ K_{3131}D_{31}^2+K_{1313}D_{13}^2 \nonumber\\
&\,+K_{1221}D_{12}D_{21}+K_{2332}D_{23}D_{32}+K_{1331}D_{13}D_{31}\Big). \label{elas_Bi}
\end{align}
We here take no account of three surface terms, such as
\begin{equation}
\partial_in_{2i}\partial_jn_{2j}-\partial_in_{2j}\partial_jn_{2i}
=2(D_{33}D_{11}-D_{31}D_{13}). \label{surfterm}
\end{equation}
The coefficients in \eqref{elas_Bi} can be derived from the coefficients in the tensor model \cite{Xu2} as
\begin{align}\label{K-coeffi}
\left\{
\begin{array}{l}
K_{1111}=J_2,\quad
K_{2222}=J_1,\quad
K_{3333}=J_1+J_2-J_3,
\vspace{0.5ex}\\
K_{1212}=K_{3232}=J_1+J_4,\quad
K_{2121}=K_{3131}=J_2+J_5,
\vspace{0.5ex}\\
K_{2323}=K_{1313}=J_1+J_2-J_3+J_4+J_5-J_6,
\vspace{0.5ex}\\
K_{1221}=-J_6,\quad
K_{2332}=J_6-2J_4,\quad
K_{1331}=J_6-2J_5,
\end{array}
\right.
\end{align}
where
\begin{align}\label{J-coefficients-6}
\left\{
\begin{array}{l}
J_1=\,2c\Big(c_{22}(s_1+b_1)^2+c_{23}(s_2+b_2)^2+2c_{24}(s_1+b_1)(s_2+b_2)\Big),
\vspace{0.5ex}\\
J_2=\,8c\Big(c_{22}b^2_1+c_{23}b^2_2+2c_{24}b_1b_2\Big),
\vspace{0.5ex}\\
J_3=\,8c\Big(c_{22}b_1(s_1+b_1)+c_{23}b_2(s_2+b_2)+c_{24}[b_1(s_2+b_2)+b_2(s_1+b_1)]\Big),
\vspace{0.5ex}\\
J_4=\,c\Big(c_{28}(s_1+b_1)^2+c_{29}(s_2+b_2)^2+2c_{2,10}(s_1+b_1)(s_2+b_2)\Big),
\vspace{0.5ex}\\
J_5=\,4c\Big(c_{28}b^2_1+c_{29}b^2_2+2c_{2,10}b_1b_2\Big),
\vspace{0.5ex}\\
J_6=\,4c\Big(c_{28}b_1(s_1+b_1)+c_{29}b_2(s_2+b_2)+c_{2,10}[b_1(s_2+b_2)+b_2(s_1+b_1)]\Big).
\end{array}
\right.
\end{align}

Using the chain rule, we deduce that
\begin{align}\label{CL-ML}
    \ML\CF_{Bi}=\ML\CF_{Bi}[\QQ^{(0)}(\Fp)]=\frac{\delta \CF_{Bi}}{\delta \QQ^{(0)}}\cdot \ML\QQ^{(0)}=ck_BT\CG(\QQ^{(0)})\cdot\ML\QQ^{(0)}=ck_BTL^T\widetilde{\Lambda}g.
\end{align}
Therefore, it is deduced from \eqref{Mv1-QQ} and \eqref{kernel-coord} that
\begin{align}\label{frame-p-equation}
L^TM^{-1}(q-V\omega)+\frac{1}{ck_BT}\ML\CF_{Bi}=0.
\end{align}
In the above, we notice that $M$ is positive definite, thus invertible.

To calculate \eqref{frame-p-equation}, we rearrange the rows and columns of the matrices so that they can be divided into blocks appropriately.
To this end, we introduce a $10\times 10$ permutation matrix
\begin{align}
    C=(\EE_1,\EE_6,\EE_2,\EE_7,\EE_3,\EE_8,\EE_4,\EE_9,\EE_5,\EE_{10}),
\end{align}
where $\EE_j$ is the $10\times 1$ unit vector with the $j$-th component equal to one.
We have $CC^T=I_{10}$, which is the $10\times 10$ identity matrix.
Then we have
\begin{align}\label{M-matrix}
C^TMC=&\,\left(\begin{array}{c|ccc}
        M_0 &\, &\, &\, \\\hline
        &\, M_1&\, &\,\\
        &\, &\, M_2&\,\\
        &\, &\, &\,M_3
    \end{array}
    \right),
\end{align}
where the blocks $M_i(i=0,1,2,3)$ are given by
\begin{align*}
M_0=&\,\left(
    \begin{array}{cccc}
        {\alpha}_{11} &\, {\beta}_{11} &\, {\alpha}_{12} &\, {\beta}_{12} \vspace{0.5ex}\\
        {\beta}_{11} &\, {\gamma}_{11} &\, {\beta}_{12} &\, {\gamma}_{12} \vspace{0.5ex}\\
        {\alpha}_{12} &\, {\beta}_{12} &\, {\alpha}_{22} &\, {\beta}_{22} \vspace{0.5ex}\\
        {\beta}_{12} &\, {\gamma}_{12} &\, {\beta}_{22} &\, {\gamma}_{22}
    \end{array}
    \right),\\
M_1=&\,\left(
    \begin{array}{cc}
        {\alpha}_{33} &\, {\beta}_{33} \vspace{0.5ex}\\
        {\beta}_{33} &\, {\gamma}_{33}
    \end{array}
    \right),\quad
M_2=\left(
    \begin{array}{cc}
        {\alpha}_{44} &\, {\beta}_{44} \vspace{0.5ex}\\
        {\beta}_{44} &\, {\gamma}_{44}
    \end{array}
    \right),\quad
M_3=\left(
    \begin{array}{cc}
        {\alpha}_{55} &\, {\beta}_{55} \vspace{0.5ex}\\
        {\beta}_{55} &\, {\gamma}_{55}
    \end{array}
    \right).
\end{align*}
These blocks are all positive definite because $M$ is.
For the matrix $V$, we have
\begin{align}\label{V-block}
C^TV=\left(
    \begin{array}{c|cccccc}
        V_0 &\, &\, &\, &\, &\, &\,\\\hline
        &\, V_1&\,0 &\,0&\, V_4&\,0&\,0 \\
        &\, 0&\,V_2&\, 0&\,0&\,V_5&\,0\\
       &\,0&\,0&\,V_3&\,0&\,0 &\,V_6
    \end{array}
    \right),
\end{align}
where the blocks $V_i(i=0,1,\cdots,6)$ are given by
\begin{align*}
V_0=&\,\left(
    \begin{array}{cc}
        \mu_{11} &\, \mu_{12} \vspace{0.5ex}\\
        \nu_{11} &\, \nu_{12} \vspace{0.5ex}\\
        \mu_{12} &\, \mu_{22} \vspace{0.5ex}\\
        \nu_{12} &\, \nu_{22}
    \end{array}
    \right),\quad
V_1=\left(
    \begin{array}{c}
        \mu_{33}  \vspace{0.5ex}\\
        \nu_{33}
    \end{array}
    \right),\quad
V_2=\left(
    \begin{array}{c}
        \mu_{44}  \vspace{0.5ex}\\
        \nu_{44}
    \end{array}
    \right),\quad
V_3=\left(
    \begin{array}{c}
        \mu_{55}  \vspace{0.5ex}\\
        \nu_{55}
    \end{array}
    \right),\\
V_4=&\,\left(
    \begin{array}{c}
        s_1-b_1 \vspace{0.5ex}\\
        s_2-b_2
    \end{array}
    \right),\quad
V_5=-\left(
    \begin{array}{c}
        s_1+b_1 \vspace{0.5ex}\\
        s_2+b_2
    \end{array}
    \right),\quad
V_6=\left(
    \begin{array}{c}
        2b_1 \vspace{0.5ex}\\
        2b_2
    \end{array}
    \right).
\end{align*}
We also rearrange the indices of the vector $q$ by $C$,
\begin{align}\label{q-block-vector}
C^Tq=\left(\begin{array}{c}0_{4\times1}
\vspace{0.5ex}\\
2(\dot{\nn}_1\cdot\nn_2)V_4\vspace{0.5ex}\\
2(\dot{\nn}_3\cdot\nn_1)V_5
\vspace{0.5ex}\\
2(\dot{\nn}_2\cdot\nn_3)V_6
\end{array}\right),
\end{align}
where we use $0_{N_1\times N_2}$ to represent an $N_1\times N_2$ zero matrix.
The matrix $L$ is rearranged as
\begin{align}\label{Lt-block}
(C^TL)^T=L^TC=(0_{3\times4}, L^T_1),
\end{align}
where
\begin{align*}
L^T_1=\left(
    \begin{array}{cccc}
        2V^T_4 &\,  &\, \vspace{0.5ex}\\
         &\,2V^T_5 &\, \vspace{0.5ex}\\
         &\,  &\, 2V^T_6
    \end{array}
    \right).
\end{align*}
Thus, from (\ref{M-matrix}) and (\ref{Lt-block}), we have
\begin{align}\label{Lt-M-1}
L^TM^{-1}C=(C^TL)^TC^TM^{-1}C=&\,\left(
    \begin{array}{c|ccc}
        &\,2V^T_4M^{-1}_1&\, &\,
        \\0_{3\times4}
        &\, &\,2V^T_5M^{-1}_2&\,\\
        &\, &\, &\,2V^T_6M^{-1}_3
    \end{array}
    \right).
\end{align}
Together with (\ref{Lt-M-1}), we deduce that
\begin{align}\label{LtM-1q}
L^TM^{-1}q=(L^TC)(C^TMC)^{-1}(C^Tq)=
   \left(
    \begin{array}{c}
       \chi_3\dot{\nn}_1\cdot\nn_2
       \vspace{0.5ex}\\
        \chi_2\dot{\nn}_3\cdot\nn_1
        \vspace{0.5ex}\\
        \chi_1\dot{\nn}_2\cdot\nn_3
    \end{array}
    \right),
\end{align}
where the coefficients $\chi_i(i=1,2,3)$ are given by
\begin{align*}
\chi_3=4V_4^TM_1^{-1}V_4,\quad
\chi_2=4V_5^TM_2^{-1}V_5,\quad
\chi_1=4V_6^TM_3^{-1}V_6.
\end{align*}

From (\ref{M-matrix}), (\ref{V-block}) and (\ref{Lt-block}), we have
\begin{align}\label{LtM-1V}
&\,L^TM^{-1}V\nonumber\\
&\,=\left(
    \begin{array}{c|ccc}
        &\,2V^T_4M^{-1}_1&\, &\,\\
        0_{3\times4}
        &\, &\,2V^T_5M^{-1}_2&\,\\
        &\, &\, &\,2V^T_6M^{-1}_3
    \end{array}
    \right)
\left(
    \begin{array}{c|cccccc}
        V_0 &\, &\, &\, &\, &\, &\,\\\hline
        &\, V_1&\,0 &\,0&\, V_4&\,0&\,0 \\
        &\, 0&\,V_2&\, 0&\,0&\,V_5&\,0\\
       &\,0&\,0&\,V_3&\,0&\,0 &\,V_6
    \end{array}
    \right)\nonumber\\
&\,=\left(
    \begin{array}{c|cccccc}
     &2V^T_4M^{-1}_1V_1&0&0&2V^T_4M^{-1}_1V_4 &0&0\\
     0_{3\times4}
        & 0&2V^T_5M^{-1}_2V_2&0&0&2V^T_5M^{-1}_2V_5&0\\
        &0 &0& 2V^T_6M^{-1}_3V_3&0&0&2V^T_6M^{-1}_3V_6
    \end{array}
    \right).
\end{align}
So we have
\begin{align}\label{LtM-1V-omega}
L^TM^{-1}V\omega=&\,\left(
    \begin{array}{c}
        2V^T_4M^{-1}_1V_1\A_0\cdot\sss_3+2V^T_4M^{-1}_1V_4\BOm_0\cdot\aaa_1 \vspace{1ex}\\
        2V^T_5M^{-1}_2V_2\A_0\cdot\sss_4+2V^T_5M^{-1}_2V_5\BOm_0\cdot\aaa_2 \vspace{1ex}\\
        2V^T_6M^{-1}_3V_3\A_0\cdot\sss_5+2V^T_6M^{-1}_3V_6\BOm_0\cdot\aaa_3
    \end{array}
    \right)\nonumber\\
=&\,\left(
    \begin{array}{c}
        \eta_3\A_0\cdot\sss_3+\frac{1}{2}\chi_3\BOm_0\cdot\aaa_1  \vspace{1ex}\\
        \eta_2\A_0\cdot\sss_4+\frac{1}{2}\chi_2\BOm_0\cdot\aaa_2  \vspace{1ex}\\
        \eta_1\A_0\cdot\sss_5+\frac{1}{2}\chi_1\BOm_0\cdot\aaa_3
    \end{array}
    \right),
\end{align}
where the coefficients $\eta_i(i=1,2,3)$ are expressed by
\begin{align*}
\eta_3=2V^T_4M^{-1}_1V_1,\quad
\eta_2=2V^T_5M^{-1}_2V_2,\quad
\eta_1=2V^T_6M^{-1}_3V_3.
\end{align*}
Therefore, using (\ref{LtM-1q}) and (\ref{LtM-1V-omega}), the equation (\ref{frame-p-equation}) can be reformulated as
\begin{align}
&\,\chi_1\dot{\nn}_2\cdot\nn_3-\frac{1}{2}\chi_1\BOm_0\cdot\aaa_3-\eta_1\A_0\cdot\sss_5+\frac{1}{ck_BT}\ML_1\CF_{Bi}=0,\label{frame-equation-n1}\\
&\,\chi_2\dot{\nn}_3\cdot\nn_1-\frac{1}{2}\chi_2\BOm_0\cdot\aaa_2-\eta_2\A_0\cdot\sss_4+\frac{1}{ck_BT}\ML_2\CF_{Bi}=0,\label{frame-equation-n2}\\
&\,\chi_3\dot{\nn}_1\cdot\nn_2-\frac{1}{2}\chi_3\BOm_0\cdot\aaa_1-\eta_3\A_0\cdot\sss_3+\frac{1}{ck_BT}\ML_3\CF_{Bi}=0,\label{frame-equation-n3}
\end{align}
where $\ML_i\CF_{Bi}\,(i=1,2,3)$ are the variational derivatives along the infinitesimal rotation round $\nn_i\,(i=1,2,3)$, respectively, which are written as
\begin{align*}
\ML_1\CF_{Bi}=&\,\nn_3\cdot\frac{\delta \CF_{Bi}}{\delta\nn_2}-\nn_2\cdot\frac{\delta \CF_{Bi}}{\delta\nn_3},\\
\ML_2\CF_{Bi}=&\,\nn_1\cdot\frac{\delta \CF_{Bi}}{\delta\nn_3}-\nn_3\cdot\frac{\delta \CF_{Bi}}{\delta\nn_1},\\
\ML_3\CF_{Bi}=&\,\nn_2\cdot\frac{\delta \CF_{Bi}}{\delta\nn_1}-\nn_1\cdot\frac{\delta \CF_{Bi}}{\delta\nn_2}.
\end{align*}

It remains to derive the equation of the fluid velocity $\vv^{(0)}$.
From \eqref{coordQdev}, we have
\begin{align}
    \tilde{q}_i^T\widetilde{\Lambda} (h+g)=&\,\big(\partial_i{\nn}_1\cdot\nn_2,\partial_i{\nn}_3\cdot\nn_1,\partial_i{\nn}_2\cdot\nn_3\big)L^T\widetilde{\Lambda} (h+g).
\end{align}
Using \eqref{CL-ML} and \eqref{kernel-coord}, we arrive at
\begin{align}
    \tilde{q}_i^T\widetilde{\Lambda} (h+g)=\frac{1}{ck_BT}\big(\ML_3\CF_{Bi}\partial_i{\nn}_1\cdot\nn_2+\ML_2\CF_{Bi}\partial_i{\nn}_3\cdot\nn_1+\ML_1\CF_{Bi}\partial_i{\nn}_2\cdot\nn_3\big)\eqdefa\mathfrak{F}_i.
\end{align}
Taking them into \eqref{Mv2-v}, we deduce that
\begin{align}
\rho_s\Big(\frac{\partial\vv^{(0)}}{\partial t}+\vv^{(0)}\cdot\nabla\vv^{(0)}\Big)_i
=&\,-\partial_i p^{(0)}+\eta\Delta v^{(0)}_i
+\partial_j\Big((\sss_1,\cdots,\sss_5)P\omega_s\Big)_{ij}\nonumber\\
&\,-ck_BT\partial_j\Big((\sss_1,\cdots,\sss_5,\aaa_1,\aaa_2,\aaa_3)NM^{-1}(q-V\omega)\Big)_{ij}\nonumber\\
&\,+ck_BT\mathfrak{F}_i.\label{Mv2-v-coordinate}
\end{align}
In the above, we recall that the matrix $P$ is given by \eqref{P-matrix}.

Noticing $N=V^T$, we obtain
\begin{align}\label{NM-1}
NM^{-1}=&\,\left(
    \begin{array}{c|ccc}
        V^T_0 &\, &\, &\, \\\hline
        &\, V^T_1&\, 0&\,0 \\
        &\, 0&\,V^T_2&\,0 \\
        &\,0 &\,0 &\,V^T_3\\
        &\,V^T_4&\,0&\,0\\
        &\,0&\,V^T_5&\,0\\
        &\,0&\,0&\,V^T_6
    \end{array}
    \right)
\left(\begin{array}{c|ccc}
        M^{-1}_0 &\, &\, &\, \\\hline
        &\, M^{-1}_1&\, &\,\\
        &\, &\, M^{-1}_2&\,\\
        &\, &\, &\,M^{-1}_3
    \end{array}
    \right)\nonumber\\
=&\,\left(
    \begin{array}{c|ccc}
        V^T_0M^{-1}_0 &\, &\, &\, \\\hline
        &\, V^T_1M^{-1}_1&\, 0&\,0 \\
        &\, 0&\,V^T_2M^{-1}_2&\, 0\\
        &\,0&\,0&\,V^T_3M^{-1}_3\\
        &\,V^T_4M^{-1}_1&\,0&\,0\\
        &\,0&\,V^T_5M^{-1}_2&\,0\\
        &\,0&\,0&\,V^T_6M^{-1}_3
    \end{array}
    \right).
\end{align}
Thus, we have
\begin{align}\label{NM-1-q}
NM^{-1}q=&\,\left(
    \begin{array}{c|ccc}
        V^T_0M^{-1}_0 &\, &\, &\, \\\hline
        &\, V^T_1M^{-1}_1&\, 0&\,0 \\
        &\, 0&\,V^T_2M^{-1}_2&\, 0\\
        &\,0&\,0&\,V^T_3M^{-1}_3\\
        &\,V^T_4M^{-1}_1&\,0&\,0\\
        &\,0&\,V^T_5M^{-1}_2&\,0\\
        &\,0&\,0&\,V^T_6M^{-1}_3
    \end{array}
    \right)
 \left(\begin{array}{c}0_{4\times1}
\vspace{0.5ex}\\
2(\dot{\nn}_1\cdot\nn_2)V_4\vspace{0.5ex}\\
2(\dot{\nn}_3\cdot\nn_1)V_5
\vspace{0.5ex}\\
2(\dot{\nn}_2\cdot\nn_3)V_6
\end{array}\right)\nonumber\\
=&\,\left(
    \begin{array}{c}
        0_{2\times1}
        \vspace{1ex}\\
       2V^T_1M^{-1}_1V_4(\dot{\nn}_1\cdot\nn_2)
       \vspace{1ex}\\
       2V^T_2M^{-1}_2V_5(\dot{\nn}_3\cdot\nn_1)
       \vspace{1ex}\\
        2V^T_3M^{-1}_3V_6(\dot{\nn}_2\cdot\nn_3)
        \vspace{1ex}\\
       2V^T_4M^{-1}_1V_4(\dot{\nn}_1\cdot\nn_2)
       \vspace{1ex}\\ 2V^T_5M^{-1}_2V_5(\dot{\nn}_3\cdot\nn_1)
       \vspace{1ex}\\
       2V^T_6M^{-1}_3V_6(\dot{\nn}_2\cdot\nn_3)
   \end{array}\right)
  =\left(
    \begin{array}{c}
        0_{2\times1}
        \vspace{1ex}\\
       \eta_3(\dot{\nn}_1\cdot\nn_2)
       \vspace{1ex}\\
       \eta_2(\dot{\nn}_3\cdot\nn_1)
       \vspace{1ex}\\
        \eta_1(\dot{\nn}_2\cdot\nn_3)
        \vspace{1ex}\\
       \frac{1}{2}\chi_3(\dot{\nn}_1\cdot\nn_2)
       \vspace{1ex}\\ \frac{1}{2}\chi_2(\dot{\nn}_3\cdot\nn_1)
       \vspace{1ex}\\
       \frac{1}{2}\chi_1(\dot{\nn}_2\cdot\nn_3)
   \end{array}\right),
\end{align}
and
\begin{align}\label{NM-1V}
&\,NM^{-1}V\nonumber\\
=&\,\left(
    \begin{array}{c|ccc}
        V^T_0M^{-1}_0 &\, &\, &\, \\\hline
        &\, V^T_1M^{-1}_1&\, 0&\,0 \\
        &\, 0&\,V^T_2M^{-1}_2&\, 0\\
        &\,0&\,0&\,V^T_3M^{-1}_3\\
        &\,V^T_4M^{-1}_1&\,0&\,0\\
        &\,0&\,V^T_5M^{-1}_2&\,0\\
        &\,0&\,0&\,V^T_6M^{-1}_3
    \end{array}
    \right)
\left(
    \begin{array}{c|cccccc}
        V_0 &\, &\, &\, &\, &\, &\,\\\hline
        &\, V_1&\,0 &\,0&\, V_4&\,0&\,0 \\
        &\, 0&\,V_2&\, 0&\,0&\,V_5&\,0\\
       &\,0&\,0&\,V_3&\,0&\,0 &\,V_6
    \end{array}
    \right)\nonumber\\
=&\,\left(
    \begin{array}{c|cccccc}
        V^T_0M^{-1}_0V_0 &\, &\, &\, \\\hline
        &\, V^T_1M^{-1}_1V_1&\,0&\,0&\, V^T_1M^{-1}_1V_4&\,0&\,0\\
        &\, 0&\, V^T_2M^{-1}_2V_2&\,0&\,0&\,V^T_2M^{-1}_2V_5&\,0\\
        &\, 0&\,0 &\,V^T_3M^{-1}_3V_3&\,0&\,0&\,V^T_3M^{-1}_3V_6\\
        &\,V^T_4M^{-1}_1V_1&\, 0&\, 0&\,V^T_4M^{-1}_1V_4&\,0&\,0\\
        &\,0&\,V^T_5M^{-1}_2V_2&\,0&\,0&\,V^T_5M^{-1}_2V_5&\,0\\
        &\,0&\,0&\,V^T_6M^{-1}_3V_1&\,0&\,0&\,V^T_6M^{-1}_3V_6
    \end{array}
    \right).
\end{align}
It further gives
\begin{align}\label{NM-1V-omega}
NM^{-1}V\omega
=\left(
    \begin{array}{c}
        V^T_0M^{-1}_0V_0\omega_0
         \vspace{1ex}\\
         V^T_1M^{-1}_1V_1\A_0\cdot\sss_3+ V^T_1M^{-1}_1V_4\BOm_0\cdot\aaa_1
          \vspace{1ex}\\
        V^T_2M^{-1}_2V_2\A_0\cdot\sss_4+ V^T_2M^{-1}_2V_5\BOm_0\cdot\aaa_2
         \vspace{1ex}\\
         V^T_3M^{-1}_3V_3\A_0\cdot\sss_5+ V^T_3M^{-1}_3V_6\BOm_0\cdot\aaa_3
          \vspace{1ex}\\
         V^T_4M^{-1}_1V_1\A_0\cdot\sss_3+ V^T_4M^{-1}_1V_4\BOm_0\cdot\aaa_1
          \vspace{1ex}\\
         V^T_5M^{-1}_2V_2\A_0\cdot\sss_4+ V^T_5M^{-1}_2V_5\BOm_0\cdot\aaa_2
          \vspace{1ex}\\
         V^T_6M^{-1}_3V_3\A_0\cdot\sss_5+ V^T_6M^{-1}_3V_6\BOm_0\cdot\aaa_3
    \end{array}
    \right)
=\left(
    \begin{array}{c}
        V^T_0M^{-1}_0V_0\omega_0
         \vspace{1ex}\\
         \beta_3\A_0\cdot\sss_3+ \frac{1}{2}\eta_3\BOm_0\cdot\aaa_1
          \vspace{1ex}\\
       \beta_4\A_0\cdot\sss_4+ \frac{1}{2}\eta_2\BOm_0\cdot\aaa_2
         \vspace{1ex}\\
         \beta_5\A_0\cdot\sss_5+ \frac{1}{2}\eta_1\BOm_0\cdot\aaa_3
          \vspace{1ex}\\
         \frac{1}{2}\eta_3\A_0\cdot\sss_3+ \frac{1}{4}\chi_3\BOm_0\cdot\aaa_1
          \vspace{1ex}\\
         \frac{1}{2}\eta_2\A_0\cdot\sss_4+ \frac{1}{4}\chi_2\BOm_0\cdot\aaa_2
          \vspace{1ex}\\
         \frac{1}{2}\eta_1\A_0\cdot\sss_5+ \frac{1}{4}\chi_1\BOm_0\cdot\aaa_3
    \end{array}
    \right),
\end{align}
where we denote
\begin{align*}
    \omega_0=(\A_0\cdot\sss_1,\A_0\cdot\sss_2)^T,
\end{align*}
and the coefficients $\beta_i(i=3,4,5)$ are given by
\begin{align}
\beta_3=V^T_1M^{-1}_1V_1,\quad \beta_4=V^T_2M^{-1}_2V_2,\quad
\beta_5=V^T_3M^{-1}_3V_3.
\end{align}
Since $M_0$ is positive definite, the $2\times 2$ matrix $V^T_0M^{-1}_0V_0$ is symmetric positive semi-definite. If we write out its components,
\begin{align*}
V^T_0M^{-1}_0V_0=\left(
    \begin{array}{cc}
        \beta_1 &\, \beta_0 \\
        \beta_0 &\, \beta_2
    \end{array}
    \right),
\end{align*}
then $\beta_i(i=0,1,2)$ satisfy
\begin{align}
\beta_i\geq 0,~i=1,2,\quad \beta^2_0\leq\beta_1\beta_2. \label{beta012}
\end{align}

Hence, combining (\ref{NM-1-q}) with (\ref{NM-1V-omega}), we deduce that
\begin{align}
NM^{-1}(q-V\omega)=\left(
    \begin{array}{c}
        -V^T_0M^{-1}_0V_0\omega_0  \vspace{1ex}\\
         -\beta_3\A_0\cdot\sss_3+ \eta_3\big(\dot{\nn}_1\cdot\nn_2-\frac{1}{2}\BOm_0\cdot\aaa_1\big)
         \vspace{1ex}\\
        -\beta_4\A_0\cdot\sss_4+ \eta_2\big(\dot{\nn}_3\cdot\nn_1-\frac{1}{2}\BOm_0\cdot\aaa_2\big)
        \vspace{1ex}\\
         -\beta_5\A_0\cdot\sss_5+ \eta_1\big(\dot{\nn}_2\cdot\nn_3-\frac{1}{2}\BOm_0\cdot\aaa_3\big)
         \vspace{1ex}\\
         -\frac{1}{2}\eta_3\A_0\cdot\sss_3+ \frac{1}{2}\chi_3\big(\dot{\nn}_1\cdot\nn_2-\frac{1}{2}\BOm_0\cdot\aaa_1\big)
         \vspace{1ex}\\
         -\frac{1}{2}\eta_2\A_0\cdot\sss_4+ \frac{1}{2}\chi_2\big(\dot{\nn}_3\cdot\nn_1-\frac{1}{2}\BOm_0\cdot\aaa_2\big)
         \vspace{1ex}\\
         -\frac{1}{2}\eta_1\A_0\cdot\sss_5+ \frac{1}{2}\chi_1\big(\dot{\nn}_2\cdot\nn_3-\frac{1}{2}\BOm_0\cdot\aaa_3\big)
    \end{array}
    \right),
\end{align}
which further implies
\begin{align}
\frac{1}{ck_BT}\sigma^{(0)}_e=&\,-(\sss_1,\cdots,\sss_5,\aaa_1,\aaa_2,\aaa_3) NM^{-1}(q-V\omega)\nonumber\\
=&\,\beta_1(\A_0\cdot\sss_1)\sss_1+\beta_0(\A_0\cdot\sss_2)\sss_1+\beta_0(\A_0\cdot\sss_1)\sss_2+\beta_2(\A_0\cdot\sss_2)\sss_2\nonumber\\
&\,+\beta_3(\A_0\cdot\sss_3)\sss_3-\eta_3\Big(\dot{\nn}_1\cdot\nn_2-\frac{1}{2}\BOm_0\cdot\aaa_1\Big)\sss_3\nonumber\\
&\,+\beta_4(\A_0\cdot\sss_4)\sss_4-\eta_2\Big(\dot{\nn}_3\cdot\nn_1-\frac{1}{2}\BOm_0\cdot\aaa_2\Big)\sss_4\nonumber\\
&\,+\beta_5(\A_0\cdot\sss_5)\sss_5-\eta_1\Big(\dot{\nn}_2\cdot\nn_3-\frac{1}{2}\BOm_0\cdot\aaa_3\Big)\sss_5\nonumber\\
&\,+\frac{1}{2}\eta_3(\A_0\cdot\sss_3)\aaa_1-\frac{1}{2}\chi_3\Big(\dot{\nn}_1\cdot\nn_2-\frac{1}{2}\BOm_0\cdot\aaa_1\Big)\aaa_1\nonumber\\
&\,+\frac{1}{2}\eta_2(\A_0\cdot\sss_4)\aaa_2-\frac{1}{2}\chi_2\Big(\dot{\nn}_3\cdot\nn_1-\frac{1}{2}\BOm_0\cdot\aaa_2\Big)\aaa_2\nonumber\\
&\,+\frac{1}{2}\eta_1(\A_0\cdot\sss_5)\aaa_3-\frac{1}{2}\chi_1\Big(\dot{\nn}_2\cdot\nn_3-\frac{1}{2}\BOm_0\cdot\aaa_3\Big)\aaa_3.
\end{align}

Therefore, from (\ref{Mv2-v-coordinate}), the equation of $\vv^{(0)}$ reads
\begin{align}
\rho_s\Big(\frac{\partial\vv^{(0)}}{\partial t}+\vv^{(0)}\cdot\nabla\vv^{(0)}\Big)_i=&\,-\partial_ip^{(0)}+\partial_j\big((\sigma^{(0)}_v)_{ij}+(\sigma^{(0)}_e)_{ij}\big)+ck_BT\mathfrak{F}_i,\label{frame-equation-v0}\\\nabla\cdot\vv^{(0)}=&\,0.\label{imcompressible-v0}
\end{align}
Here, the viscous stress $\sigma^{(0)}_v$ is denoted by
\begin{align}\label{sigma-v}
\sigma^{(0)}_v=&\,\eta\A_0+(\sss_1,\cdots,\sss_5)P\omega_s\nonumber\\
=&\,(\sss_1,\cdots,\sss_5)(\eta\Lambda^{-1}+P)\omega_s,
\end{align}
where we have used the following fact
\begin{align*}
\A_0=\sum^5_{i=1}\frac{1}{|\sss_i|^2}(\A_0\cdot\sss_i)\sss_i=(\sss_1,\cdots,\sss_5)\Lambda^{-1}\omega_s,\quad \Lambda^{-1}={\rm diag}\Big(\frac{3}{2},\frac{1}{2},2,2,2\Big).
\end{align*}

To sum up, the frame hydrodynamics for the biaxial nematic phase is given by (\ref{frame-equation-n1})-(\ref{frame-equation-n3}), (\ref{frame-equation-v0}) and (\ref{imcompressible-v0}).

\subsection{Energy dissipation}

Taking the derivative about $t$ of the biaxial elastic energy (\ref{elas_Bi}), we deduce that
\begin{align}\label{energy-time-t}
\frac{\ud \CF_{Bi}}{\ud t}
=&\,\int\Big(\frac{\delta\CF_{Bi}}{\delta\nn_1}\cdot\frac{\partial\nn_1}{\partial t}+\frac{\delta\CF_{Bi}}{\delta\nn_2}\cdot\frac{\partial\nn_2}{\partial t}+\frac{\delta\CF_{Bi}}{\delta\nn_3}\cdot\frac{\partial\nn_3}{\partial t}\Big)\ud\xx\nonumber\\
=&\,\int\Big(\frac{\delta\CF_{Bi}}{\delta\nn_1}\cdot\big(\nn_2(\nn_2\cdot\partial_t\nn_1)+\nn_3(\nn_3\cdot\partial_t\nn_1)\big)+\frac{\delta\CF_{Bi}}{\delta\nn_2}\cdot\big(\nn_1(\nn_1\cdot\partial_t\nn_2)+\nn_3(\nn_3\cdot\partial_t\nn_2)\big)\nonumber\\
&\,+\frac{\delta\CF_{Bi}}{\delta\nn_3}\cdot\big(\nn_1(\nn_1\cdot\partial_t\nn_3)+\nn_2(\nn_2\cdot\partial_t\nn_3)\big)\Big)\ud\xx\nonumber\\
=&\,\int\bigg[ n_{3k}\partial_tn_{2k}\Big(\nn_3\cdot\frac{\delta \CF_{Bi}}{\delta\nn_2}-\nn_2\cdot\frac{\delta \CF_{Bi}}{\delta\nn_3}\Big)+n_{1k}\partial_tn_{3k}\Big(\nn_1\cdot\frac{\delta \CF_{Bi}}{\delta\nn_3}-\nn_3\cdot\frac{\delta \CF_{Bi}}{\delta\nn_1}\Big)\nonumber\\
&\,+n_{2k}\partial_tn_{1k}\Big(\nn_2\cdot\frac{\delta \CF_{Bi}}{\delta\nn_1}-\nn_1\cdot\frac{\delta \CF_{Bi}}{\delta\nn_2}\Big)\bigg]\ud\xx\nonumber\\
=&\,\int\Big(n_{3k}\partial_tn_{2k}\ML_1\CF_{Bi}+n_{1k}\partial_tn_{3k}\ML_2\CF_{Bi}+n_{2k}\partial_tn_{1k}\ML_3\CF_{Bi}\Big)\ud\xx.
\end{align}
Taking the inner product on the equation (\ref{frame-equation-v0}) with $\vv^{(0)}$ and using $\nabla\cdot\vv^{(0)}=0$, we derive that
\begin{align}\label{energy-v-time}
\frac{\rho_s}{2}\frac{\ud}{\ud t}\int|\vv^{(0)}|^2\ud\xx=&\,-\langle\sigma^{(0)}_v,\A_0\rangle-\langle\sigma^{(0)}_e,\nabla\vv^{(0)}\rangle+ck_BT\langle\mathfrak{F},\vv^{(0)}\rangle,
\end{align}
where
\begin{align*}
ck_BT\langle\mathfrak{F},\vv^{(0)}\rangle=&\,\int v^{(0)}_i\big(n_{3k}\partial_in_{2k}\ML_1\CF_{Bi}+n_{1k}\partial_in_{3k}\ML_2\CF_{Bi}
+n_{2k}\partial_in_{1k}\ML_3\CF_{Bi}\big)\ud\xx.
\end{align*}
Combining (\ref{energy-time-t}) with (\ref{energy-v-time}), and using the equations \eqref{frame-equation-n1}--\eqref{frame-equation-n3}, we obtain the energy dissipation law,
\begin{align}
&\,\frac{\ud}{\ud t}\Big(\frac{\rho_s}{2}\int|\vv^{(0)}|^2\ud\xx+\CF_{Bi}(\Fp)\Big)\nonumber\\
&\,=-\langle\sigma^{(0)}_v,\A_0\rangle-\langle\sigma^{(0)}_e,\nabla\vv^{(0)}\rangle\nonumber\\
&\,\quad+\int\Big((\dot{\nn}_2\cdot\nn_3)\ML_1\CF_{Bi}+(\dot{\nn}_3\cdot\nn_1)\ML_2\CF_{Bi}+(\dot{\nn}_1\cdot\nn_2)\ML_3\CF_{Bi}\Big)\ud\xx\nonumber\\
&\,=-\int\omega^T_s(\eta\Lambda^{-1}+P)\omega_s\ud\xx+ck_BT\Bigg(-\beta_1\|\A_0\cdot\sss_1\|^2_{L^2}-2\beta_0\int(\A_0\cdot\sss_1)(\A_0\cdot\sss_2)\ud\xx\nonumber\\
&\,\quad-\beta_2\|\A_0\cdot\sss_2\|^2_{L^2}-\beta_3\|\A_0\cdot\sss_3\|^2_{L^2}+\eta_3\int\Big(\dot{\nn}_1\cdot\nn_2-\frac{1}{2}\BOm_0\cdot\aaa_1\Big)(\A_0\cdot\sss_3)\ud\xx\nonumber\\
&\,\quad-\beta_4\|\A_0\cdot\sss_4\|^2_{L^2}+\eta_2\int\Big(\dot{\nn}_3\cdot\nn_1-\frac{1}{2}\BOm_0\cdot\aaa_2\Big)(\A_0\cdot\sss_4)\ud\xx\nonumber\\
&\,\quad-\beta_5\|\A_0\cdot\sss_5\|^2_{L^2}+\eta_1\int\Big(\dot{\nn}_2\cdot\nn_3-\frac{1}{2}\BOm_0\cdot\aaa_3\Big)(\A_0\cdot\sss_5)\ud\xx\nonumber\\
&\,\quad-\frac{1}{2}\eta_3\int(\A_0\cdot\sss_3)\BOm_0\cdot\aaa_1\ud\xx+\frac{1}{2}\chi_3\int\Big(\dot{\nn}_1\cdot\nn_2-\frac{1}{2}\BOm_0\cdot\aaa_1\Big)\BOm_0\cdot\aaa_1\ud\xx\nonumber\\
&\,\quad-\frac{1}{2}\eta_2\int(\A_0\cdot\sss_4)\BOm_0\cdot\aaa_2\ud\xx+\frac{1}{2}\chi_2\int\Big(\dot{\nn}_3\cdot\nn_1-\frac{1}{2}\BOm_0\cdot\aaa_2\Big)\BOm_0\cdot\aaa_2\ud\xx\nonumber\\
&\,\quad-\frac{1}{2}\eta_1\int(\A_0\cdot\sss_5)\BOm_0\cdot\aaa_3\ud\xx+\frac{1}{2}\chi_1\int\Big(\dot{\nn}_2\cdot\nn_3-\frac{1}{2}\BOm_0\cdot\aaa_3\Big)\BOm_0\cdot\aaa_3\ud\xx\nonumber\\
&\,\quad+\int(\dot{\nn}_2\cdot\nn_3)\Big[-\chi_1\Big(\dot{\nn}_2\cdot\nn_3-\frac{1}{2}\BOm_0\cdot\aaa_3\Big)+\eta_1\A_0\cdot\sss_5\Big]\ud\xx\nonumber\\
&\,\quad+\int(\dot{\nn}_3\cdot\nn_1)\Big[-\chi_2\Big(\dot{\nn}_3\cdot\nn_1-\frac{1}{2}\BOm_0\cdot\aaa_2\Big)+\eta_2\A_0\cdot\sss_4\Big]\ud\xx\nonumber\\
&\,\quad+\int(\dot{\nn}_1\cdot\nn_2)\Big[-\chi_3\Big(\dot{\nn}_1\cdot\nn_2-\frac{1}{2}\BOm_0\cdot\aaa_1\Big)+\eta_3\A_0\cdot\sss_3\Big]\ud\xx\Bigg)\nonumber\\
&\,=-\int\omega^T_s(\eta\Lambda^{-1}+P)\omega_s\ud\xx +ck_BT\Bigg(-\beta_1\|\A_0\cdot\sss_1\|^2_{L^2}-2\beta_0\int(\A_0\cdot\sss_1)(\A_0\cdot\sss_2)\ud\xx\nonumber\\
&\,\quad{-\beta_2\|\A_0\cdot\sss_2\|^2_{L^2}}-\beta_3\|\A_0\cdot\sss_3\|^2_{L^2}-\beta_4\|\A_0\cdot\sss_4\|^2_{L^2}-\beta_5\|\A_0\cdot\sss_5\|^2_{L^2}\nonumber\\
&\,\quad+2\eta_3\int\Big(\dot{\nn}_1\cdot\nn_2-\frac{1}{2}\BOm_0\cdot\aaa_1\Big)(\A_0\cdot\sss_3)\ud\xx\nonumber\\
&\,\quad+2\eta_2\int\Big(\dot{\nn}_3\cdot\nn_1-\frac{1}{2}\BOm_0\cdot\aaa_2\Big)(\A_0\cdot\sss_4)\ud\xx\nonumber\\
&\,\quad+2\eta_1\int\Big(\dot{\nn}_2\cdot\nn_3-\frac{1}{2}\BOm_0\cdot\aaa_3\Big)(\A_0\cdot\sss_5)\ud\xx\nonumber\\
&\,\quad-\chi_1\Big\|\dot{\nn}_2\cdot\nn_3-\frac{1}{2}\BOm_0\cdot\aaa_3\Big\|^2_{L^2}-\chi_2\Big\|\dot{\nn}_3\cdot\nn_1-\frac{1}{2}\BOm_0\cdot\aaa_2\Big\|^2_{L^2}\nonumber\\
&\,\quad-\chi_3\Big\|\dot{\nn}_1\cdot\nn_2-\frac{1}{2}\BOm_0\cdot\aaa_1\Big\|^2_{L^2}\Bigg). \label{EngDissip}
\end{align}
The dissipation can be recognized by noticing the following facts:
\begin{itemize}
    \item $\Lambda$ and $P$ are positive definite.
    \item $\beta_1,\beta_2\ge 0$ and $\beta_0^2\le \beta_1\beta_2$. This comes from \eqref{beta012}.
    \item $\beta_3,\chi_3\ge 0$ and $\eta_3^2\le \beta_3\chi_3$. To realize this, we use the expressions $\beta_3=V_1^TM_1^{-1}V_1$, $\chi_3=4V_4^TM_1^{-1}V_4$ and $\eta_3=2V_4^TM_1^{-1}V_1$ and the fact that $M_1$ is positive definite.
\end{itemize}

\section{Reduction to uniaxial dynamics}\label{unaxial-dynamics}

In the tensor model, the minimum of the bulk energy \eqref{free-energy-bulk} might be uniaxial in the form
\begin{align}
  Q_i=s_i\Big(\nn_1^2-\frac{\Fi}{3}\Big),~i=1,2. \label{Q-uni}
\end{align}
In this case, the local anisotropy is axisymmetric, and the corresponding hydrodynamics is reduced to the Ericksen--Leslie theory, which we derive in the following.

The most important thing is that the form of tensors in Theorem \ref{Q-biaxial-theorem} will be reduced.
\begin{theorem}\label{Q-unixial-theorem}
Assume that $Q_1$ and $Q_2$ have the uniaxial form \eqref{Q-uni}.
Then, the high-order symmetric traceless tensors obtained from closure by the original entropy or the quasi-entropy have the following form,
\begin{align*}\quad
  &\,\langle\mm_1\mm_2\mm_3\rangle=0,\quad
  \langle (\mm_1^4)_0\rangle=a_1(\nn_1^4)_0,\\
  &\,\langle (\mm_2^4)_0\rangle=\tilde{a}_1(\nn_1^4)_0,\quad
  \langle (\mm_1^2\mm_2^2)_0\rangle=\bar{a}_1(\nn_1^4)_0.
\end{align*}
\end{theorem}
The proof is left to Appendix \ref{Apprendix-D}.

For the elastic energy, we notice that when $b_1=b_2=0$, it holds $J_2=J_3=J_5=J_6=0$ in (\ref{J-coefficients-6}).
According to \eqref{elest-bi-11}, the elastic energy only depends on $\nn_1$,
\begin{align}\label{uni-elast-energy+app}
\CF_{Un}=&\,\int\ud\xx\frac{1}{2}\Big(\widetilde{J}_1(\partial_in_{1j})^2+\widetilde{J}_4(|\nabla\cdot\nn_1|^2+n_{1i}n_{1j}\partial_in_{1k}\partial_jn_{1k})\Big)\ud\xx.
\end{align}
Such an $\CF_{Un}$ can be written in the form of the Oseen-Frank energy (which we omit the derivation here), and the coefficients are given by
\begin{align*}
\widetilde{J}_1=&\,2k_BTc^2(c_{22}s^2_1+c_{23}s^2_2+2c_{24}s_1s_2),\\
\widetilde{J}_4=&\,k_BTc^2(c_{28}s^2_1+c_{29}s^2_2+2c_{2,10}s_1s_2).
\end{align*}
From (\ref{uni-elast-energy+app}) we immediately get
\begin{align*}
\frac{\delta\CF_{Un}}{\delta\nn_2}=\frac{\delta\CF_{Un}}{\delta\nn_3}=0,
\end{align*}
which implies that
\begin{align}\label{variation-n2-n3=0}
\ML_1\CF_{Bi}=&\,\nn_3\cdot\frac{\delta \CF_{Bi}}{\delta\nn_2}-\nn_2\cdot\frac{\delta \CF_{Bi}}{\delta\nn_3}
=0,\\
\ML_2\CF_{Bi}=&\,\nn_1\cdot\frac{\delta \CF_{Bi}}{\delta\nn_3}-\nn_3\cdot\frac{\delta \CF_{Bi}}{\delta\nn_1}
=-\nn_3\cdot\frac{\delta \CF_{Un}}{\delta\nn_1},\label{uniaxial-CL-ML2}\\
\ML_3\CF_{Bi}=&\,\nn_2\cdot\frac{\delta \CF_{Bi}}{\delta\nn_1}-\nn_1\cdot\frac{\delta \CF_{Bi}}{\delta\nn_2}
=\nn_2\cdot\frac{\delta
\CF_{Un}}{\delta\nn_1}.\label{uniaxial-CL-ML3}
\end{align}

By $b_1,b_2=0$ and Theorem \ref{Q-unixial-theorem}, in the matrices $M_{ij}$, $N_i^u$ and $P$ (see \eqref{M11block}--\eqref{P-matrix}), the coefficients of $X_3, X_5, X_6$ are all zero.
The matrices $X_1, X_2$ and $X_4$ are all diagonal matrices with their elements satisfying the following relations:
\begin{align}\label{X-relation-elem}
 (X_{i})_{33}=(X_i)_{44},~ (X_i)_{55}=4(X_i)_{22},\quad i=1,2,4.
\end{align}

Thus, the blocks in \eqref{M-matrix} become
\begin{align*}
M_0=&\,\left(
    \begin{array}{cccc}
        {\alpha}_{11} &\, {\beta}_{11} &\, 0 &\, 0 \vspace{0.5ex}\\
        {\beta}_{11} &\, {\gamma}_{11} &\, 0 &\, 0 \vspace{0.5ex}\\
        0 &\, 0 &\, {\alpha}_{22} &\, {\beta}_{22} \vspace{0.5ex}\\
       0 &\, 0 &\, {\beta}_{22} &\, {\gamma}_{22}
    \end{array}
    \right)
    \eqdefa\left(
    \begin{array}{cc}
        M_{01} &\,  \vspace{0.5ex}\\
         &\, M_{02}
    \end{array}
    \right),\\
M_1=&M_2=\left(
    \begin{array}{cc}
        {\alpha}_{33} &\, {\beta}_{33} \vspace{0.5ex}\\
        {\beta}_{33} &\, {\gamma}_{33}
    \end{array}
    \right),\
M_3=4\left(
    \begin{array}{cc}
        {\alpha}_{22} &\, {\beta}_{22} \vspace{0.5ex}\\
        {\beta}_{22} &\, {\gamma}_{22}
    \end{array}
    \right).
\end{align*}
Similarly, the blocks in \eqref{V-block}
are reduced to
\begin{align*}
V_0=&\,\left(
    \begin{array}{cc}
        \mu_{11} &\, 0
        \vspace{0.5ex}\\
        \nu_{11} &\, 0
        \vspace{0.5ex}\\
        0 &\, \mu_{22}
        \vspace{0.5ex}\\
        0 &\, \nu_{22}
    \end{array}
    \right)\eqdefa
    \left(
    \begin{array}{cc}
        V_{01} &\,  \vspace{0.5ex}\\
         &\, V_{02}
    \end{array}
    \right),\quad
V_1=V_2=\left(
    \begin{array}{c}
        \mu_{33}
        \vspace{0.5ex}\\
        \nu_{33}
    \end{array}
    \right),\\
V_3=&\,4\left(
    \begin{array}{c}
        \mu_{22}
        \vspace{0.5ex}\\
        \nu_{22}
    \end{array}
    \right),\quad
V_4=\,\left(
    \begin{array}{c}
        s_1
        \vspace{0.5ex}\\
        s_2
    \end{array}
    \right),\quad
V_5=-\left(
    \begin{array}{c}
        s_1
        \vspace{0.5ex}\\
        s_2
    \end{array}
    \right)=-V_4,\quad V_6=\left(
    \begin{array}{c}
        0
        \vspace{0.5ex}\\
        0
    \end{array}
    \right).
\end{align*}

We know from (\ref{variation-n2-n3=0}) and $V_6$ is a zero vector that the equation (\ref{frame-equation-n1}) disappears.
Meanwhile, noting the following relations between coefficients,
\begin{align*}
\chi_3=\chi_2=\,4V^T_4M_1^{-1}V_4>0,\quad \eta_3=-\eta_2=2V^T_4M_1^{-1}V_1.
\end{align*}
we could simplify the equations (\ref{frame-equation-n2}) and (\ref{frame-equation-n3}) as
\begin{align}\label{two-uniaxal-equations}
\left\{
\begin{array}{l}
\chi_2\big(\dot{\nn}_3\cdot\nn_1-\frac{1}{2}\BOm_0\cdot\aaa_2\big)-\eta_2\A_0\cdot\sss_4-\frac{1}{ck_BT}\nn_3\cdot\frac{\delta\CF_{Un}}{\delta\nn_1}=0,
\vspace{1.5ex}\\
\chi_2\big(\dot{\nn}_1\cdot\nn_2-\frac{1}{2}\BOm_0\cdot\aaa_1\big)+\eta_2\A_0\cdot\sss_3+\frac{1}{ck_BT}\nn_2\cdot\frac{\delta\CF_{Un}}{\delta\nn_1}=0.
\end{array}
\right.
\end{align}

Denote the rotational derivative of the director and the molecular field as
\begin{align*}
\NN_1=\dot{\nn}_1-\Omega^{(0)}_{ij}n_{1j},\quad \hh_1=-\frac{1}{ck_BT}\frac{\delta\CF_{Un}}{\delta\nn_1}.
\end{align*}
Then, based on the facts that
\begin{align*}
&\,\dot{\nn}_3\cdot\nn_1-\frac{1}{2}\BOm_0\cdot\aaa_2=-\dot{\nn}_1\cdot\nn_3+\Omega^{(0)}_{ij}n_{1i}n_{3j}=-\NN_1\cdot\nn_3,\\
&\,\A_0\cdot\sss_3=A^{(0)}_{ij}n_{1i}n_{2j},\quad A_0\cdot\sss_4=A^{(0)}_{ij}n_{1i}n_{3j},
\end{align*}
the equations (\ref{two-uniaxal-equations}) can be rewritten as
\begin{align*}
\left\{
\begin{array}{l}
n_{3j}\big(h_{1j}-\chi_2N_{1j}-\eta_2A^{(0)}_{ij}n_{1i}\big)=0,
\vspace{1.5ex}\\
n_{2j}\big(h_{1j}-\chi_2N_{1j}-\eta_2A^{(0)}_{ij}n_{1i}\big)=0,
\end{array}
\right.
\end{align*}
which further implies
\begin{align}\label{uniaxal-equation-n1}
\nn_1\times\big(\hh_1-\chi_2\NN_1-\eta_2\A_0\nn_1\big)=0.
\end{align}

It remains to reduce the equation (\ref{frame-equation-v0}) to the unixial case. It follows from the derivations above about the blocks in $M$ and $V$ that
\begin{align*}
&\,\beta_0=0,\quad\eta_1=0,\quad\chi_1=0,\\
&\,\chi_2=\chi_3,\quad\eta_2=-\eta_3,\quad\beta_3=\beta_4,\quad\beta_5=4\beta_2,
\end{align*}
and the coefficients $\beta_1, \beta_2, \beta_3$ are given by
\begin{align*}
\beta_1=V^T_{01}M^{-1}_{01}V_{01},\quad\beta_2=V^T_{02}M^{-1}_{02}V_{02},\quad \beta_3=V^T_1M_1^{-1}V_1.
\end{align*}

To shorten the notations, we define five tensors as follows:
\begin{align}\label{five-tensor-GG}
\left\{
\begin{array}{l}
    \GG_1=(\A_0\cdot\sss_2)\sss_2+4(\A_0\cdot\sss_5)\sss_5,\\
    \GG_2=(\A_0\cdot\sss_3)\sss_3+(\A_0\cdot\sss_4)\sss_4,\\
    \GG_3=(\A_0\cdot\sss_3)\aaa_1-(\A_0\cdot\sss_4)\aaa_2,\\
    \GG_4=(\nn_2\cdot\NN_1)\sss_3+(\nn_3\cdot\NN_1)\sss_4,\\
    \GG_5=(\nn_2\cdot\NN_1)\aaa_1-(\nn_3\cdot\NN_1)\aaa_2.
\end{array}
\right.
\end{align}
Consequently, the viscosity stress $\sigma^{(0)}_v$ and the elastic stress $\sigma^{(0)}_e$ can be reduced to, respectively,
\begin{align}
\sigma^{(0)}_v=&\Big(\frac{3}{2}\eta+\vartheta_{11}\Big)(\A_0\cdot\sss_1)\sss_1+\Big(\frac{1}{2}\eta+\vartheta_{22}\Big)\GG_1+\Big(2\eta+\vartheta_{33}\Big)\GG_2,\label{uniaxial-sigma-v}\\
\frac{1}{ck_BT}\sigma^{(0)}_e=&\beta_1(\A_0\cdot\sss_1)\sss_1+\beta_2\GG_1+\beta_3\GG_2-\eta_2\GG_4-\frac{1}{2}\eta_2\GG_3-\frac{1}{2}\chi_2\GG_5.\label{uniaxial-sigma-e}
\end{align}

A direct calculation shows that
\begin{align*}
&\,(\nn^2_2-\nn^2_3)\otimes(\nn^2_2-\nn^2_3)+4\nn_2\nn_3\otimes\nn_2\nn_3\\
&\,=(\nn^2_2+\nn^2_3)\otimes(\nn^2_2+\nn^2_3)-2\nn^2_2\otimes\nn^2_3-2\nn^2_3\otimes\nn^2_2
+\nn_2\otimes\nn_3\otimes\nn_2\otimes\nn_3\\
&\,\quad+\nn_2\otimes\nn_3\otimes\nn_3\otimes\nn_2+\nn_3\otimes\nn_2\otimes\nn_2\otimes\nn_3+\nn_3\otimes\nn_2\otimes\nn_3\otimes\nn_2\\
&\,=(\Fi-\nn^2_1)\otimes(\Fi-\nn^2_1)-(n_{2i}n_{3l}-n_{3i}n_{2l})(n_{2j}n_{3k}-n_{3j}n_{2k})\\
&\,\quad-(n_{2i}n_{3k}-n_{3i}n_{2k})(n_{2j}n_{3l}-n_{3j}n_{2l})\\
&\,=(\Fi-\nn^2_1)\otimes(\Fi-\nn^2_1)-(\epsilon^{il\alpha}\epsilon^{jk\beta}+\epsilon^{ik\alpha}\epsilon^{jl\beta})n_{1\alpha}n_{1\beta}\\
&\,=(\Fi-\nn^2_1)\otimes(\Fi-\nn^2_1)-2\delta_{ij}\delta_{kl}+\delta_{ik}\delta_{jl}+\delta_{il}\delta_{jk}\\
&\,\quad+2\delta_{ij}(\nn^2_1)_{kl}+2\delta_{kl}(\nn^2_1)_{ij}-\delta_{ik}(\nn^2_1)_{jl}-\delta_{il}(\nn^2_1)_{jk}-\delta_{jk}(\nn^2_1)_{il}-\delta_{jl}(\nn^2_1)_{ik}.
\end{align*}
From this, $\GG_1$ can be expressed as
\begin{align}\label{GG_1}
\GG_1=(\A_0\cdot\nn^2_1)(\nn^2_1-\Fi)+2\A_0-2(A^{(0)}_{ik}n_{1k}n_{1j}+n_{1i}n_{1k}A^{(0)}_{kj}).
\end{align}
For the tensor $\GG_2$, it follows that
\begin{align}\label{GG_2}
\GG_2
&\,=\frac{1}{4}A^{(0)}_{ij}(n_{1i}n_{2j}+n_{2i}n_{1j})(n_{1k}n_{2l}+n_{2k}n_{1l})\nonumber\\
&\,\quad+\frac{1}{4}A^{(0)}_{ij}(n_{1i}n_{3j}+n_{3i}n_{1j})(n_{1k}n_{3l}+n_{3k}n_{1l})\nonumber\\
&\,=\frac{1}{4}\Big[A^{(0)}_{ij}\big(n_{1i}n_{2j}n_{1k}n_{2l}+n_{2i}n_{1j}n_{1k}n_{2l}+n_{1i}n_{2j}n_{2k}n_{1l}+n_{2i}n_{1j}n_{2k}n_{1l}\big)\nonumber\\
&\,\quad+A^{(0)}_{ij}\big(n_{1i}n_{3j}n_{1k}n_{3l}+n_{3i}n_{1j}n_{1k}n_{3l}+n_{1i}n_{3j}n_{3k}n_{1l}+n_{3i}n_{1j}n_{3k}n_{1l}\big)\Big]\nonumber\\
&\,=\frac{1}{4}\Big[A^{(0)}_{ij}\Big((\delta_{jl}-n_{1j}n_{1l})n_{1i}n_{1k}+(\delta_{il}-n_{1i}n_{1l})n_{1j}n_{1k}\Big)\nonumber\\
&\,\quad+A^{(0)}_{ij}\Big((\delta_{jk}-n_{1j}n_{1k})n_{1i}n_{1l}+(\delta_{ik}-n_{1i}n_{1k})n_{1j}n_{1l}\Big)\Big]\nonumber\\
&\,=\frac{1}{2}(A^{(0)}_{ik}n_{1k}n_{1j}+n_{1i}n_{1k}A^{(0)}_{kj})-(\A_0\cdot\nn^2_1)\nn^2_1.
\end{align}
Using the relation
\begin{align*}
&\,\Big((\nn_1\otimes\nn_2-\nn_2\otimes\nn_1)\otimes\nn_1\nn_2+(\nn_1\otimes\nn_3-\nn_3\otimes\nn_1)\otimes\nn_1\nn_3\Big)_{ijkl}\\
&\,=\frac{1}{2}\Big[\Big(\nn^2_1-\frac{\Fi}{3}\Big)_{ki}\delta_{jl}-\Big(\nn^2_1-\frac{\Fi}{3}\Big)_{kj}\delta_{il}
+\Big(\nn^2_1-\frac{\Fi}{3}\Big)_{li}\delta_{jk}-\Big(\nn^2_1-\frac{\Fi}{3}\Big)_{lj}\delta_{ik}\Big],
\end{align*}
we obtain
\begin{align}\label{GG_3}
\GG_3=&\,A^{(0)}_{kl}\Big((\nn_1\otimes\nn_2-\nn_2\otimes\nn_1)\otimes\nn_1\nn_2+(\nn_1\otimes\nn_3-\nn_3\otimes\nn_1)\otimes\nn_1\nn_3\Big)_{ijkl}\nonumber\\
=&\,\frac{1}{2}\Big((\nn^2_1)_{ki}A_{kj}-\frac{1}{3}A_{ij}-(\nn^2_1)_{kj}A_{ki}+\frac{1}{3}A_{ij}\nonumber\\
&\,+(\nn^2_1)_{li}A_{jl}-\frac{1}{3}A_{ij}-(\nn^2_1)_{lj}A_{il}+\frac{1}{3}A_{ij}\Big)\nonumber\\
=&\,(\nn^2_1)_{ki}A_{kj}-(\nn^2_1)_{kj}A_{ki}.
\end{align}
In addition, by virtue of the relation $\Fi=\nn^2_1+\nn^2_2+\nn^2_3$, $\GG_4$ and $\GG_5$ can be calculated as, respectively,
\begin{align}
\GG_4=&\,(\nn_2\cdot\NN_1)\nn_1\nn_2+(\nn_3\cdot\NN_1)\nn_1\nn_3\nonumber\\
=&\,\frac{1}{2}n_{2i}N_{1i}(n_{1j}n_{2k}+n_{2j}n_{1k})+\frac{1}{2}n_{3i}N_{1i}(n_{1j}n_{3k}+n_{3j}n_{1k})\nonumber\\
=&\,\frac{1}{2}(\delta_{ik}-n_{1i}n_{1k})N_{1i}n_{1j}+\frac{1}{2}(\delta_{ij}-n_{1i}n_{1k})N_{1i}n_{1j}\nonumber\\
=&\,\frac{1}{2}(\nn_1\otimes\NN_1+\NN_1\otimes\nn_1),\label{GG_4}\\
\GG_5=&\,(\nn_2\cdot\NN_1)(\nn_1\otimes\nn_2-\nn_2\otimes\nn_1)+(\nn_3\cdot\NN_1)(\nn_1\otimes\nn_3-\nn_3\otimes\nn_1)\nonumber\\
=&\,n_{2i}N_{1i}n_{1j}n_{2k}+n_{3i}N_{1i}n_{1j}n_{3k}-(n_{2i}N_{1i}n_{2j}n_{1k}+n_{3i}N_{1i}n_{3j}n_{1k})\nonumber\\
=&\,(\delta_{ik}-n_{1i}n_{1k})N_{1i}n_{1j}-(\delta_{ij}-n_{1i}n_{1j})N_{1i}n_{1k}\nonumber\\
=&\,\nn_1\otimes\NN_1-\NN_1\otimes\nn_1.\label{GG_5}
\end{align}
Hence, from (\ref{GG_1}) and (\ref{GG_2}), the viscous stress $\sigma^{(0)}_v$ can be expressed by
\begin{align}
\sigma^{(0)}_v=&\,(\vartheta_{11}+\vartheta_{22}-\vartheta_{33})(\A_0\cdot\nn^2_1)\nn^2_1+(\eta+2\vartheta_{22})\A_0\nonumber\\
&+\Big(\frac{1}{2}\vartheta_{33}-2\vartheta_{22}\Big)(A^{(0)}_{ik}n_{1k}n_{1j}+n_{1i}n_{1k}A^{(0)}_{kj}),\label{uniaxial-sigma-v1}
\end{align}
where we have neglected the term $(\A_0\cdot\nn^2_1)\Fi$, since it can be absorbed into the pressure.
Furthermore, from (\ref{GG_1})-(\ref{GG_5}), the elastic stress $\sigma^{(0)}_e$ can be expressed by
\begin{align}
\frac{1}{ck_BT}\sigma^{(0)}_e=&(\beta_1+\beta_2-\beta_3)(\A_0\cdot\nn^2_1)\nn^2_1+2\beta_2\A_0\nonumber\\
&+\Big(\frac{1}{2}\beta_3-2\beta_2\Big)\big(A^{(0)}_{ik}n_{1k}n_{1j}+n_{1i}n_{1k}A^{(0)}_{kj}\big)\nonumber\\
&-\frac{1}{2}\eta_2\big(n_{1i}n_{1k}A^{(0)}_{kj}-A^{(0)}_{ik}n_{1k}n_{1j}\big)\nonumber\\
&-\frac{1}{2}\eta_2(\nn_1\otimes\NN_1+\NN_1\otimes\nn_1)-\frac{1}{2}\chi_2(\nn_1\otimes\NN_1-\NN_1\otimes\nn_1).\label{uniaxial-sigma-e1}
\end{align}

Using $\frac{\delta\CF_{Un}}{\delta\nn_2}=\frac{\delta\CF_{Un}}{\delta\nn_3}=0$, the body force can be simplified as
\begin{align}
ck_BT\mathfrak{F}_i=&\,\ML_3\CF_{Bi}\partial_i{\nn}_1\cdot\nn_2+\ML_2\CF_{Bi}\partial_i{\nn}_3\cdot\nn_1+\ML_1\CF_{Bi}\partial_i{\nn}_2\cdot\nn_3\nonumber\\
=&\,\partial_in_{1k}n_{3k}\Big(\nn_3\cdot\frac{\delta \CF_{Un}}{\delta\nn_1}\Big)+n_{2k}\partial_in_{1k}\Big(\nn_2\cdot\frac{\delta \CF_{Un}}{\delta\nn_1}\Big)\nonumber\\
=&\,\partial_in_{1k}\frac{\delta \CF_{Un}}{\delta n_{1\alpha}}(\delta_{k\alpha}-n_{1k}n_{1\alpha})\nonumber\\
=&\,\partial_in_{1k}\frac{\delta \CF_{Un}}{\delta n_{1k}}
=\partial_j\sigma^E_{ij},\label{sigma-E}
\end{align}
where $\sigma^E_{ij}=-\frac{\partial\CF_{Un}}{\partial(\partial_jn_{1k})}\partial_in_{1k}$ is called the Ericksen stress, and the calculation of last step in (\ref{sigma-E}) is similar to that of Lemma 3.5 in \cite{WZZ3}.

Therefore, from (\ref{uniaxial-sigma-v1}), (\ref{uniaxial-sigma-e1}) and (\ref{sigma-E}), the equation of the fluid velocity $\vv^{(0)}$ for the uniaxial case is given by
\begin{align}\label{uniaxal-equation-v1}
\rho_s\Big(\frac{\partial\vv^{(0)}}{\partial t}+\vv^{(0)}\cdot\nabla\vv^{(0)}\Big)_i=-\partial_ip^{(0)}+\partial_j(\sigma^L_{ij}+\sigma^E_{ij}).
\end{align}
Here, the Leslie stress $\sigma^L$ is written as
\begin{align}
\sigma^L=&\sigma^{(0)}_v+\sigma^{(0)}_e\nonumber\\
=&\alpha_1(\A_0\cdot\nn^2_1)\nn^2_1+\alpha_2\nn_1\otimes\NN_1+\alpha_3\NN_1\otimes\nn_1+\alpha_4\A_0+\alpha_5n_{1i}n_{1k}A^{(0)}_{kj}+\alpha_6A^{(0)}_{ik}n_{1k}n_{1j},
\end{align}
where the coefficients $\alpha_i(i=1,\cdots,6)$ are given by
\begin{align*}
\alpha_1=&\,\vartheta_{11}+\vartheta_{22}-\vartheta_{33}+ck_BT(\beta_1+\beta_2-\beta_3),\\
\alpha_2=&\,-\frac{1}{2}ck_BT(\chi_2+\eta_2),\quad \alpha_3=\frac{1}{2}ck_BT(\chi_2-\eta_2),\\
\alpha_4=&\,\eta+2\vartheta_{22}+2ck_BT\beta_2,\\ \alpha_5=&\,\frac{1}{2}\Big(\vartheta_{33}+ck_BT(\beta_3-\eta_2)\Big)-2(\vartheta_{22}+ck_BT\beta_2),\\
\alpha_6=&\,\frac{1}{2}\Big(\vartheta_{33}+ck_BT(\beta_3+\eta_2)\Big)-2(\vartheta_{22}+ck_BT\beta_2),
\end{align*}
which satisfy the following relations:
\begin{align*}
&\alpha_2+\alpha_3=\alpha_6-\alpha_5,\\
&ck_BT\chi_2=\alpha_3-\alpha_2,\quad ck_BT\eta_2=\alpha_6-\alpha_5.
\end{align*}

The equations (\ref{uniaxal-equation-n1}) and (\ref{uniaxal-equation-v1}) can be called the Ericksen-Leslie system, which also keep the following energy dissipation:
\begin{align}\label{unixial-energy-dissip}
&\,\frac{\ud}{\ud t}\Big(\frac{\rho_s}{2}\int|\vv^{(0)}|^2\ud\xx+\CF_{Un}(\nn_1)\Big)\nonumber\\
=&\,-\int\bigg(\Big(\alpha_1+ck_BT\frac{\eta^2_2}{\chi_2}\Big)|\A_0\cdot\nn^2_1|^2+\alpha_4|\A_0|^2\nonumber\\
&+\Big(\alpha_5+\alpha_6-ck_BT\frac{\eta^2_2}{\chi_2}\Big)|\A_0\nn_1|^2+ck_BT\frac{1}{\chi_2}|\nn_1\times\hh_1|^2\bigg)\ud\xx.
\end{align}

We denote
\begin{align*}
\alpha'_1=\alpha_1+ck_BT\frac{\eta^2_2}{\chi_2},\quad \alpha'_2=\alpha_4,\quad \alpha'_3=\alpha_5+\alpha_6-ck_BT\frac{\eta^2_2}{\chi_2}.
\end{align*}
It can be seen from Proposition 2.2 in \cite{WZZ1} that the first three terms in (\ref{unixial-energy-dissip}) are negative semi-definite if and only if
\begin{align*}
    \alpha_2'\geq0,\quad2\alpha_2'+\alpha_3'\geq0,\quad\frac{3}{2}\alpha_2'+\alpha_3'+\alpha_1'\ge 0.
\end{align*}
From our derivation, it holds
\begin{align*}
&\alpha'_2=\eta+2\vartheta_{22}+2ck_BT\beta_2,\\
&2\alpha'_2+\alpha'_3=2\eta+\vartheta_{33}+ck_BT\left(\beta_3-\frac{\eta^2_2}{\chi_2}\right),\\
&\frac{3}{2}\alpha'_2+\alpha'_3+\alpha'_1=\frac{3}{2}\eta+\vartheta_{11}+ck_BT\beta_{1}.
\end{align*}
They are indeed nonnegative, since we have $\eta>0$, $\vartheta_{11},\vartheta_{22},\vartheta_{33}> 0$ from the positive definiteness of $P$, and $\beta_1,\beta_2,\beta_3>0,\beta_3\chi_2-\eta_2^2>0$ from the positive definiteness of $M$.

\section{Comparison with other models}\label{comparison}

In previous works, the discussion of biaxial hydrodynamics focused on the dissipation function, i.e. \eqref{EngDissip}.
If the dissipation function is determined, the hydrodynamics can be established by deriving the forces from it and apply the Newton's law.
For this reason, we compare the dissipation function in this work and those in previous works.
Although the dissipation function has different expressions previously, they turn out to be equivalent as claimed in \cite{GV2}.
Thus, we choose the expression in \cite{GV2} for comparison.

We use the expressions of $\Lambda$ and $P$ in \eqref{Lambda} and \eqref{P-matrix}, respectively.
Then, the dissipation function in \eqref{EngDissip} can be rewritten as (to simplify the presentation, we omit $ck_BT$ below)
\begin{align*}
&\,\quad-(\beta_1+\vartheta_{11}+\frac{3}{2}\eta)\|\A_0\cdot\sss_1\|^2_{L^2}-2(\beta_0+\vartheta_{12})\int(\A_0\cdot\sss_1)(\A_0\cdot\sss_2)\ud\xx\nonumber\\
&\,\quad-(\beta_2+\vartheta_{22}+\frac{1}{2}\eta)\|\A_0\cdot\sss_2\|^2_{L^2}-(\beta_3+\vartheta_{33}+2\eta)\|\A_0\cdot\sss_3\|^2_{L^2}\nonumber\\
&\,\quad-(\beta_4+\vartheta_{44}+2\eta)\|\A_0\cdot\sss_4\|^2_{L^2}-(\beta_5+\vartheta_{55}+2\eta)\|\A_0\cdot\sss_5\|^2_{L^2}\nonumber\\
&\,\quad+2\eta_3\int\Big(\dot{\nn}_1\cdot\nn_2-\frac{1}{2}\BOm_0\cdot\aaa_1\Big)(\A_0\cdot\sss_3)\ud\xx\nonumber\\
&\,\quad+2\eta_2\int\Big(\dot{\nn}_3\cdot\nn_1-\frac{1}{2}\BOm_0\cdot\aaa_2\Big)(\A_0\cdot\sss_4)\ud\xx\nonumber\\
&\,\quad+2\eta_1\int\Big(\dot{\nn}_2\cdot\nn_3-\frac{1}{2}\BOm_0\cdot\aaa_3\Big)(\A_0\cdot\sss_5)\ud\xx\nonumber\\
&\,\quad-\chi_1\Big\|\dot{\nn}_2\cdot\nn_3-\frac{1}{2}\BOm_0\cdot\aaa_3\Big\|^2_{L^2}-\chi_2\Big\|\dot{\nn}_3\cdot\nn_1-\frac{1}{2}\BOm_0\cdot\aaa_2\Big\|^2_{L^2}\nonumber\\
&\,\quad-\chi_3\Big\|\dot{\nn}_1\cdot\nn_2-\frac{1}{2}\BOm_0\cdot\aaa_1\Big\|^2_{L^2}.
\end{align*}
It has twelve terms that are exactly those given in \cite{GV2}.
While the form is identical, we manage to derive the coefficients from the physical parameters, which is not attained previously.

\section{Conclusion}\label{concl}

Using the Hilbert expansion, we derive a frame hydrodynamics for the biaxial nematic phase from a molecular-theory-based tensor model.
Its coefficients are all expressed as those in the tensor model, and the energy dissipation is maintained.
The model is further reduced to the Ericksen--Leslie model if the bulk energy minimum becomes uniaxial.

The key ingredient is to recognize the form of the high-order tensors from the properties of the original entropy or the quasi-entropy.
This technique is also applicable to other mesoscopic symmetries.
It calls for expressions of tensors under other symmetries \cite{Xu3,XC,Xu4}, which we aim to investigate in future works.

\appendix

\section{Symmetric traceless tensors}\label{Apprendix-A}

As we have mentioned, any tensor can be decomposed into symmetric traceless tensors.
To carry out calculations of high-order tensors, it is necessary to discuss some fundamental ingredients of symmetric traceless tensors.

For a tensor $U$ expressed in the basis generated by $\Fq=(\mm_1,\mm_2,\mm_3)$, let us denote it as a function of $\Fq$, i.e. $U(\Fq)$, to allow $\Fq$ to vary.
For example, let us consider a tensor $U(\Fq)=3\mm_1\otimes\mm_3-\mm_3\otimes\mm_2$.
For another orthonormal frame $\Fq'=(\mm_1',\mm_2',\mm_3')$, we mean  $U(\Fq')=3\mm_1'\otimes\mm_3'-\mm_3'\otimes\mm_2'$.

\subsection{Basis of symmetric traceless tensors}\label{Apprendix-A1}

Any symmetric tensor can generate a symmetric traceless tensor in the form \eqref{symtrlsnot}.
To write down a basis of symmetric traceless tensors of certain order, we could choose those generated by monomials.
Their expressions are derived previously \cite{Xu1}.
Below, we list the third-order and fourth-order tensors that we will make use of.

A basis of third-order symmetric traceless tensors can be given by
\begin{align*}
    &\,(\mm_1^3)_0,\ (\mm_1^2\mm_2)_0,\ (\mm_1\mm_2^2)_0,\ (\mm_2^3)_0, \\
    &\,(\mm_1^2\mm_3)_0,\ (\mm_1\mm_2\mm_3)_0,\ (\mm_2^2\mm_3)_0.
\end{align*}
Their expressions are given by
\begin{align*}
    &\,(\mm_1\mm_2\mm_3)_0=\mm_1\mm_2\mm_3, \\
    &\,(\mm_1^3)_0=\mm_1^3-\frac 35\mm_1\Fi,\\
    &\,(\mm_1^2\mm_2)_0=\mm_1^2\mm_2-\frac 15\mm_2\Fi.
\end{align*}
The others can be written down by changing the indices.
A basis of fourth-order symmetric traceless tensors can be given by
\begin{align*}
    &\,(\mm_1^4)_0,\ (\mm_1^3\mm_2)_0,\ (\mm_1^2\mm_2^2)_0,\ (\mm_1\mm_2^3)_0,\ (\mm_2^4)_0, \\
    &\,(\mm_1^3\mm_3)_0,\ (\mm_1^2\mm_2\mm_3)_0,\ (\mm_1\mm_2^2\mm_3)_0,\ (\mm_2^3\mm_3)_0.
\end{align*}
Their expressions are given by
\begin{align*}
    &\,(\mm_1^4)_0=\mm_1^4-\frac{6}{7}\mm_1^2\Fi+\frac{3}{35}\Fi^2, \\
    &\,(\mm_1^3\mm_2)_0=\mm_1^3\mm_2-\frac{3}{7}\mm_1\mm_2\Fi,\\
    &\,(\mm_1^2\mm_2^2)_0=\mm_1^2\mm_2^2-\frac{1}{7}(\mm_1^2+\mm_2^2)\Fi+\frac{1}{35}\Fi^2,\\
    &\,(\mm_1^2\mm_2\mm_3)_0=\mm_1^2\mm_2\mm_3-\frac{1}{7}\mm_2\mm_3\Fi.
\end{align*}
An additional note is that for a monomial with the power of $\mm_3$ not less than two, we could substitute it by $\mm_3^2=\Fi-\mm_1^2-\mm_2^2$ to obtain equations such as (cf. the uniqueness of $W$ in \eqref{symtrlsnot})
\begin{align*}
    (\mm_3^2)_0=(\Fi-\mm_1^2-\mm_2^2)_0=(-\mm_1^2-\mm_2^2)_0,\quad (\mm_3^4)_0=\big((\mm_1^2+\mm_2^2)^2\big)_0.
\end{align*}

For our discussion afterwards, we introduce the group $\MD_2$ that has four elements,
$$
\Fi=\mathrm{diag}(1,1,1),\ \Fb_1=\mathrm{diag}(1,-1,-1),\ \Fb_2=\mathrm{diag}(-1,1,-1),\  \Fb_3=\mathrm{diag}(-1,-1,1).
$$
The tensor $U$ is called invariant of $\MD_2$ if $U(\Fq\Fb_i)=U(\Fq)$ (recall the notation at the beginning of Appendix).
All the invariant tensors of the order $n$ form a linear subspace of $n$th-order symmetric traceless tensors, denoted by $\BA^{\MD_2,n}$.
According to a short discussion in \cite{Xu1}, its orthogonal complement $(\BA^{\MD_2,n})^{\perp}$ consists of all the $n$th-order symmetric traceless tensors $U$ such that $U(\Fq)+U(\Fq\Fb_1)+U(\Fq\Fb_2)+U(\Fq\Fb_3)=0$.

It is evident that $\Fq\Fb_i$ transforms two of $\mm_1$, $\mm_2$, $\mm_3$ to their opposites.
From the expressions of symmetric traceless tensors written above, we can easily identify the decomposition $\BA^{\MD_2,n}$ and $(\BA^{\MD_2,n})^{\perp}$.
For $n=1,2,3,4$, they are listed below,
\begin{align}
    &\,\BA^{\MD_2,1}=\{0\},\quad (\BA^{\MD_2,1})^{\perp}=\mathrm{span}\{\mm_1,\mm_2,\mm_3\},\nonumber\\
    &\,\BA^{\MD_2,2}=\mathrm{span}\{(\mm_1^2)_0,(\mm_2^2)_0\},\quad (\BA^{\MD_2,2})^{\perp}=\mathrm{span}\{\mm_1\mm_2,\mm_1\mm_3,\mm_2\mm_3\},\nonumber\\
    &\,\BA^{\MD_2,3}=\mathrm{span}\{\mm_1\mm_2\mm_3\},\nonumber\\ &\,(\BA^{\MD_2,3})^{\perp}=\mathrm{span}\{(\mm_1^3)_0,(\mm_1^2\mm_2)_0,(\mm_1\mm_2^2)_0,(\mm_2^3)_0,(\mm_1^2\mm_3)_0,(\mm_2^2\mm_3)_0\},\nonumber\\
    &\,\BA^{\MD_2,4}=\mathrm{span}\{(\mm_1^4)_0,(\mm_1^2\mm_2^2)_0,(\mm_2^4)_0\},\nonumber\\
    &\,(\BA^{\MD_2,4})^{\perp}=\mathrm{span}\{(\mm_1^3\mm_2)_0,(\mm_1\mm_2^3)_0,(\mm_1^3\mm_3)_0,(\mm_1^2\mm_2\mm_3)_0,(\mm_1\mm_2^2\mm_3)_0,(\mm_2^3\mm_3)_0\}. \label{invrnt-decomp}
\end{align}

Let us write down some equalities to be used later.
Define
\begin{align}
\SSS_1=\mm_1^2-\Fi/3,\ \SSS_2=\mm_2^2-\mm_3^2,\ \SSS_3=\mm_1\mm_2,\ \SSS_4=\mm_1\mm_3,\ \SSS_5=\mm_2\mm_3.
\end{align}
For third-order tensors, we have
\begin{align}
    \mm_1\mm_2\mm_3\cdot \mm_i\otimes\SSS_j=&\,0, \quad \text{ if }(i,j)\ne\{(1,5),(2,4),(3,3)\},\nonumber\\
    \epsilon^{ils}(\mm_1\mm_2\mm_3)_{jks}(\SSS_{\nu}\otimes\SSS_{\nu'})_{ijkl}=&\,0,\quad \text{ if }i\ne j\text{ and }\{\nu,\nu'\}\ne\{1,2\}. \label{third-orth1}
\end{align}
If $U\in(\BA^{\MD_2,3})^{\perp}$, then
\begin{align}
    U\cdot \mm_i\otimes\SSS_j=&\,0,\quad \text{ if }(i,j)=\{(1,5),(2,4),(3,3)\},\nonumber\\
    \epsilon^{ils}U_{jks}(\SSS_{\nu}\otimes\SSS_{\nu})_{ijkl}=&\,\epsilon^{ils}U_{jks}(\SSS_1\otimes\SSS_2)_{ijkl}=\epsilon^{ils}U_{jks}(\SSS_2\otimes\SSS_1)_{ijkl}=0. \label{third-orth2}
\end{align}
For fourth-order tensors, if $U\in \BA^{\MD_2,4}$, then
\begin{align}
    &\,U\cdot\SSS_i\otimes\SSS_j=0,\quad \text{ if }i\ne j\text{ and }\{i,j\}\ne\{1,2\}. \label{fourth-orth1}
\end{align}
If $U\in(\BA^{\MD_2,4})^{\perp}$, then
\begin{align}
    &\,U\cdot \SSS_i\otimes\SSS_i=(\mm_1^3\mm_2)_0\cdot \SSS_1\otimes\SSS_2=0. \label{fourth-orth2}
\end{align}
The equalities \eqref{third-orth1}--\eqref{fourth-orth2} can be recognized straightforwardly by expanding the tensors into several terms of tensor products of $\mm_i$.
The Levi-Civita symbol can be expanded as
\begin{align*}
    \epsilon^{ijk}=&\,(\mm_1\otimes\mm_2\otimes\mm_3+\mm_2\otimes\mm_3\otimes\mm_1+\mm_3\otimes\mm_1\otimes\mm_2\\
    &\,-\mm_1\otimes\mm_3\otimes\mm_2-\mm_3\otimes\mm_2\otimes\mm_1-\mm_2\otimes\mm_1\otimes\mm_3)_{ijk}.
\end{align*}

The equations above hold independent of the orthonormal frame we choose.
In particular, they are valid if we substitute $\mm_i$ with $\nn_i$ and correspondingly $\SSS_i$ with $\sss_i$ (recall \eqref{sss-five}).

\subsection{Expressing tensors by symmetric traceless tensors}\label{symtrls-decomp}
There are complicated linear relations between high-order tensors.
To figure out the linear relations, we shall express them by symmetric traceless tensors that completely give linearly independent components.
These linear relations are inherited by averaged high-order tensors.
The special forms of averaged high-order tensors are also revealed in this way.

To express a general tensor $U$ by symmetric traceless tensors, we first decompose it as $U=U_{\rm sym}+(U-U_{\rm sym})$. The anti-symmetric part $U-U_{\rm sym}$ can be written as the sum of several terms of the form
\begin{align*}
U_{\ldots i\ldots j\ldots}-U_{\ldots j\ldots i\ldots}=\epsilon^{ijk}W_{k\ldots},
\end{align*}
where $W$ is an $(n-1)$th-order tensor.
Carry out this action repeatedly until we express $U$ by symmetric tensors.
Then, each symmetric tensor can be expressed by a symmetric traceless tensor and several symmetric tensors of lower order.
Also do it repeatedly to finally express $U$ by symmetric traceless tensors.

In what follows, we use this procedure to deal with the tensors
\begin{align*}
    &(\mm_1^2-\frac{\Fi}{3})\otimes(\mm_1^2-\frac{\Fi}{3}),\ (\mm_1^2-\frac{\Fi}{3})\otimes(\mm_2^2-\mm_3^2),\ (\mm_2^2-\mm_3^2)\otimes (\mm_2^2-\mm_3^2),\\
    &\mm_1\mm_2\otimes\mm_1\mm_2,\ \mm_1\mm_3\otimes\mm_1\mm_3,\ \mm_2\mm_3\otimes\mm_2\mm_3.
\end{align*}
From the expressions of symmetric traceless tensors, we can derive that
\begin{align}
\mm^4_1=&\,(\mm^4_1)_0+\frac{6}{7}\mm^2_1\Fi-\frac{3}{35}\Fi^2\nonumber\\
=&\,(\mm^4_1)_0+\frac{6}{7}\Big(\mm^2_1-\frac{\Fi}{3}\Big)\Fi+\Big(\frac{2}{7}-\frac{3}{35}\Big)\Fi^2\nonumber\\
=&\,(\mm^4_1)_0+\frac{6}{7}(\mm^2_1)_0\Fi+\frac{1}{5}\Fi^2,
\label{m1-4}\\
\mm^2_1\mm^2_2=&\,(\mm^2_1\mm^2_2)_0+\frac{1}{7}(\mm^2_1+\mm^2_2)\Fi-\frac{1}{35}\Fi^2\nonumber\\
=&\,(\mm^2_1\mm^2_2)_0+\frac{1}{7}\Big(\frac{\Fi}{3}-\mm^2_3\Big)\Fi+\Big(\frac{2}{21}-\frac{1}{35}\Big)\Fi^2\nonumber\\
=&\,(\mm^2_1\mm^2_2)_0-\frac{1}{7}(\mm^2_3)_0\Fi+\frac{1}{15}\Fi^2.
\label{m12-m22}
\end{align}

For a second-order tensor $U$, define
 \begin{align*}
\CA(U)_{ijkl}\eqdefa&\,\delta_{kl}U_{ij}+\delta_{ij}U_{kl}-\frac{3}{4}\big(\delta_{ik}U_{jl}+\delta_{jl}U_{ik}+\delta_{il}U_{jk}+\delta_{jk}U_{il}\big),\\
\CB(U)_{ijkl}\eqdefa&\,U_{ki}\delta_{jl}-U_{kj}\delta_{il}+U_{li}\delta_{jk}-U_{lj}\delta_{ik}.
 \end{align*}
Using the expressions of symmetric traceless tensors and \eqref{m1-4}, it follows that
\begin{align}\label{m1-block-traceless}
&\,\Big(\mm^2_1-\frac{1}{3}\Fi\Big)_{ij}\Big(\mm^2_1-\frac{1}{3}\Fi\Big)_{kl}\nonumber\\
&\,=\Big((\mm^4_1)_0+\frac{6}{7}(\mm^2_1)_0\Fi+\frac{1}{5}\Fi^2\Big)_{ijkl}-\frac{1}{3}\delta_{ij}(\mm^2_1)_{kl}-\frac{1}{3}\delta_{kl}(\mm^2_1)_{ij}+\frac{1}{9}\delta_{ij}\delta_{kl}\nonumber\\
&\,=\Big((\mm^4_1)_0+\frac{6}{7}(\mm^2_1)_0\Fi+\frac{1}{5}\Fi^2\Big)_{ijkl}-\frac{1}{3}\delta_{ij}\big((\mm^2_1)_0\big)_{kl}-\frac{1}{3}\delta_{kl}\big((\mm^2_1)_0\big)_{ij}-\frac{1}{9}\delta_{ij}\delta_{kl}\nonumber\\
&\,=\big((\mm^4_1)_0\big)_{ijkl}-\frac{4}{21}\CA\big((\mm^2_1)_0\big)_{ijkl}-\frac{1}{45}(2\delta_{ij}\delta_{kl}-3\delta_{ik}\delta_{jl}-3\delta_{il}\delta_{jk}).
\end{align}

The symmetric tensor $\mm^2_1\mm^2_2$ is expressed by
\begin{align}\label{m1-2-m2-2-sym}
(\mm^2_1\mm^2_2)_{ijkl}=&\,\frac{1}{6}\Big(m_{1i}m_{1j}m_{2k}m_{2l}+m_{2i}m_{2j}m_{1k}m_{1l}\nonumber\\
&\,+(m_{1i}m_{2j}+m_{2i}m_{1j})(m_{1k}m_{2l}+m_{2k}m_{1l})\Big).
\end{align}
Then we obtain from (\ref{m1-2-m2-2-sym}) that
\begin{align}\label{m1m2-sym}
&\,(\mm_1\mm_2\otimes\mm_1\mm_2)_{ijkl}-(\mm_1^2\mm_2^2)_{ijkl}\nonumber\\
&\,\quad=-\frac{1}{12}\Big(2m_{1i}m_{1j}m_{2k}m_{2l}+2m_{2i}m_{2j}m_{1k}m_{1l}
-(m_{1i}m_{2j}+m_{2i}m_{1j})(m_{1k}m_{2l}+m_{2k}m_{1l})\Big).
\end{align}
Using the equality
\begin{align}\label{m1m2-m2m1-e}
m_{1i}m_{2j}-m_{2i}m_{1j}=\epsilon^{ijs}m_{3s},
\end{align}
the tensor in the big parenthesis of (\ref{m1m2-sym}) can be calculated as
\begin{align}\label{F-ijkl}
&\,m_{1i}m_{2l}(m_{1j}m_{2k}-m_{2j}m_{1k})+m_{1i}m_{2k}(m_{1j}m_{2l}-m_{2j}m_{1l})\nonumber\\
&\,+m_{2i}m_{1l}(m_{2j}m_{1k}-m_{1j}m_{2k})+m_{2i}m_{1k}(m_{2j}m_{1l}-m_{1j}m_{2l})\nonumber\\
=&\,(m_{1i}m_{2l}-m_{2i}m_{1l})(m_{1j}m_{2k}-m_{2j}m_{1k})+(m_{1i}m_{2k}-m_{2i}m_{1k})(m_{1j}m_{2l}-m_{2j}m_{1l})\nonumber\\
=&\,(\epsilon^{ils}\epsilon^{jkt}+\epsilon^{iks}\epsilon^{jlt})(\mm^2_3)_{st}\nonumber\\
=&\,2\delta_{ij}\delta_{kl}-\delta_{ik}\delta_{jl}-\delta_{il}\delta_{jk}\nonumber\\
&\,-2\delta_{ij}(\mm^2_3)_{kl}-2\delta_{kl}(\mm^2_3)_{ij}+\delta_{ik}(\mm^2_3)_{jl}+\delta_{il}(\mm^2_3)_{jk}+\delta_{jk}(\mm^2_3)_{il}+\delta_{jl}(\mm^2_3)_{ik}\nonumber\\
=&\,\frac{1}{3}\big(2\delta_{ij}\delta_{kl}-\delta_{ik}\delta_{jl}-\delta_{il}\delta_{jk}\big)
-2\delta_{ij}\big((\mm^2_3)_0\big)_{kl}-2\delta_{kl}\big((\mm^2_3)_0\big)_{ij}\nonumber\\
&\,+\delta_{ik}\big((\mm^2_3)_0\big)_{jl}+\delta_{il}\big((\mm^2_3)_0\big)_{jk}+\delta_{jk}\big((\mm^2_3)_0\big)_{il}+\delta_{jl}\big((\mm^2_3)_0\big)_{ik}.
\end{align}
Thus, combining (\ref{m12-m22}), (\ref{m1m2-sym}) and (\ref{F-ijkl}) yields
\begin{align}\label{m1m2-m1m2-sym}
&\,(\mm_1\mm_2\otimes\mm_1\mm_2)_{ijkl}\nonumber\\
&\,\quad=\big((\mm_1^2\mm_2^2)_0\big)_{ijkl}
 +\frac{1}{7}\CA\big((\mm^2_3)_0\big)_{ijkl}-\frac{1}{60}(2\delta_{ij}\delta_{kl}-3\delta_{ik}\delta_{jl}-3\delta_{il}\delta_{jk}).
\end{align}
Similar to the calculation of (\ref{m1m2-m1m2-sym}), we obtain
\begin{align}
&\,(\mm_1\mm_3\otimes\mm_1\mm_3)_{ijkl}\nonumber\\
&\,\quad=\big((\mm_1^2\mm_3^2)_0\big)_{ijkl}
  +\frac{1}{7}\CA\big((\mm^2_2)_0\big)_{ijkl}-\frac{1}{60}(2\delta_{ij}\delta_{kl}-3\delta_{ik}\delta_{jl}-3\delta_{il}\delta_{jk}), \label{m1m3-m1m3-sym}\\
 &\,(\mm_2\mm_3\otimes\mm_2\mm_3)_{ijkl}\nonumber\\
 &\,\quad=\big((\mm_2^2\mm_3^2)_0\big)_{ijkl}
  +\frac{1}{7}\CA\big((\mm^2_1)_0\big)_{ijkl}-\frac{1}{60}(2\delta_{ij}\delta_{kl}-3\delta_{ik}\delta_{jl}-3\delta_{il}\delta_{jk}).\label{m2m3-m2m3-sym}
\end{align}

In the same way, we have
\begin{align}\label{mm2-mm3-otimes-2}
&\,\big((\mm^2_2-\mm^2_3)\otimes(\mm^2_2-\mm^2_3)\big)_{ijkl}-(\mm^4_2-2\mm^2_2\mm^2_3+\mm^4_3)_{ijkl}
\nonumber\\
=&\,-\frac{1}{3}\Big(2m_{2i}m_{2j}m_{3k}m_{3l}+2m_{3i}m_{3j}m_{2k}m_{2l}-(m_{2i}m_{3j}+m_{3i}m_{2j})(m_{2k}m_{3l}+m_{3k}m_{2l})\Big).
\end{align}
Similar to the calculation of (\ref{F-ijkl}), we obtain
\begin{align}\label{overline-F-ijkl}
&\,2m_{2i}m_{2j}m_{3k}m_{3l}+2m_{3i}m_{3j}m_{2k}m_{2l}-(m_{2i}m_{3j}+m_{3i}m_{2j})(m_{2k}m_{3l}+m_{3k}m_{2l})\nonumber\\
=&\,m_{2i}m_{3l}(m_{2j}m_{3k}-m_{3j}m_{2k})+m_{2i}m_{3k}(m_{2j}m_{3l}-m_{3j}m_{2l})\nonumber\\
&\,+m_{3i}m_{2l}(m_{3j}m_{2k}-m_{2j}m_{3k})+m_{3i}m_{2k}(m_{3j}m_{2l}-m_{2j}m_{3l})\nonumber\\
=&\,(m_{2i}m_{3l}-m_{3i}m_{2l})(m_{2j}m_{3k}-m_{3j}m_{2k})+(m_{2i}m_{3k}-m_{3i}m_{2k})(m_{2j}m_{3l}-m_{3j}m_{2l})\nonumber\\
=&\,(\epsilon^{ils}\epsilon^{jkt}+\epsilon^{iks}\epsilon^{jlt})(\mm^2_1)_{st}\nonumber\\
=&\,2\delta_{ij}\delta_{kl}-\delta_{ik}\delta_{jl}-\delta_{il}\delta_{jk}-2\delta_{ij}(\mm^2_1)_{kl}-2\delta_{kl}(\mm^2_1)_{ij}\nonumber\\
&\,+\delta_{ik}(\mm^2_1)_{jl}+\delta_{il}(\mm^2_1)_{jk}+\delta_{jk}(\mm^2_1)_{il}+\delta_{jl}(\mm^2_1)_{ik}\nonumber\\
=&\,\frac{1}{3}\big(2\delta_{ij}\delta_{kl}-\delta_{ik}\delta_{jl}-\delta_{il}\delta_{jk}\big)
-2\delta_{ij}\big((\mm^2_1)_0\big)_{kl}-2\delta_{kl}\big((\mm^2_1)_0\big)_{ij}\nonumber\\
&\,+\delta_{ik}\big((\mm^2_1)_0\big)_{jl}+\delta_{il}\big((\mm^2_1)_0\big)_{jk}+\delta_{jk}\big((\mm^2_1)_0\big)_{il}+\delta_{jl}\big((\mm^2_1)_0\big)_{ik}.
\end{align}
Note that
\begin{align*}
&\,\mm^4_2-2\mm^2_2\mm^2_3+\mm^4_3\\
&\,=(\mm^4_2-2\mm^2_2\mm^2_3+\mm^4_3)_0+\frac{6}{7}(\mm^2_2+\mm^2_3)_0\Fi+\frac{2}{7}(\mm^2_1)_0\Fi+\frac{4}{15}\Fi^2\\
&\,=(\mm^4_2-2\mm^2_2\mm^2_3+\mm^4_3)_0-\frac{4}{7}(\mm^2_1)_0\Fi+\frac{4}{15}\Fi^2.
\end{align*}
Then, from (\ref{mm2-mm3-otimes-2}) and (\ref{overline-F-ijkl}), we deduce
\begin{align}
\big((\mm^2_2-\mm^2_3)\otimes(\mm^2_2-\mm^2_3)\big)_{ijkl}
 &\,=\big((\mm^4_2-2\mm^2_2\mm^2_3+\mm^4_3)_0\big)_{ijkl}
  +\frac{4}{7}\CA\big((\mm_1^2)_0\big)_{ijkl}\nonumber\\
  &\,\quad-\frac{1}{15}(2\delta_{ij}\delta_{kl}-3\delta_{ik}\delta_{jl}-3\delta_{il}\delta_{jk}).\label{m2m2-m3m3-sym}
\end{align}

The next task is to calculate the fourth-order tensor $(\mm^2_1)_0\otimes(\mm^2_2-\mm^2_3)$. A direct calculation gives
\begin{align}\label{m1m1m2m2-m2m2m1m1}
 &\,m_{1i}m_{1j}m_{2k}m_{2l}-m_{2i}m_{2j}m_{1k}m_{1l}\nonumber\\
&\,=m_{1i}m_{2l}(m_{1j}m_{2k}-m_{2j}m_{1k})+m_{2j}m_{1k}(m_{1i}m_{2l}-m_{2i}m_{1l})\nonumber\\
&\,=\epsilon^{jks}m_{1i}m_{2l}m_{3s}+\epsilon^{ils}m_{1k}m_{2j}m_{3s}.
\end{align}
Furthermore, the asymmetric part of  (\ref{m1m1m2m2-m2m2m1m1}) can be calculated as follows:
\begin{align}\label{m1m2m3+2-asym}
&\,\epsilon^{jks}m_{1i}m_{2l}m_{3s}+\epsilon^{ils}m_{1k}m_{2j}m_{3s}-\epsilon^{jks}(\mm_1\mm_2\mm_3)_{ils}-\epsilon^{ils}(\mm_1\mm_2\mm_3)_{kjs}\nonumber\\
&\,=\frac{1}{6}\epsilon^{jks}\Big(m_{1i}m_{2l}m_{3s}-m_{2i}m_{3l}m_{1s}+m_{1i}m_{2l}m_{3s}-m_{3i}m_{1l}m_{2s}\nonumber\\
&\,\quad+m_{1i}(m_{2l}m_{3s}-m_{3l}m_{2s})+(m_{1i}m_{2l}-m_{2i}m_{1l})m_{3s}+(m_{1i}m_{3s}-m_{3i}m_{1s})m_{2l}\Big)\nonumber\\
&\,\quad+\frac{1}{6}\epsilon^{ils}\Big(m_{1k}m_{2j}m_{3s}-m_{2k}m_{3j}m_{1s}+m_{1k}m_{2j}m_{3s}-m_{3k}m_{1j}m_{2s}\nonumber\\
&\,\quad+m_{1k}(m_{2j}m_{3s}-m_{3j}m_{2s})+(m_{1k}m_{2j}-m_{2k}m_{1j})m_{3s}+(m_{1k}m_{3s}-m_{3k}m_{1s})m_{2j}\Big)\nonumber\\
&\,=\frac{1}{6}\epsilon^{jks}\Big(3\epsilon^{ilt}m_{3t}m_{3s}+\epsilon^{slt}m_{2i}m_{2t}+\epsilon^{ist}m_{1t}m_{1l}+\epsilon^{lst}m_{1i}m_{1t}+\epsilon^{sit}m_{2t}m_{2l}\Big)\nonumber\\
&\,\quad+\frac{1}{6}\epsilon^{ils}\Big(3\epsilon^{kjt}m_{3t}m_{3s}+\epsilon^{sjt}m_{2k}m_{2t}+\epsilon^{kst}m_{1t}m_{1j}+\epsilon^{jst}m_{1k}m_{1t}+\epsilon^{skt}m_{2t}m_{2j}\Big)\nonumber\\
&\,=\frac{1}{2}\epsilon^{jks}\epsilon^{ilt}(\mm^2_3)_{ts}+\frac{1}{6}\Big((\delta_{jl}\delta_{kt}-\delta_{jt}\delta_{kl})m_{2i}m_{2t}+(\delta_{jt}\delta_{ki}-\delta_{ji}\delta_{kt})m_{1t}m_{1l}\nonumber\\
&\,\quad+(\delta_{jt}\delta_{kl}-\delta_{jl}\delta_{kt})m_{1i}m_{1t}+(\delta_{ji}\delta_{kt}-\delta_{jt}\delta_{ki})m_{2t}m_{2l}\Big)\nonumber\\
&\,\quad+\frac{1}{2}\epsilon^{ils}\epsilon^{kjt}(\mm^2_3)_{ts}+\frac{1}{6}\Big((\delta_{ij}\delta_{lt}-\delta_{it}\delta_{lj})m_{2k}m_{2t}+(\delta_{it}\delta_{lk}-\delta_{ik}\delta_{lt})m_{1t}m_{1j}\nonumber\\
&\,\quad+(\delta_{it}\delta_{lj}-\delta_{ij}\delta_{lt})m_{1k}m_{1t}+(\delta_{ik}\delta_{lt}-\delta_{it}\delta_{lk})m_{2t}m_{2j}\Big)\nonumber\\
&\,=\frac{1}{2}(\epsilon^{jks}\epsilon^{ilt}+\epsilon^{ils}\epsilon^{kjt})(\mm^2_3)_{ts}+\frac{1}{3}\Big(\delta_{kl}(\mm^2_1)_{ij}-\delta_{ij}(\mm^2_1)_{kl}+\delta_{ij}(\mm^2_2)_{kl}-\delta_{kl}(\mm^2_2)_{ij}\Big)\nonumber\\
&\,=\frac{1}{3}\Big(\delta_{kl}(\mm^2_1)_{ij}-\delta_{ij}(\mm^2_1)_{kl}+\delta_{ij}(\mm^2_2)_{kl}-\delta_{kl}(\mm^2_2)_{ij}\Big),
\end{align}
where we have used the fact that
\begin{align*}
 &\,(\epsilon^{jks}\epsilon^{ilt}+\epsilon^{ils}\epsilon^{kjt})(\mm^2_3)_{ts}\\
 &\,=\Big(\delta_{ji}\delta_{kl}\delta_{st}+\delta_{jl}\delta_{kt}\delta_{si}+\delta_{jt}\delta_{ki}\delta_{sl}-\delta_{ji}\delta_{sl}\delta_{kt}-\delta_{jl}\delta_{ki}\delta_{st}-\delta_{jt}\delta_{kl}\delta_{si}\\
 &\,\quad+\delta_{ik}\delta_{lj}\delta_{st}+\delta_{ij}\delta_{lt}\delta_{sk}+\delta_{it}\delta_{lk}\delta_{sj}-\delta_{ik}\delta_{sj}\delta_{lt}-\delta_{ij}\delta_{lk}\delta_{st}-\delta_{it}\delta_{lj}\delta_{sk}\Big)(\mm^2_3)_{ts}\\
 &\,=\delta_{jl}(\mm^2_3)_{ik}+\delta_{ik}(\mm^2_3)_{jl}-\delta_{ij}(\mm^2_3)_{kl}-\delta_{kl}(\mm^2_3)_{ij}\\
 &\,\quad+\delta_{ij}(\mm^2_3)_{kl}+\delta_{kl}(\mm^2_3)_{ij}-\delta_{ik}(\mm^2_3)_{jl}-\delta_{jl}(\mm^2_3)_{ik}\\
 &\,=0.
\end{align*}
Then, using (\ref{m1-2-m2-2-sym}), (\ref{m1-4})-(\ref{m12-m22}), (\ref{m1m2-m1m2-sym}), (\ref{m1m1m2m2-m2m2m1m1})-(\ref{m1m2m3+2-asym}) and the relation $\mm^2_1+\mm^2_2+\mm^2_3=\Fi$,
we obtain
\begin{align}\label{m1-0-m2-m3-block}
&\,\Big(\mm^2_1-\frac{\Fi}{3}\Big)_{ij}(\mm^2_2-\mm^2_3)_{kl}=\Big(\mm^2_1-\frac{\Fi}{3}\Big)_{ij}(2\mm^2_2+\mm^2_1-\Fi)_{kl}\nonumber\\
&\,=2m_{1i}m_{1j}m_{2k}m_{2l}+(\mm^4_1)_{ijkl}-\delta_{kl}(\mm^2_1)_{ij}-\frac{2}{3}\delta_{ij}(\mm^2_2)_{kl}-\frac{1}{3}\delta_{ij}(\mm^2_1)_{kl}+\frac{1}{3}\delta_{ij}\delta_{kl}\nonumber\\
&\,=m_{1i}m_{1j}m_{2k}m_{2l}-m_{2i}m_{2j}m_{1k}m_{1l}+6(\mm^2_1\mm^2_2)_{ijkl}-4(\mm_1\mm_2\otimes\mm_1\mm_2)_{ijkl}\nonumber\\
&\,\quad+(\mm^4_1)_{ijkl}-\delta_{kl}(\mm^2_1)_{ij}-\frac{2}{3}\delta_{ij}(\mm^2_2)_{kl}-\frac{1}{3}\delta_{ij}(\mm^2_1)_{kl}+\frac{1}{3}\delta_{ij}\delta_{kl}\nonumber\\
&\,=\epsilon^{jks}(\mm_1\mm_2\mm_3)_{ils}+\epsilon^{ils}(\mm_1\mm_2\mm_3)_{kjs}\nonumber\\
&\,\quad+\frac{1}{3}\Big(\delta_{kl}\big((\mm^2_1)_0\big)_{ij}-\delta_{ij}\big((\mm^2_1)_0\big)_{kl}+\delta_{ij}\big((\mm^2_2)_0\big)_{kl}-\delta_{kl}\big((\mm^2_2)_0\big)_{ij}\Big)\nonumber\\
&\,\quad+6\big((\mm^2_1\mm^2_2)_0\big)_{ijkl}-\frac{6}{7}\big((\mm^2_3)_0\Fi\big)_{ijkl}+\frac{2}{5}(\Fi^2)_{ijkl}\nonumber\\
&\,\quad-4\big((\mm^2_1\mm^2_2)_0\big)_{ijkl}-\frac{4}{7}\CA\big((\mm_3^2)_0\big)_{ijkl}
 +\frac{1}{15}(2\delta_{ij}\delta_{kl}-3\delta_{ik}\delta_{jl}-3\delta_{il}\delta_{jk})\nonumber\\
 &\,\quad+\big((\mm^4_1)_0\big)_{ijkl}+\frac{6}{7}\big((\mm^2_1)_0\Fi\big)_{ijkl}+\frac{1}{5}(\Fi^2)_{ijkl}\nonumber\\
 &\,\quad-\delta_{kl}\big((\mm^2_1)_0\big)_{ij}-\frac{2}{3}\delta_{ij}\big((\mm^2_2)_0\big)_{kl}-\frac{1}{3}\delta_{ij}\big((\mm^2_1)_0\big)_{kl}-\frac{1}{3}\delta_{ij}\delta_{kl}\nonumber\\
 &\,=\epsilon^{jks}(\mm_1\mm_2\mm_3)_{ils}+\epsilon^{ils}(\mm_1\mm_2\mm_3)_{kjs}+2\big((\mm^2_1\mm^2_2)_0\big)_{ijkl}+\big((\mm^4_1)_0\big)_{ijkl}\nonumber\\
 &\,\quad+\frac{4}{21}\CA\big((\mm^2_1)_0\big)_{ijkl}+\frac{8}{21}\CA\big((\mm^2_2)_0\big)_{ijkl},
\end{align}
where we have also used the following fact
\begin{align*}
&\,\frac{1}{3}\Big(\delta_{kl}\big((\mm^2_1)_0\big)_{ij}-\delta_{ij}\big((\mm^2_1)_0\big)_{kl}+\delta_{ij}\big((\mm^2_2)_0\big)_{kl}-\delta_{kl}\big((\mm^2_2)_0\big)_{ij}\Big)\\
&\,-\delta_{kl}\big((\mm^2_1)_0\big)_{ij}-\frac{2}{3}\delta_{ij}\big((\mm^2_2)_0\big)_{kl}-\frac{1}{3}\delta_{ij}\big((\mm^2_1)_0\big)_{kl}-\frac{6}{7}\big((\mm^2_3)_0\Fi\big)_{ijkl}+\frac{2}{5}(\Fi^2)_{ijkl}\\
&\,+\frac{6}{7}\big((\mm^2_1)_0\Fi\big)_{ijkl}+\frac{1}{5}(\Fi^2)_{ijkl}-\frac{1}{3}\delta_{ij}\delta_{kl}\\
&\,=-\frac{1}{3}\Big(2\delta_{kl}\big((\mm^2_1)_0\big)_{ij}+2\delta_{ij}\big((\mm^2_1)_0\big)_{kl}+\delta_{kl}\big((\mm^2_2)_0\big)_{ij}+\delta_{ij}\big((\mm^2_2)_0\big)_{kl}\Big)\\
&\,\quad+\frac{12}{7}\big((\mm^2_1)_0\Fi\big)_{ijkl}+\frac{6}{7}\big((\mm^2_2)_0\Fi\big)_{ijkl}-\frac{1}{15}(2\delta_{ij}\delta_{kl}-3\delta_{ik}\delta_{jl}-3\delta_{il}\delta_{jk})\\
&\,=-\frac{8}{21}\CA\big((\mm^2_1)_0\big)_{ijkl}-\frac{4}{21}\CA\big((\mm^2_2)_0\big)_{ijkl}-\frac{1}{15}(2\delta_{ij}\delta_{kl}-3\delta_{ik}\delta_{jl}-3\delta_{il}\delta_{jk}).
\end{align*}

Note that $\mm^2_2-\mm^2_3=(2\mm^2_2+\mm^2_1)_0$. It follows that
\begin{align}\label{m2-0-otimes-m2-0}
(\mm^2_2)_0\otimes(\mm^2_2)_0=&\,\frac{1}{4}\big(\mm^2_2-\mm^2_3-(\mm^2_1)_0\big)\otimes\big(\mm^2_2-\mm^2_3-(\mm^2_2)_0\big)\nonumber\\
=&\,\frac{1}{4}\Big((\mm^2_2-\mm^2_3)\otimes(\mm^2_2-\mm^2_3)-(\mm^2_2-\mm^2_3)\otimes(\mm^2_1)_0\nonumber\\
&\,-(\mm^2_1)_0\otimes(\mm^2_2-\mm^2_3)+(\mm^2_1)_0\otimes(\mm^2_1)_0\Big).
\end{align}
Thus, combining (\ref{m2-0-otimes-m2-0}) with  (\ref{m1-block-traceless}), (\ref{m2m2-m3m3-sym}) and (\ref{m1-0-m2-m3-block}), we obtain
\begin{align}\label{m2-block-traceless}
&\,\Big(\mm^2_2-\frac{1}{3}\Fi\Big)_{ij}\Big(\mm^2_2-\frac{1}{3}\Fi\Big)_{kl}\nonumber\\
&\,=\frac{1}{4}\Big(\big((\mm^4_2-2\mm^2_2\mm^2_3+\mm^4_3)_0\big)_{ijkl}-\frac{4}{7}\CA\big((\mm^2_1)_0\big)_{ijkl}-\frac{1}{15}(2\delta_{ij}\delta_{kl}-3\delta_{ik}\delta_{jl}-3\delta_{il}\delta_{jk})\nonumber\\
&\,\quad+\big((\mm^4_1)_0\big)_{ijkl}-\frac{4}{21}\CA\big((\mm^2_1)_0\big)_{ijkl}-\frac{1}{45}(2\delta_{ij}\delta_{kl}-3\delta_{ik}\delta_{jl}-3\delta_{il}\delta_{jk})\nonumber\\
&\,\quad-4\big((\mm^2_1\mm^2_2)_0\big)_{ijkl}-2\big((\mm^4_1)_0\big)_{ijkl}-\frac{8}{21}\CA\big((\mm^2_1)_0\big)_{ijkl}-\frac{16}{21}\CA\big((\mm^2_2)_0\big)_{ijkl}\Big)\nonumber\\
&\,=\big((\mm^4_2)_0\big)_{ijkl}-\frac{4}{21}\CA\big((\mm^2_2)_0\big)_{ijkl}-\frac{1}{45}(2\delta_{ij}\delta_{kl}-3\delta_{ik}\delta_{jl}-3\delta_{il}\delta_{jk}),
\end{align}
where we have employed the cancellation relation
\begin{align*}
&\,\epsilon^{jks}(\mm_1\mm_2\mm_3)_{ils}+\epsilon^{ils}(\mm_1\mm_2\mm_3)_{kjs}\\
&\,\quad+\epsilon^{lis}(\mm_1\mm_2\mm_3)_{kjs}+\epsilon^{kjs}(\mm_1\mm_2\mm_3)_{ils}=0.
\end{align*}

Next, we calculate the three tensors
\begin{align}
    \mm_1\otimes\mm_2\mm_3,\quad \mm_2\otimes\mm_1\mm_3,\quad \mm_3\otimes\mm_1\mm_2. \nonumber
\end{align}
It turns out that
\begin{align}
&\,\frac{1}{2}m_{1i}(m_{2j}m_{3k}+m_{3j}m_{2k})\nonumber\\
&\,\quad=(\mm_1\mm_2\mm_3)_{ijk}+\frac{1}{6}\Big((m_{1i}m_{2j}-m_{2i}m_{1j})m_{3k}+(m_{1i}m_{3j}-m_{3i}m_{1j})m_{2k}\nonumber\\
&\,\qquad+(m_{1i}m_{2k}-m_{2i}m_{1k})m_{3j}+(m_{1i}m_{3k}-m_{3i}m_{1k})m_{3j}\Big)\nonumber\\
&\,\quad=(\mm_1\mm_2\mm_3)_{ijk}+\frac{1}{6}\Big(\epsilon^{ijs}\big((\mm^2_3)_0-(\mm^2_2)_0\big)_{ks}+\epsilon^{iks}\big((\mm^2_3)_0-(\mm^2_2)_0\big)_{js}\Big),\label{m1-m2m3-block}\\
&\,\frac{1}{2}m_{2i}(m_{1j}m_{3k}+m_{3j}m_{1k})\nonumber\\
&\,\quad=(\mm_1\mm_2\mm_3)_{ijk}+\frac{1}{6}\Big(\epsilon^{ijs}\big((\mm^2_1)_0-(\mm^2_3)_0\big)_{ks}+\epsilon^{iks}\big((\mm^2_1)_0-(\mm^2_3)_0\big)_{js}\Big),\label{m2-m1m3-block}\\
&\,\frac{1}{2}m_{3i}(m_{1j}m_{2k}+m_{2j}m_{1k})\nonumber\\
&\,\quad=(\mm_1\mm_2\mm_3)_{ijk}+\frac{1}{6}\Big(\epsilon^{ijs}\big((\mm^2_2)_0-(\mm^2_1)_0\big)_{ks}+\epsilon^{iks}\big((\mm^2_2)_0-(\mm^2_1)_0\big)_{js}\Big).\label{m3-m1m2-block}
\end{align}
The equations \eqref{m1-m2m3-block}--\eqref{m3-m1m2-block} also hold if we replace $\mm_i$ with $\nn_i$, which we also need to use later.

To deal with $\CV^{(0)}$ in \eqref{V0+N0}, we need to calculate
\begin{align*}
&\,(\mm_1\otimes\mm_2)\otimes\mm_1\mm_2,\quad (\mm_2\otimes\mm_1)\otimes\mm_1\mm_2,\\
&\,(\mm_1\otimes\mm_3)\otimes\mm_1\mm_3,\quad (\mm_2\otimes\mm_3)\otimes\mm_2\mm_3.
\end{align*}
Using the definition of $\CR^{(0)}_i(i=3,4,5)$, it follows that
\begin{align*}
2\langle m_{1i}m_{2j}(\mm_1\mm_2)_{kl}\rangle=&\,\frac{1}{2}\CR^{(0)}_3
+\big\langle(m_{1i}m_{2j}-m_{2i}m_{1j})(\mm_1\mm_2)_{kl}\big\rangle,\\
2\langle m_{2i}m_{1j}(\mm_1\mm_2)_{kl}\rangle=&\,\frac{1}{2}\CR^{(0)}_3-\big\langle(m_{1i}m_{2j}-m_{2i}m_{1j})(\mm_1\mm_2)_{kl}\big\rangle,\\
2\langle m_{1i}m_{3j}(\mm_1\mm_3)_{kl}\rangle=&\,\frac{1}{2}\CR^{(0)}_4
+\big\langle(m_{1i}m_{3j}-m_{3i}m_{1j})(\mm_1\mm_3)_{kl}\big\rangle,\\
2\langle m_{2i}m_{3j}(\mm_2\mm_3)_{kl}\rangle=&\,\frac{1}{2}\CR^{(0)}_5
+\big\langle(m_{2i}m_{3j}-m_{3i}m_{2j})(\mm_2\mm_3)_{kl}\big\rangle.
\end{align*}
In order to calculate the above tensor moments, it
is desirable to utilize the following relation:
\begin{align}\label{epsilon+2}
\epsilon^{ijk}\epsilon^{ist}=\delta_{js}\delta_{kt}-\delta_{jt}\delta_{ks}.
\end{align}
Then, using (\ref{m1m2-m2m1-e}) and (\ref{epsilon+2}), we derive that
\begin{align}\label{anti-m1m2+m1m2}
&\,(m_{1i}m_{2j}-m_{2i}m_{1j})(\mm_1\mm_2)_{kl}\nonumber\\
&\,=\frac{1}{2}\epsilon^{ijs}m_{3s}(m_{1k}m_{2l}+m_{2k}m_{1l})\nonumber\\
&\,=\epsilon^{ijs}\Big((\mm_1\mm_2\mm_3)_{kls}+\frac{1}{6}m_{1k}(m_{2l}m_{3s}-m_{3l}m_{2s})
+\frac{1}{6}m_{2l}(m_{1k}m_{3s}-m_{3k}m_{1s})\nonumber\\
&\,\quad+\frac{1}{6}m_{2k}(m_{1l}m_{3s}-m_{3l}m_{1s})
+\frac{1}{6}m_{1l}(m_{2k}m_{3s}-m_{3k}m_{2s})\Big)\nonumber\\
&\,=\epsilon^{ijs}\Big((\mm_1\mm_2\mm_3)_{kls}+\frac{1}{6}\big(m_{1k}m_{1t}\epsilon^{lst}+m_{2l}m_{2t}\epsilon^{skt}+m_{2k}m_{2t}\epsilon^{slt}+m_{1l}m_{1t}\epsilon^{kst}\big)\Big)\nonumber\\
&\,=\epsilon^{ijs}(\mm_1\mm_2\mm_3)_{kls}
+\frac{1}{6}\Big(m_{1k}m_{1t}(\delta_{it}\delta_{jl}-\delta_{il}\delta_{jt})+m_{2l}m_{2t}(\delta_{ik}\delta_{jt}-\delta_{it}\delta_{jk})\nonumber\\
&\,\quad+m_{2k}m_{2t}(\delta_{il}\delta_{jt}-\delta_{it}\delta_{jl})
+m_{1l}m_{1t}(\delta_{it}\delta_{jk}-\delta_{ik}\delta_{jt})
\Big)\nonumber\\
&\,=\epsilon^{ijs}(\mm_1\mm_2\mm_3)_{kls}
+\frac{1}{6}\Big(\CB\big((\mm^2_1)_0\big)_{ijkl}+\CB\big((\mm^2_2)_0\big)_{ijkl}\Big).
\end{align}

In the above, we have encountered several symmetric traceless tensors. They have the following relations.
\begin{align*}
  &\,(\mm_3^2)_0=-(\mm_1^2)_0-(\mm_2^2)_0, \\
  &\,(\mm_1^2\mm_3^2)_0=-(\mm_1^4)_0-(\mm_1^2\mm_2^2)_0,\\
  &\,(\mm_2^2\mm_3^2)_0=-(\mm_2^4)_0-(\mm_1^2\mm_2^2)_0.
\end{align*}
When averaged tensors are considered, the linear relations obtained above still hold. Therefore, we only need to focus on the following tensors that are the linearly independent:
\begin{align*}
\langle(\mm^2_1\mm^2_2)_0\rangle,~\langle(\mm^2_1\mm^2_3)_0\rangle,~\langle(\mm^2_2\mm^2_3)_0\rangle,~\langle(\mm^2_i)_0\rangle,~i=1,2,3.
\end{align*}

\subsection{Expression involving low-order tensors}\label{Apprendix-A3}
When $Q_i=Q_i^{(0)}$, we will encounter a few terms only involving second-order tensors, which we provide alternative expressions below.
They will be useful for matrix manipulations in the main text, and the discussion afterwards.

Let us look into the last tensor in (\ref{six-tensors}). Using the relation $\Fi=\nn^2_1+\nn^2_2+\nn^2_3$, we deduce that
\begin{align*}
  &\,2\delta_{ij}\delta_{kl}-3\delta_{ik}\delta_{jl}-3\delta_{il}\delta_{jk}\\
  &\,=2(\nn_1^2+\nn_2^2+\nn_3^2)_{ij}(\nn_1^2+\nn_2^2+\nn_3^2)_{kl}-3(\nn_1^2+\nn_2^2+\nn_3^2)_{ik}(\nn_1^2+\nn_2^2+\nn_3^2)_{jl}\\
  &\,\quad-3(\nn_1^2+\nn_2^2+\nn_3^2)_{il}(\nn_1^2+\nn_2^2+\nn_3^2)_{jk}\\
  &\,=2\sum_{\alpha\ne\beta}\nn_{\alpha}^2\otimes\nn^2_{\beta}-4\sum^3_{\alpha=1}\nn_{\alpha}^4\\
  &\,\quad-12(\nn_1\nn_2\otimes\nn_1\nn_2+\nn_1\nn_3\otimes\nn_1\nn_3+\nn_2\nn_3\otimes\nn_2\nn_3),
\end{align*}
where we have used the fact that $\nn_1\nn_2=\frac{1}{2}(\nn_1\otimes\nn_2+\nn_2\otimes\nn_1)$.
The terms in the second line are expressed linearly by $\sss_i\otimes \sss_j$
We would also like to express the first line in this form.
Note that
\begin{align}\label{2nn-4nn4}
  \nn_1^2-\frac{\mathfrak{i}}{3}=\frac{1}{3}(2\nn_1^2-\nn_2^2-\nn_3^2).
\end{align}
Thus, fitting with a term $\nn^2-\Fi/3$ in (\ref{2nn-4nn4}) yields
\begin{align}\label{n1-traceless}
  -9\sss_1\otimes\sss_1=&\,-9(\nn_1^2-\frac{\mathfrak{i}}{3})\otimes(\nn_1^2-\frac{\mathfrak{i}}{3})\nonumber\\
  =&\,-(2\nn_1^2-\nn_2^2-\nn_3^2)\otimes(2\nn_1^2-\nn_2^2-\nn_3^2)\nonumber\\
  =&\,-4\nn_1^4+2\big(\nn_1^2\otimes(\nn_2^2+\nn_3^2)+(\nn_2^2+\nn_3^2)\otimes\nn_1^2\big)-(\nn_2^2+\nn_3^2)\otimes (\nn_2^2+\nn_3^2).
\end{align}
Then the remaining terms are given by
\begin{align*}
  &\,-4\nn_2^4-4\nn_3^4+2(\nn_2^2\otimes\nn_3^2+\nn_3^2\otimes\nn_2^2)+(\nn_2^2+\nn_3^2)\otimes (\nn_2^2+\nn_3^2)\\
  &\,=-3(\nn_2^2-\nn_3^2)\otimes (\nn_2^2-\nn_3^2)=-3\sss_2\otimes\sss_2.
\end{align*}
Therefore, we arrive at
\begin{align}\label{233delta}
  2\delta_{ij}\delta_{kl}-3\delta_{ik}\delta_{jl}-3\delta_{il}\delta_{jk}=-9\sss_1\otimes\sss_1-3\sss_2\otimes\sss_2-12(\sss_3\otimes\sss_3+\sss_4\otimes\sss_4+\sss_5\otimes\sss_5),
\end{align}
where the corresponding coordinate $X_1$ is given by (\ref{X1-X2}).

We note that
\begin{align}\label{n2n3-traceless}
\sss_2\otimes\sss_2=&\,
(\nn^2_2-\nn^2_3)\otimes(\nn^2_2-\nn^2_3)\nonumber\\
=&\,\nn^4_2+\nn^4_3-(\nn^2_2\otimes\nn^2_3+\nn^2_3\otimes\nn^2_2).
\end{align}
Then $\CA\big((\nn^2_1)_0\big)_{ijkl}$ can be
calculated as follows:
\begin{align*}
\CA\big((\nn^2_1)_0\big)_{ijkl}=&\,(\nn^2_1+\nn^2_2+\nn^2_3)_{kl}\Big(\nn_1^2-\frac{\mathfrak{i}}{3}\Big)_{ij}+(\nn^2_1+\nn^2_2+\nn^2_3)_{ij}\Big(\nn_1^2-\frac{\mathfrak{i}}{3}\Big)_{kl}\\
&\,-\frac{3}{4}(\nn_1^2+\nn_2^2+\nn_3^2)_{ik}\Big(\nn_1^2-\frac{\mathfrak{i}}{3}\Big)_{jl}-\frac{3}{4}(\nn_1^2+\nn_2^2+\nn_3^2)_{jl}\Big(\nn_1^2-\frac{\mathfrak{i}}{3}\Big)_{ik}\\
&\,-\frac{3}{4}(\nn_1^2+\nn_2^2+\nn_3^2)_{il}\Big(\nn_1^2-\frac{\mathfrak{i}}{3}\Big)_{jk}-\frac{3}{4}(\nn_1^2+\nn_2^2+\nn_3^2)_{jk}\Big(\nn_1^2-\frac{\mathfrak{i}}{3}\Big)_{il}\\
=&\,\frac13\Big((2\nn^2_1-\nn^2_2-\nn^2_3)\otimes(\nn^2_1+\nn^2_2+\nn^2_3)+(\nn^2_1+\nn^2_2+\nn^2_3)\otimes(2\nn^2_1-\nn^2_2-\nn^2_3)\Big)\\
&\,-(2\nn^4_1-\nn^4_2-\nn^4_3)
-(\nn_1\nn_2\otimes\nn_1\nn_2+\nn_1\nn_3\otimes\nn_1\nn_3-2\nn_2\nn_3\otimes\nn_2\nn_3),
\end{align*}
where we have used the relation $\Fi=\nn^2_1+\nn^2_2+\nn^2_3$.
Using (\ref{n1-traceless}) and (\ref{n2n3-traceless}), we get
\begin{align*}
&\,\frac13\Big((2\nn^2_1-\nn^2_2-\nn^2_3)\otimes(\nn^2_1+\nn^2_2+\nn^2_3)+(\nn^2_1+\nn^2_2+\nn^2_3)\otimes(2\nn^2_1-\nn^2_2-\nn^2_3)\Big)\\
&\,\quad-(2\nn^4_1-\nn^4_2-\nn^4_3)\\
&\,=\nn^4_2+\nn^4_3+\frac{1}{3}\Big(-2\nn^4_1+\nn^2_1\otimes(\nn^2_2+\nn^2_3)+(\nn^2_2+\nn^2_3)\otimes\nn^2_1-2(\nn^2_2+\nn^2_3)\otimes(\nn^2_2+\nn^2_3)\Big)\\
&\,=-\frac{3}{2}(\nn_1^2-\frac{\mathfrak{i}}{3})\otimes(\nn_1^2-\frac{\mathfrak{i}}{3})+\frac{1}{2}(\nn^2_2-\nn^2_3)\otimes(\nn^2_2-\nn^2_3)\\
&\,=-\frac{3}{2}\sss_1\otimes\sss_1+\frac{1}{2}\sss_2\otimes\sss_2.
\end{align*}
Consequently, we obtain
\begin{align}\label{T1ijkl}
\CA\big((\nn^2_1)_0\big)_{ijkl}=-\frac{3}{2}\sss_1\otimes\sss_1+\frac{1}{2}\sss_2\otimes\sss_2
-(\sss_3\otimes\sss_3+\sss_4\otimes\sss_4-2\sss_5\otimes\sss_5),
\end{align}
where the corresponding coordinate $X_2$ is written by (\ref{X1-X2}).

Next we deal with the term $\CA\big(\nn^2_2-\nn^2_3\big)_{ijkl}$. Note that
\begin{align*}
\frac{3}{2}(\sss_1\otimes\sss_2+\sss_2\otimes\sss_1)=
&\,\frac{3}{2}\Big((\nn^2_1-\frac{\Fi}{3})\otimes(\nn^2_2-\nn^2_3)+(\nn^2_2-\nn^2_3)\otimes(\nn^2_1-\frac{\Fi}{3})\Big)\\
=&\,-(\nn^4_2-\nn^4_3)+(\nn^2_1\otimes\nn^2_2+\nn^2_2\otimes\nn^2_1)-(\nn^2_1\otimes\nn^2_3+\nn^2_3\otimes\nn^2_1).
\end{align*}
Then, $\CA\big(\nn^2_2-\nn^2_3\big)_{ijkl}$ can be calculated as
\begin{align}\label{T2ijkl}
\CA\big(\nn^2_2-\nn^2_3\big)_{ijkl}=&\,(\nn^2_1+\nn^2_2+\nn^2_3)_{kl}(\nn_2^2-\nn^2_3)_{ij}+(\nn^2_1+\nn^2_2+\nn^2_3)_{ij}(\nn_2^2-\nn^2_3)_{kl}\nonumber\\
  &\,-\frac{3}{4}(\nn^2_1+\nn^2_2+\nn^2_3)_{ik}(\nn_2^2-\nn^2_3)_{jl}-\frac{3}{4}(\nn^2_1+\nn^2_2+\nn^2_3)_{jl}(\nn_2^2-\nn^2_3)_{ik}\nonumber\\
  &\,-\frac{3}{4}(\nn^2_1+\nn^2_2+\nn^2_3)_{il}(\nn_2^2-\nn^2_3)_{jk}-\frac{3}{4}(\nn^2_1+\nn^2_2+\nn^2_3)_{jk}(\nn_2^2-\nn^2_3)_{il}\nonumber\\
 =&\,2(\nn^4_2-\nn^4_3)+(\nn^2_1\otimes\nn^2_2+\nn^2_2\otimes\nn^2_1)-(\nn^2_1\otimes\nn^2_3+\nn^2_3\otimes\nn^2_1)\nonumber\\
 &\,-3\big(\nn^4_2-\nn^4_3+\nn_1\nn_2\otimes\nn_1\nn_2-\nn_1\nn_3\otimes\nn_1\nn_3\big)\nonumber\\
=&\,\frac{3}{2}\Big(\Big(\nn^2_1-\frac{\Fi}{3}\Big)\otimes(\nn^2_2-\nn^2_3)+(\nn^2_2-\nn^2_3)\otimes\Big(\nn^2_1-\frac{\Fi}{3}\Big)\Big)\nonumber\\
&\,-3(\nn_1\nn_2\otimes\nn_1\nn_2-\nn_1\nn_3\otimes\nn_1\nn_3)\nonumber\\
 =&\,\frac{3}{2}(\sss_1\otimes\sss_2+\sss_2\otimes\sss_1)-3(\sss_3\otimes\sss_3-\sss_4\otimes\sss_4),
\end{align}
where the corresponding coordinate $X_3$ is written by (\ref{X3-X4}).

\section{Closure approximation: Theorem \ref{Q-biaxial-theorem}}\label{Apprendix-B}

We discuss Theorem \ref{Q-biaxial-theorem} that recognizes the form of high-order tensors.
Theorem \ref{Q-biaxial-theorem} is stated for the original entropy and the quasi-entropy.
So we need to consider them separately.

Theorem \ref{Q-biaxial-theorem} is actually a special case in previous works:
for the original entropy, it is a special case of Theorem 5.2 in \cite{Xu1};
for quasi-entropy, it is a special case of Theorem 4.8 in \cite{Xu3}.
Nevertheless, both of them were shown for general cases of symmetry and the explicit form \eqref{biaxial_high} is not provided.
For this reason, we shall explain how those theorems are applied to the current work to obtain Theorem \ref{Q-biaxial-theorem}, and at places show some results.


\subsection{Original entropy}\label{Apprendix-B1}
We first discuss the closure by the original entropy. The following result has been shown in Appendix in \cite{XYZ}.
\begin{lemma}\label{existence-bi}
If $s_i,b_i$ satisfy \eqref{range-2nd}, then there exists a unique density function
\begin{align*}
    \rho(\Fq)=\frac 1Z \exp(\sum_{i,j=1,2}\lambda_{ij}(\mm_i\cdot\nn_j)^2),
\end{align*}
such that $\langle(\mm_i^2)_0\rangle=s_i(\nn_1^2)_0+b_i(\nn_2^2-\nn_3^2)$.
It minimizes $\int_{SO(3)}\rho\ln\rho\ud\Fq$ when $Q_i$ is fixed.
\end{lemma}

Recall that $\Fq=(\mm_1,\mm_2,\mm_3)$ and $\Fp=(\nn_1,\nn_2,\nn_3)$.
The density function satisfies $\rho(\Fp\Fb_k\Fp^T\Fq)=\rho(\Fq)$ for $k=1,2,3$.
This can be seen by noticing that $\mm_i\cdot\nn_j$ is the $(j,i)$ element of $\Fp^T\Fq$.
Thus, when $\Fq$ is replaced by $\Fp\Fb_k\Fp^T\Fq$, the dot product $\mm_i\cdot\nn_j$ becomes the $(j,i)$ element of $\Fp^T(\Fp\Fb_k\Fp^T\Fq)=\Fb_k\Fp^T\Fq=(\Fp\Fb_k^T)^T\Fq$.
It suffices to notice the equalities like $\Fp\Fb_1^T=(\nn_1,-\nn_2,-\nn_3)$.

By Theorem 5.2 in \cite{Xu1} and the related discussions before the theorem, when an $n$th-order  symmetric traceless tensor is calculated from the density function above, it could be expressed as $W(\Fp)$ for some $W\in\BA^{\MD_2,n}$.
Using the decomposition written down in \eqref{invrnt-decomp}, we arrive at the expression in Theorem \ref{Q-biaxial-theorem}.

The positive-definiteness of the averaged tensors $\CR_i$ in \eqref{seven-tensors} is obvious because they are calculated from a positive density function.

\subsection{Quasi-entropy}\label{Apprendix-B2}

To illustrate some ideas, let us start from the second-order quasi-entropy $\Xi_2$.
Denote by $\rr_1$ the vector formed by $1$ and $\mm_i\cdot\nn_j,\;1\le i,j\le 3$, which is a $10\times 1$ vector.
For a first-order tensor $U$, we define a row vector as
\begin{align*}
    \Phi_1(U)_j=(U\cdot\nn_j).
\end{align*}
For a second-order tensor $U$, we define a matrix as
\begin{align}
    \Psi_2(U)_{ij}=(U\cdot\nn_i\otimes\nn_j).
\end{align}
The general second-order quasi-entropy, denoted by $\widetilde{\Xi}_2$, is defined as the minus log-determinant of the second moment of $\rr_1$ (hereafter we omit the free parameter $\nu$ introduced in \eqref{qent-2nd}),
\begin{align}
    \widetilde{\Xi}_2=&\,-\ln\det\langle\rr_1\rr_1^T\rangle\nonumber\\
    =&\,-\ln\det\left(\begin{array}{cccc}
        1 & \Phi_1(\langle\mm_1\rangle) & \Phi_1(\langle\mm_2\rangle) & \Phi_1(\langle\mm_3\rangle) \\
        \Phi_1(\langle\mm_1\rangle)^T & \Psi_2(\langle\mm_1^2\rangle) & \Psi_2(\langle\mm_1\otimes\mm_2\rangle) & \Psi_2(\langle\mm_1\otimes\mm_3\rangle) \\
        \Phi_1(\langle\mm_2\rangle)^T & \Psi_2(\langle\mm_2\otimes\mm_1\rangle) & \Psi_2(\langle\mm_2^2\rangle) & \Psi_2(\langle\mm_2\otimes\mm_3\rangle) \\
        \Phi_1(\langle\mm_3\rangle)^T & \Psi_2(\langle\mm_3\otimes\mm_1\rangle) & \Psi_2(\langle\mm_3\otimes\mm_2\rangle) & \Psi_2(\langle\mm_3^2\rangle)
    \end{array}
    \right).
\end{align}
In \cite{Xu3}, the second moment of $\rr_1$ is replaced by the covariance matrix of the $9\times 1$ vector formed by the last nine components of $\rr_1$.
It can be seen that these two formulations are equivalent.

Here, we need to emphasize that the notation $\langle\cdot\rangle$ does not assume that they are averaged by certain positive density function, but only implies that the tensors obey linear relations such as what we have obtained in Appendix \ref{symtrls-decomp}.
For second-order tensors not symmetric, we express them using symmetric traceless tensors, such as  $\langle\mm_1\otimes\mm_2\rangle_{ij}=\langle\mm_1\mm_2\rangle_{ij}+\epsilon^{ijk}\langle\mm_3\rangle_k$.
Thus, $\widetilde{\Xi}_2$ is a function of symmetric traceless tensors up to second order.
If we choose a basis of symmetric traceless tensors, their `average' are independent variables in $\widetilde{\Xi}_2$.

Now, for our problem, the tensors specified are $Q_1$ and $Q_2$, which determine $\langle\mm_1^2 \rangle=Q_1+\Fi/3$, $\langle\mm_2^2 \rangle=Q_2+\Fi/3$, $\langle\mm_3^2 \rangle=-Q_1-Q_2+\Fi/3$.
To obtain the quasi-entropy $\Xi_2$ about $Q_1$ and $Q_2$ only, we shall minimize $\widetilde{\Xi}_2$ with $Q_1$ and $Q_2$ fixed.
At the minimizer many tensors vanish, because we have the following lemma.

\begin{lemma}\label{det-est1}
    For a symmetric positive-definite matrix $K$, suppose that it is given in blocks as
    \begin{align}
        K=\left(\begin{array}{cc}
            K_1 &\, A \vspace{0.5ex}\\
            A^T &\, K_2
        \end{array}\right).
    \end{align}
    Then we have
    \begin{align}
        \det K\le\det K_1\det K_2.
    \end{align}
    The equality holds if and only if $A=0$.
\end{lemma}

Notice that off-diagonal blocks are functions of $\langle\mm_i\rangle$ and $\langle\mm_i\mm_j\rangle$ for $i\ne j$, which are independent of $Q_1$ and $Q_2$.
Using this lemma, we immediately deduce that the minimizer is attained when all the off-diagonal blocks are zero.
In this way, we obtain the quasi-entropy $\Xi_2$ in \eqref{qent-2nd}.

It is worthy noting that for $Q_1$ and $Q_2$, $\mm_1^2-\Fi/3$ and $\mm_2^2-\Fi/3$ are invariant under $\MD_2$.
On the other hand, the off-diagonal blocks vanish when averaged over $\MD_2$, since in these blocks the times of $\mm_1$, $\mm_2$, $\mm_3$ appearing are not all odd or not all even.
This result actually holds for quasi-entropy up to arbitrary order, as indicated by Theorem 4.8 in \cite{Xu3}.

The ideas above are also useful when discussing the fourth-order quasi-entropy $\Xi_4$.
Denote by $\rr_2$ the vector formed by $1$, $\mm_i\cdot\nn_j,\;1\le i,j\le 3$ and $\SSS_i\cdot\sss_j\;1\le i,j\le 5$, which has the size $35\times 1$.
The fourth-order quasi-entropy is defined as the minus
log-determinant of $\langle\rr_2\rr_2^T\rangle$.
It is a function of symmetric traceless tensors up to fourth order.

The closure approximation minimizes the quasi-entropy with $Q_1$ and $Q_2$ fixed.
Still, if the times of $\mm_1$, $\mm_2$, $\mm_3$ are not all odd or not all even, then the tensor vanishes when averaged over $\MD_2$.
Theorem 4.8 in \cite{Xu3} guarantees that when seeking the minimizer with $Q_1$ and $Q_2$ fixed, these tensors are zero.
After setting these tensors as zero in the quasi-entropy, we could get a reduced expression, which we write down below.

For a second-order tensor $U$, we define a $1\times 5$ row vector as
\begin{align*}
    \Phi_2(U)_j
    =(U\cdot\sss_j).
\end{align*}
For a third-order tensor $U$, we define a $3\times 5$ matrix,
\begin{align*}
  \Psi_3(U)_{ij}
  =(U\cdot \nn_i\otimes\sss_j).
\end{align*}
For a fourth order tensor $U$, we define a $5\times 5$ matrix,
\begin{align*}
  \Psi_4(U)_{ij}
  =(U\cdot \sss_i\otimes\sss_j).
\end{align*}
The reduced quasi-entropy is given by
\begin{align}
    &\,\Xi_4=\nonumber\\
    &\,-\ln\det\left(
    \begin{array}{ccc}
        1 &\, \Phi_2(\langle\mm_1^2-\frac\Fi 3\rangle) &\, \Phi_2(\langle\mm_2^2-\mm_3^2\rangle)
        \vspace{1ex}\\
        \Phi_2(\langle\mm_1^2-\frac\Fi 3\rangle)^T &\,
        \Psi_4\big(\langle(\mm_1^2-\frac\Fi 3)\otimes(\mm_1^2-\frac\Fi 3)\rangle\big) &\,  \Psi_4\big(\langle(\mm_2^2-\mm_3^2)\otimes(\mm_1^2-\frac{\Fi}{3})\rangle\big)
        \vspace{1ex}\\
        \Phi_2(\langle\mm_2^2-\mm_3^2\rangle)^T &\,
        \Psi_4\big(\langle(\mm_2^2-\mm_3^2)\otimes(\mm_1^2-\frac\Fi 3)\rangle\big)^T &\,
        \Psi_4\big(\langle(\mm_2^2-\mm_3^2)\otimes(\mm_2^2-\mm_3^2)\rangle\big)
    \end{array}
    \right)\nonumber\\
    &\,-\ln\det\left(
    \begin{array}{cc}
        \Psi_2(\langle\mm_1^2\rangle) &\, \Psi_3(\langle\mm_1\otimes\mm_2\mm_3\rangle)
        \vspace{1ex}\\
        \Psi_3(\langle\mm_1\otimes\mm_2\mm_3\rangle)^T &\, \Psi_4(\langle\mm_2\mm_3\otimes\mm_2\mm_3\rangle)
    \end{array}
    \right)\nonumber\\
    &\,-\ln\det\left(
    \begin{array}{cc}
        \Psi_2(\langle\mm_2^2\rangle) &\, \Psi_3(\langle\mm_2\otimes\mm_1\mm_3\rangle)
        \vspace{1ex}\\
        \Psi_3(\langle\mm_2\otimes\mm_1\mm_3\rangle)^T &\, \Psi_4(\langle\mm_1\mm_3\otimes\mm_1\mm_3\rangle)
    \end{array}
    \right)\nonumber\\
    &\,-\ln\det\left(
    \begin{array}{cc}
        \Psi_2(\langle\mm_3^2\rangle) &\, \Psi_3(\langle\mm_3\otimes\mm_1\mm_2\rangle)
        \vspace{1ex}\\
        \Psi_3(\langle\mm_3\otimes\mm_1\mm_2\rangle)^T &\, \Psi_4(\langle\mm_1\mm_2\otimes\mm_1\mm_2\rangle)
    \end{array}
    \right).\label{qent-4th}
\end{align}
The first matrix is $11\times 11$, while the other three are $8\times 8$.
The blocks can be expressed by symmetric traceless tensors as we have calculated in Appendix \ref{symtrls-decomp}.

The quasi-entropy $\Xi_4$ is defined on the domain such that the four matrices in $\Xi_4$ are positive definite.
Thus, we conclude that if the high-order tensors are calculated from the constrained minimization of $\Xi_4$, the tensors $\CR_i$ in \eqref{CVQ-CNQ} are positive definite in the sense of \eqref{P-spd}.
This is because that many of $\CR_1,\CR_3,\CR_4,\CR_5$ are diagonal blocks of $\Xi_4$, and for $\CR_2$ we use \eqref{m2-0-otimes-m2-0}.

Now, let us assume that $Q_i$ has the biaxial form \eqref{Qi-biaxial}.
First, we claim that the domain of quasi-entropy $\Xi_4$ is not empty when $s_i,b_i$ are fixed with the conditions \eqref{range-2nd}.
This is because that the high-order tensors calculated from any positive density function must make the covariance matrix positive definite.
Such a density function exists because of Lemma \ref{existence-bi}.

We are now ready to show Theorem \ref{Q-biaxial-theorem}.
By \eqref{m1-m2m3-block}--\eqref{m3-m1m2-block}, \eqref{233delta}, \eqref{T1ijkl} and \eqref{T2ijkl}, the zeroth- and second-order tensors could fill the following entries in the quasi-entropy.
In the $11\times 11$ matrix, they are labelled as
\begin{align}
    \left(\begin{array}{c|ccccc|ccccc}
         1 &\, * &\, * &\, &\, &\, &\, * &\, * &\, &\, &\, \\\hline
         * &\, * &\, * &\, &\, &\, &\, * &\, * &\, &\, &\, \\
         * &\, * &\, * &\, &\, &\, &\, * &\, * &\, &\, &\, \\
         &\, &\, &\, * &\, &\, &\, &\, &\, * &\, &\, \\
         &\, &\, &\, &\, * &\, &\, &\, &\, &\, * &\, \\
         &\, &\, &\, &\, &\, * &\, &\, &\, &\, &\, *\\\hline
         * &\, * &\, * &\, &\, &\, &\, * &\, * &\, &\, &\, \\
         * &\, * &\, * &\, &\, &\, &\, * &\, * &\, &\, &\, \\
         &\, &\, &\, * &\, &\, &\, &\, &\, * &\, &\, \\
         &\, &\, &\, &\, * &\, &\, &\, &\, &\, * &\, \\
         &\, &\, &\, &\, &\, * &\, &\, &\, &\, &\, *
    \end{array}
    \right).
\end{align}
In the three $8\times 8$ matrices, they are labelled as
\begin{align}
    \left(\begin{array}{ccc|ccccc}
        * &\,  &\,  &\,  &\,  &\,  &\,  &\, *\\
         &\, * &\,  &\,  &\,  &\,  &\, * &\, \\
         &\,  &\, * &\,  &\,  &\, * &\,  &\, \\\hline
         &\,  &\,  &\, * &\, * &\,  &\,  &\, \\
         &\,  &\,  &\, * &\, * &\,  &\,  &\, \\
         &\,  &\, * &\,  &\,  &\, * &\,  &\, \\
         &\, * &\,  &\,  &\,  &\,  &\, * &\, \\
        * &\,  &\,  &\,  &\,  &\,  &\,  &\, *
    \end{array}\right).
\end{align}
The third-order and fourth-order symmetric traceless tensors are expressed by the bases,
\begin{align}
  &\,\langle\mm_1\mm_2\mm_3\rangle=z\nn_1\nn_2\nn_3+z'_1(\nn_1^3)_0+z'_2(\nn_1^2\nn_2)_0+z'_3(\nn_1\nn_2^2)_0+z'_4(\nn_2^3)_0+z'_5(\nn_1^2\nn_3)_0+z'_6(\nn_2^2\nn_3)_0,\nonumber\\
  &\,\langle (\mm_1^4)_0\rangle=a_1(\nn_1^4)_0+a_2(\nn_2^4)_0+a_3(\nn_1^2\nn_2^2)_0\nonumber\\
  &\,\qquad +a_4(\nn_1^3\nn_2)_0+a_5(\nn_1^3\nn_3)_0+a_6(\nn_1^2\nn_2\nn_3)_0
    +a_7(\nn_1\nn_2^3)_0+a_8(\nn_1\nn_2^2\nn_3)_0+a_9(\nn_2^3\nn_3)_0, \nonumber\\
  &\,\langle (\mm_2^4)_0\rangle=\tilde{a}_1(\nn_1^4)_0+\tilde{a}_2(\nn_2^4)_0+\tilde{a}_3(\nn_1^2\nn_2^2)_0\nonumber\\
  &\,\qquad +\tilde{a}_4(\nn_1^3\nn_2)_0+\tilde{a}_5(\nn_1^3\nn_3)_0+\tilde{a}_6(\nn_1^2\nn_2\nn_3)_0
    +\tilde{a}_7(\nn_1\nn_2^3)_0+\tilde{a}_8(\nn_1\nn_2^2\nn_3)_0+\tilde{a}_9(\nn_2^3\nn_3)_0,\nonumber\\
  &\,\langle (\mm_1^2\mm_2^2)_0\rangle=\bar{a}_1(\nn_1^4)_0+\bar{a}_2(\nn_2^4)_0+\bar{a}_3(\nn_1^2\nn_2^2)_0\nonumber\\
  &\,\qquad +\bar{a}_4(\nn_1^3\nn_2)_0+\bar{a}_5(\nn_1^3\nn_3)_0+\bar{a}_6(\nn_1^2\nn_2\nn_3)_0
    +\bar{a}_7(\nn_1\nn_2^3)_0+\bar{a}_8(\nn_1\nn_2^2\nn_3)_0+\bar{a}_9(\nn_2^3\nn_3)_0.
\end{align}
Using \eqref{third-orth1}--\eqref{fourth-orth2}, the terms $a_i,\tilde{a}_i,\bar{a}_i$ for $i=1,2,3$ and $z$ contribute only to the starred entries, while the terms $z'_i$ and $a_j,\tilde{a}_j,\bar{a}_j$ for $4\le j\le 9$ contribute only to the non-starred entries.
Meanwhile, as long as the starred entries form a positive definite matrix, the determinant reaches its unique maximum when the non-starred entries are zero.
This can be observed by rearranging the rows and columns of the four matrices in $\Xi_4$.
In the $11\times 11$ matrix, we group the indices as $\{1,2,3,7,8\}$, $\{4,9\}$, $\{5,10\}$ and $\{6,11\}$.
In the three $8\times 8$ matrices, we group the indices as $\{1,8\}$, $\{2,7\}$, $\{3,6\}$, $\{4,5\}$.
After rearrangement, these matrices become block diagonal.
Thus, the determinant must be no less than that the off-diagonal blocks are zero.
Therefore, at the minimizer of $\Xi_4$ we must have $z'_i=0$ and $a_i=\tilde{a}_i=\bar{a}_i=0,\, i=4,\cdots,9$.

\section{Explicit expression with biaxial $Q_i$}\label{Apprendix-C}

Next, we calculate the blocks in \eqref{qent-4th} when the tensors take \eqref{biaxial_high}.
They also give the matrices in Section \ref{matvec-local}.

Using \eqref{biaxial_high} and the average of (\ref{m1m2-m1m2-sym}) with respect to the density function,  we derive that
\begin{align}\label{average-m1m2m1m2}
&\,\langle\mm_1\mm_2\otimes\mm_1\mm_2\rangle_{ijkl}\nonumber\\
&\,\quad=\big(\bar{a}_1(\nn^4_1)_0+\bar{a}_2(\nn^4_2)_0+\bar{a}_3(\nn^2_1\nn^2_2)_0\big)_{ijkl}
-\frac{1}{7}\CA\big(Q^{(0)}_1+Q^{(0)}_2\big)_{ijkl}\nonumber\\
&\,\qquad-\frac{1}{60}(2\delta_{ij}\delta_{kl}-3\delta_{ik}\delta_{jl}-3\delta_{il}\delta_{jk})\nonumber\\
&\,\quad=\big(\bar{a}_1(\nn^4_1)_0+\bar{a}_2(\nn^4_2)_0+\bar{a}_3(\nn^2_1\nn^2_2)_0\big)_{ijkl}
-\frac{1}{7}\Big((s_1+s_2)\CA\big((\nn^2_1)_0\big)_{ijkl}\nonumber\\
&\,\qquad+(b_1+b_2)\CA\big(\nn^2_2-\nn^2_3\big)_{ijkl}\Big)-\frac{1}{60}(2\delta_{ij}\delta_{kl}-3\delta_{ik}\delta_{jl}-3\delta_{il}\delta_{jk}).
\end{align}
From here, we can see that we shall need to express the six tensors below in the basis of $\sss_i\otimes \sss_j$,
\begin{align}\label{six-tensors}
&2\delta_{ij}\delta_{kl}-3\delta_{ik}\delta_{jl}-3\delta_{il}\delta_{jk},~\CA\big((\nn^2_1)_0\big)_{ijkl},~
\CA\big(\nn^2_2-\nn^2_3\big)_{ijkl},~(\nn_1^4)_0,~ (\nn_2^4)_0,~ (\nn_1^2\nn_2^2)_0.
\end{align}
Actually, we will see that they only have the following terms:
\begin{align*}
\sss_1\otimes\sss_1,\quad
\sss_1\otimes\sss_2,\quad
\sss_2\otimes\sss_1,\quad
\sss_2\otimes\sss_2,\quad
\sss_3\otimes\sss_3,\quad
\sss_4\otimes\sss_4,\quad
\sss_5\otimes\sss_5.
\end{align*}
The first three tensors in \eqref{six-tensors} have been discussed in Appendix \ref{Apprendix-A3}.
In what follows, we calculate the other three tensors.

For the calculation of  the term $(\nn^4_1)_0$, employing (\ref{n2n3-traceless}) and the following equality
\begin{align}\label{-9EE1-3EE4}
2\sum_{\alpha\ne\beta}\nn_{\alpha}^2\otimes\nn^2_{\beta}-4\sum^3_{\alpha=1}\nn_{\alpha}^4=-9\sss_1\otimes\sss_1-3\sss_2\otimes\sss_2,
\end{align}
we deduce that
\begin{align}\label{trace-n14}
(\nn^4_1)_0=&\,\nn^4_1-\frac{6}{7}\nn^2_1\Fi+\frac{3}{35}\Fi^2\nonumber\\
=&\,\nn^4_1-\frac{1}{7}\Big(n_{1i}n_{1j}\delta_{kl}+n_{1i}n_{1k}\delta_{jl}
+n_{1i}n_{1l}\delta_{jk}+n_{1j}n_{1k}\delta_{il}+n_{1j}n_{1l}\delta_{ik}+n_{1k}n_{1l}\delta_{ij}\Big)\nonumber\\
&\,+\frac{1}{35}\big(\delta_{ij}\delta_{kl}+\delta_{ik}\delta_{jl}+\delta_{il}\delta_{jk}\big)\nonumber\\
=&\,\frac{1}{7}\Big(\nn^4_1-(\nn^2_1\otimes\nn^2_2+\nn^2_2\otimes\nn^2_1+\nn^2_1\otimes\nn^2_3+\nn^2_3\otimes\nn^2_1)
-4(\nn_1\nn_2\otimes\nn_1\nn_2+\nn_1\nn_3\otimes\nn_1\nn_3)\Big)\nonumber\\
&\,+\frac{1}{35}\Big(\sum_{\alpha\ne\beta}\nn_{\alpha}^2\otimes\nn^2_{\beta}+3\sum^3_{\alpha=1}\nn_{\alpha}^4
+4(\nn_1\nn_2\otimes\nn_1\nn_2+\nn_1\nn_3\otimes\nn_1\nn_3
+\nn_2\nn_3\otimes\nn_2\nn_3)\Big)\nonumber\\
=&\,\frac{18}{35}\sss_1\otimes\sss_1+\frac{1}{35}\sss_2\otimes\sss_2-\frac{16}{35}(\sss_3\otimes\sss_3+\sss_4\otimes\sss_4)+\frac{4}{35}\sss_5\otimes\sss_5,
\end{align}
where the corresponding matrix $X_4$ is given by (\ref{X3-X4}).

Similarly, from (\ref{-9EE1-3EE4}) and (\ref{n2n3-traceless}), we derive that
\begin{align}\label{trace-n24}
(\nn^4_2)_0=&\,\nn^4_2-\frac{6}{7}\nn^2_2\Fi+\frac{3}{35}\Fi^2\nonumber\\
=&\,\nn^4_2-\frac{1}{7}\Big(n_{2i}n_{2j}\delta_{kl}+n_{2i}n_{2k}\delta_{jl}
+n_{2i}n_{2l}\delta_{jk}+n_{2j}n_{2k}\delta_{il}+n_{2j}n_{2l}\delta_{ik}+n_{2k}n_{2l}\delta_{ij}\Big)\nonumber\\
&\,+\frac{1}{35}\big(\delta_{ij}\delta_{kl}+\delta_{ik}\delta_{jl}+\delta_{il}\delta_{jk}\big)\nonumber\\
=&\,\frac{1}{7}\Big(\nn^4_2-(\nn^2_1\otimes\nn^2_2+\nn^2_2\otimes\nn^2_1+\nn^2_2\otimes\nn^2_3+\nn^2_3\otimes\nn^2_2)
-4(\nn_1\nn_2\otimes\nn_1\nn_2+\nn_2\nn_3\otimes\nn_2\nn_3)\Big)\nonumber\\
&\,+\frac{1}{35}\Big(\sum_{\alpha\ne\beta}\nn_{\alpha}^2\otimes\nn^2_{\beta}+3\sum^3_{\alpha=1}\nn_{\alpha}^4
+4(\nn_1\nn_2\otimes\nn_1\nn_2+\nn_1\nn_3\otimes\nn_1\nn_3
+\nn_2\nn_3\otimes\nn_2\nn_3)\Big)\nonumber\\
=&\,\frac{27}{140}\sss_1\otimes\sss_1+\frac{19}{140}\sss_2\otimes\sss_2-\frac{3}{28}(\sss_1\otimes\sss_2+\sss_2\otimes\sss_1)\nonumber\\
&\,-\frac{16}{35}(\sss_3\otimes\sss_3+\sss_5\otimes\sss_5)+\frac{4}{35}\sss_4\otimes\sss_4,
\end{align}
where the corresponding matrix $X_5$ is given by (\ref{X5-X6}).

We may now proceed to deal with the term $(\nn^2_1\nn^2_2)_0$.
Analogously, we have
\begin{align}\label{trace-n12n22}
(\nn^2_1\nn^2_2)_0=&\,\nn^2_1\nn^2_2-\frac{1}{7}(\nn^2_1+\nn^2_2)\Fi+\frac{1}{35}\Fi^2\nonumber\\
=&\,\frac{1}{6} \Big(n_{1i}n_{1j}n_{2k}n_{2l}+n_{1i}n_{2j}n_{1k}n_{2l}+n_{1i}n_{2j}n_{2k}n_{1l}\nonumber\\
&\,+n_{2i}n_{1j}n_{1k}n_{2l}+n_{2i}n_{1j}n_{2k}n_{1l}+n_{2i}n_{2j}n_{1k}n_{1l}\Big)\nonumber\\
&\,-\frac{1}{42}\Big(n_{1i}n_{1j}\delta_{kl}+n_{1i}n_{1k}\delta_{jl}
+n_{1i}n_{1l}\delta_{jk}+n_{1j}n_{1k}\delta_{il}+n_{1j}n_{1l}\delta_{ik}+n_{1k}n_{1l}\delta_{ij}\Big)\nonumber\\
&\,-\frac{1}{42}\Big(n_{2i}n_{2j}\delta_{kl}+n_{2i}n_{2k}\delta_{jl}
+n_{2i}n_{2l}\delta_{jk}+n_{2j}n_{2k}\delta_{il}+n_{2j}n_{2l}\delta_{ik}+n_{2k}n_{2l}\delta_{ij}\Big)\nonumber\\
&\,+\frac{1}{105}\big(\delta_{ij}\delta_{kl}+\delta_{ik}\delta_{jl}+\delta_{il}\delta_{jk}\big)\nonumber\\
=&\,\frac{1}{6}\big(\nn^2_1\otimes\nn^2_2+\nn^2_2\otimes\nn^2_1+4\nn_1\nn_2\otimes\nn_1\nn_2\big)\nonumber\\
&\,-\frac{1}{42}\Big(6\nn^4_1+\sum_{\alpha\ne\beta}\nn_{\alpha}^2\otimes\nn^2_{\beta}-(\nn^2_2\otimes\nn^2_3+\nn^2_3\otimes\nn^2_2)
+4(\nn_1\nn_2\otimes\nn_1\nn_2+\nn_1\nn_3\otimes\nn_1\nn_3)\Big)\nonumber\\
&\,-\frac{1}{42}\Big(6\nn^4_2+\sum_{\alpha\ne\beta}\nn_{\alpha}^2\otimes\nn^2_{\beta}-(\nn^2_1\otimes\nn^2_3+\nn^2_3\otimes\nn^2_1)
+4(\nn_1\nn_2\otimes\nn_1\nn_2+\nn_2\nn_3\otimes\nn_2\nn_3)\Big)\nonumber\\
&\,+\frac{1}{105}\Big(\sum_{\alpha\ne\beta}\nn_{\alpha}^2\otimes\nn^2_{\beta}+3\sum^3_{\alpha=1}\nn_{\alpha}^4
+4(\nn_1\nn_2\otimes\nn_1\nn_2+\nn_1\nn_3\otimes\nn_1\nn_3
+\nn_2\nn_3\otimes\nn_2\nn_3)\Big)\nonumber\\
=&\,-\frac{9}{35}\sss_1\otimes\sss_1-\frac{1}{70}\sss_2\otimes\sss_2+\frac{3}{28}(\sss_1\otimes\sss_2+\sss_2\otimes\sss_1)\nonumber\\
&\,+\frac{18}{35}\sss_3\otimes\sss_3-\frac{2}{35}(\sss_4\otimes\sss_4+\sss_5\otimes\sss_5),
\end{align}
where the corresponding matrix $X_6$ is given by (\ref{X5-X6}).

A direct calculation leads to
\begin{align}
    \epsilon^{ilt}(\nn_1\nn_2\nn_3)_{jkt}+\epsilon^{jkt}(\nn_1\nn_2\nn_3)_{ilt}=\frac 32(\sss_1\otimes\sss_2-\sss_2\otimes\sss_1),
\end{align}
where the associated coefficient matrix $\Pi$ is given by
\begin{align}
    \Pi=\left(
    \begin{array}{ccccc}
        0 &\, \frac 32 &\,  &\,  &\,
        \vspace{0.5ex}\\
        -\frac 32 &\, 0 &\,  &\,  &\,
        \vspace{1ex}\\
        &\, &\, 0 &\, &\,
        \vspace{0.5ex}\\
        &\, &\, &\, 0 &\,
        \vspace{0.5ex}\\
        &\, &\, &\, &\, 0
    \end{array}
    \right).
\end{align}

Based on the above calculations, we immediately give the expressions of $\CR^{(0)}_i(i=1,\cdots,6)$ under the basis $\sss_i\otimes\sss_j$.
Using \eqref{biaxial_high} and the averages of (\ref{m1-block-traceless}) with respect to the density function, we deduce that
\begin{align*}
\CR^{(0)}_1=&\,\Big\langle\Big(\mm^2_1-\frac{1}{3}\Fi\Big)\otimes\Big(\mm^2_1-\frac{1}{3}\Fi\Big)\Big\rangle_{ijkl}\\
=&\,\big\langle(\mm^4_1)_0\big\rangle_{ijkl}-\frac{4}{21}\CA\big(Q^{(0)}_1\big)_{ijkl}-\frac{1}{45}(2\delta_{ij}\delta_{kl}-3\delta_{ik}\delta_{jl}-3\delta_{il}\delta_{jk}).
\end{align*}
In the light of Theorem \ref{Q-biaxial-theorem}, and from (\ref{233delta}), (\ref{trace-n14})-(\ref{trace-n12n22}), we deduce that
\begin{align}
\CR^{(0)}_1=&\,a_1(\nn_1^4)_0+a_2(\nn_2^4)_0+a_3(\nn^2_1\nn^2_2)_0-\frac{4}{21}\Big(s_1\CA\big((\nn^2_1)_0\big)_{ijkl}+b_1\CA\big(\nn^2_2-\nn^2_3\big)_{ijkl}\Big)\nonumber\\
&\,-\frac{1}{45}(2\delta_{ij}\delta_{kl}-3\delta_{ik}\delta_{jl}-3\delta_{il}\delta_{jk}),
\label{RR1+app}
\end{align}
for which the matrix $R_1$ is written as
\begin{align}\label{R1}
R_1=-\frac{1}{45}X_1-\frac{4}{21}(s_1X_2+b_1X_3)+a_1X_4+a_2X_5+a_3X_6.
\end{align}
Similarly, for $\CR^{(0)}_2$, it follows that
\begin{align}\label{RR2+app}
\CR^{(0)}_2=&\,\Big\langle\Big(\mm^2_2-\frac{1}{3}\Fi\Big)\otimes\Big(\mm^2_2-\frac{1}{3}\Fi\Big)\Big\rangle\nonumber\\
=&\,\big\langle(\mm^4_2)_0\big\rangle-\frac{4}{21}\CA\big(Q^{(0)}_2\big)-\frac{1}{45}(2\delta_{ij}\delta_{kl}-3\delta_{ik}\delta_{jl}-3\delta_{il}\delta_{jk})\nonumber\\
=&\,\tilde{a}_1(\nn_1^4)_0+\tilde{a}_2(\nn_2^4)_0+\tilde{a}_3(\nn^2_1\nn^2_2)_0-\frac{4}{21}\Big(s_2\CA\big((\nn^2_1)_0\big)_{ijkl}+b_2\CA\big(\nn^2_2-\nn^2_3\big)_{ijkl}\Big)\nonumber\\
&\,-\frac{1}{45}(2\delta_{ij}\delta_{kl}-3\delta_{ik}\delta_{jl}-3\delta_{il}\delta_{jk}),
\end{align}
for which the matrix $R_2$ is written as
\begin{align}\label{R2}
R_2
=-\frac{1}{45}X_1-\frac{4}{21}(s_2X_2+b_2X_3)+\tilde{a}_1X_4+\tilde{a}_2X_5+\tilde{a}_3X_6.
\end{align}

Combining (\ref{average-m1m2m1m2}) with (\ref{trace-n14})-(\ref{trace-n12n22}), the tensor $\CR^{(0)}_3$ is expressed by
\begin{align}\label{RR0_3}
\CR^{(0)}_{3}=&\,4\big(\bar{a}_1(\nn^4_1)_0+\bar{a}_2(\nn^4_2)_0+\bar{a}_3(\nn^2_1\nn^2_2)_0\big)_{ijkl}
-\frac{4}{7}\Big((s_1+s_2)\CA\big((\nn^2_1)_0\big)_{ijkl}\nonumber\\
&\,+(b_1+b_2)\CA\big(\nn^2_2-\nn^2_3\big)_{ijkl}\Big)-\frac{1}{15}(2\delta_{ij}\delta_{kl}-3\delta_{ik}\delta_{jl}-3\delta_{il}\delta_{jk}),
\end{align}
for which the matrix $R_3$ is denoted by
\begin{align}\label{R3}
R_3=-\frac{1}{15}X_1-\frac{4}{7}\big((s_1+s_2)X_2+(b_1+b_2)X_3\big)+4\big(\bar{a}_1X_4+\bar{a}_2X_5+\bar{a}_3X_6\big).
\end{align}

Analogously, the tensor moment $\CR^{(0)}_4$ can be expressed by
\begin{align}\label{RR0_4}
\CR^{(0)}_{4}=&\,-4\Big((a_1+\bar{a}_1)(\nn^4_1)_0
  +(a_2+\bar{a}_2)(\nn^4_2)_0
  +(a_3+\bar{a}_3)(\nn^2_1\nn^2_2)_0\Big)_{ijkl}\nonumber\\
&\,+\frac{4}{7}\Big(s_2\CA\big((\nn^2_1)_0\big)_{ijkl}+b_2\CA\big(\nn^2_2-\nn^2_3\big)_{ijkl}\Big)-\frac{1}{15}(2\delta_{ij}\delta_{kl}-3\delta_{ik}\delta_{jl}-3\delta_{il}\delta_{jk}),
\end{align}
for which the matrix $R_4$ is denoted by
\begin{align}\label{R4}
R_4=-\frac{1}{15}X_1+\frac{4}{7}(s_2X_2+b_2X_3)-4\big((a_1+\bar{a}_1)X_4+(a_2+\bar{a}_2)X_5+(a_3+\bar{a}_3)X_6\big).
\end{align}
In the same way, we obtain
\begin{align}\label{RR0_5}
\CR^{(0)}_{5}=&\,-4\Big((\tilde{a}_1+\bar{a}_1)(\nn^4_1)_0
 +(\tilde{a}_2+\bar{a}_2)(\nn^4_2)_0
  +(\tilde{a}_3+\bar{a}_3)(\nn^2_1\nn^2_2)_0\Big)_{ijkl}\nonumber\\
&\,+\frac{4}{7}\Big(s_1\CA\big((\nn^2_1)_0\big)_{ijkl}+b_1\CA\big(\nn^2_2-\nn^2_3\big)_{ijkl}\Big)-\frac{1}{15}(2\delta_{ij}\delta_{kl}-3\delta_{ik}\delta_{jl}-3\delta_{il}\delta_{jk}),
\end{align}
for which the matrix $R_5$ is denoted as
\begin{align}\label{R5}
R_5=-\frac{1}{15}X_1+\frac{4}{7}(s_1X_2+b_1X_3)-4\big((\tilde{a}_1+\bar{a}_1)X_4+(\tilde{a}_2+\bar{a}_2)X_5+(\tilde{a}_3+\bar{a}_3)X_6\big).
\end{align}

By
\begin{align*}
(\mm^4_3)_0=(\mm^4_1)_0+(\mm^4_2)_0+2(\mm^2_1\mm^2_2)_0,\quad (\mm^2_2\mm^2_3)_0=-(\mm^4_2)_0-(\mm^2_1\mm^2_2)_0,
\end{align*}
we derive from (\ref{m2m2-m3m3-sym}) that
\begin{align}\label{RR6+app}
\CR_6=&\,\langle(\mm^2_2-\mm^2_3)\otimes(\mm^2_2-\mm^2_3)\rangle\nonumber\\
=&\,\langle(\mm^4_1)_0\rangle+4\langle(\mm^4_2)_0\rangle+4\langle(\mm^2_1\mm^2_2)_0\rangle
  +\frac{4}{7}\CA\big(Q^{(0)}_1\big)\nonumber\\
  &\,-\frac{1}{15}(2\delta_{ij}\delta_{kl}-3\delta_{ik}\delta_{jl}-3\delta_{il}\delta_{jk})\nonumber\\
=&\,(a_1+4\tilde{a}_1+4\bar{a}_1)(\nn^4_1)_0+(a_2+4\tilde{a}_2+4\bar{a}_2)(\nn^4_2)_0+(a_3+4\tilde{a}_3+4\bar{a}_3)(\nn^2_1\nn^2_2)_0\nonumber\\
&\,+\frac{4}{7}\Big(s_1\CA\big((\nn^2_1)_0\big)_{ijkl}+b_1\CA\big(\nn^2_2-\nn^2_3\big)_{ijkl}\Big)-\frac{1}{15}(2\delta_{ij}\delta_{kl}-3\delta_{ik}\delta_{jl}-3\delta_{il}\delta_{jk}),
\end{align}
for which the matrix $R_6$ is given by
\begin{align}\label{R6}
R_6=&\,-\frac{1}{15}X_1+\frac{4}{7}(s_1X_2+b_1X_3)+(a_1+4\tilde{a}_1+4\bar{a}_1)X_4\nonumber\\
&\,+(a_2+4\tilde{a}_2+4\bar{a}_2)X_5+(a_3+4\tilde{a}_3+4\bar{a}_3)X_6.
\end{align}

We turn to the term $\langle(\mm^2_1)_0\otimes(\mm^2_2-\mm^2_3)\rangle$.
By (\ref{m1-0-m2-m3-block}), we have
\begin{align}
\CS=&\,\langle(\mm^2_1)_0\otimes(\mm^2_2-\mm^2_3)\rangle\nonumber\\
=&\,\epsilon^{jks}\langle\mm_1\mm_2\mm_3\rangle_{ils}+\epsilon^{ils}\langle\mm_1\mm_2\mm_3\rangle_{kjs}+2\langle(\mm^2_1\mm^2_2)_0\rangle+\langle(\mm^4_1)_0\rangle\nonumber\\
 &\,+\frac{4}{21}\CA\big(Q^{(0)}_1\big)+\frac{8}{21}\CA\big(Q^{(0)}_2\big)\nonumber\\
 =&\,z\epsilon^{jks}(\nn_1\nn_2\nn_3)_{ils}+z\epsilon^{ils}(\nn_1\nn_2\nn_3)_{kjs}\nonumber\\
 &+(a_1+2\bar{a}_1)(\nn^4_1)_0+(a_2+2\bar{a}_2)(\nn^4_2)_0+(a_3+2\bar{a}_3)(\nn^2_1\nn^2_2)_0\nonumber\\
 &+\frac{4}{21}\Big((s_1+2s_2)\CA\big((\nn^2_1)_0\big)_{ijkl}+(b_1+2b_2)\CA\big(\nn^2_2-\nn^2_3\big)_{ijkl}\Big),
\end{align}
where the coefficient matrix $S$ is given by
\begin{align*}
S=&z\Pi+\frac{4}{21}\Big((s_1+2s_2)X_2+(b_1+2b_2)X_3\Big)\nonumber\\
&+(a_1+2\bar{a}_1)X_4+(a_2+2\bar{a}_2)X_5+(a_3+2\bar{a}_3)X_6.
\end{align*}

We are now able to give the matrices $M$ and $P$.
By \eqref{dissip-operator} and \eqref{P0},
the corresponding coordinates $M_{11}, M_{12}$ and $M_{22}$ are
\begin{align}
M_{11}=&\,\Gamma_2R_4+\Gamma_3R_3,\quad M_{12}=-\Gamma_3R_3,\quad M_{22}=\Gamma_1R_5+\Gamma_3R_3,\label{M-components}\\
    P=&\,c\zeta(I_{22}R_1+I_{11}R_2+I_{11}e_1R_3). \label{CP0-operator}
\end{align}
Using the expressions of $R_i$, we arrive at \eqref{M11block}--\eqref{M22block} and \eqref{P-matrix}.

The remaining part is to express averages of fourth-order antisymmetric traceless tensors and third-order tensors. From (\ref{anti-m1m2+m1m2}), we deduce that
\begin{align}\label{m1m2-m2m1+sym}
&\,\big\langle(m_{1i}m_{2j}-m_{2i}m_{1j})(\mm_1\mm_2)_{kl}\big\rangle\nonumber\\
&\,=z\epsilon^{ijs}(\nn_1\nn_2\nn_3)_{kls}
+\frac{1}{6}\Big(\CB\big(Q^{(0)}_1\big)_{ijkl}+\CB\big(Q^{(0)}_2\big)_{ijkl}
\Big)\nonumber\\
&\,=z\epsilon^{ijs}(\nn_1\nn_2\nn_3)_{kls}
+\frac{1}{6}\Big((s_1-s_2)\CB\big((\nn^2_1)_0\big)_{ijkl}+(b_1-b_2)\CB\big(\nn^2_2-\nn^2_3\big)_{ijkl}\Big).
\end{align}
We would like to express the above tensors linearly by the three tensors below,
\begin{align*}
&\,\aaa_1\otimes\sss_3=(\nn_1\otimes\nn_2-\nn_2\otimes\nn_1)\otimes\nn_1\nn_2,\\ &\,\aaa_2\otimes\sss_4=(\nn_3\otimes\nn_1-\nn_1\otimes\nn_3)\otimes\nn_1\nn_3,\\
&\,\aaa_3\otimes\sss_5=(\nn_2\otimes\nn_3-\nn_3\otimes\nn_2)\otimes\nn_2\nn_3.
\end{align*}
Direct calculations lead to
\begin{align}
\CB\big((\nn^2_1)_0\big)_{ijkl}=&\,\frac{1}{3}\Big((2\nn^2_1-\nn^2_2-\nn^2_3)_{ki}(\nn^2_1+\nn^2_2+\nn^2_3)_{jl}-(2\nn^2_1-\nn^2_2-\nn^2_3)_{kj}(\nn^2_1+\nn^2_2+\nn^2_3)_{il}\nonumber\\
&\,+(2\nn^2_1-\nn^2_2-\nn^2_3)_{li}(\nn^2_1+\nn^2_2+\nn^2_3)_{jk}-(2\nn^2_1-\nn^2_2-\nn^2_3)_{lj}(\nn^2_1+\nn^2_2+\nn^2_3)_{ik}\Big)\nonumber\\
=&\,2(\nn_1\otimes\nn_2-\nn_2\otimes\nn_1)\otimes\nn_1\nn_2+2(\nn_1\otimes\nn_3-\nn_3\otimes\nn_1)\otimes\nn_1\nn_3\nonumber\\
=&\,2(\aaa_1\otimes\sss_3-\aaa_2\otimes\sss_4),\label{T3ijkl}\\
\CB\big(\nn^2_2-\nn^2_3\big)_{ijkl}=&\,\big(\nn^2_2-\nn^2_3\big)_{ki}(\nn^2_1+\nn^2_2+\nn^2_3)_{jl}-\big(\nn^2_2-\nn^2_3\big)_{kj}(\nn^2_1+\nn^2_2+\nn^2_3)_{il}\nonumber\\
&\,+\big(\nn^2_2-\nn^2_3\big)_{li}(\nn^2_1+\nn^2_2+\nn^2_3)_{jk}-\big(\nn^2_2-\nn^2_3\big)_{lj}(\nn^2_1+\nn^2_2+\nn^2_3)_{ik}\nonumber\\
=&\,-2(\nn_1\otimes\nn_2-\nn_2\otimes\nn_1)\otimes\nn_1\nn_2+2(\nn_1\otimes\nn_3-\nn_3\otimes\nn_1)\otimes\nn_1\nn_3\nonumber\\
&\,+4(\nn_2\otimes\nn_3-\nn_3\otimes\nn_2)\otimes\nn_2\nn_3\nonumber\\
=&\,-2(\aaa_1\otimes\sss_3+\aaa_2\otimes\sss_4)+4\aaa_3\otimes\sss_5.\label{T4ijkl}
\end{align}
By virtue of the definition of symmetric tensors and (\ref{m1m2-m2m1-e}), it follows that
\begin{align}\label{eps-n123-1}
&\,\epsilon^{ijs}(\nn_1\nn_2\nn_3)_{kls}\nonumber\\
&\,=\frac{1}{6}\epsilon^{ijs}\Big(n_{1k}n_{2l}n_{3s}+n_{2k}n_{3l}n_{1s}+n_{3k}n_{1l}n_{2s}+n_{1k}n_{3l}n_{2s}+n_{2k}n_{1l}n_{3s}+n_{3k}n_{2l}n_{1s}\Big)\nonumber\\
&\,=\frac{1}{6}\Big((n_{1i}n_{2j}-n_{2i}n_{1j})n_{1k}n_{2l}+(n_{2i}n_{3j}-n_{3i}n_{2j})n_{2k}n_{3l}+(n_{3i}n_{1j}-n_{1i}n_{3j})n_{3k}n_{1l}\nonumber\\
&\,\quad+(n_{3i}n_{1j}-n_{1i}n_{3j})n_{1k}n_{3l}
+(n_{1i}n_{2j}-n_{2i}n_{1j})n_{2k}n_{1l}+(n_{2i}n_{3j}-n_{3i}n_{2j})n_{3k}n_{2l}\Big)\nonumber\\
&\,=\frac{1}{3}(\aaa_1\otimes\sss_3+\aaa_2\otimes\sss_4+\aaa_3\otimes\sss_5).
\end{align}
Rotation of the indices leads to
\begin{align}
\epsilon^{jis}(\nn_1\nn_2\nn_3)_{ksl}=&\,-\frac{1}{3}(\aaa_1\otimes\sss_3+\aaa_2\otimes\sss_4+\aaa_3\otimes\sss_5),\label{eps-n123-2}\\
\epsilon^{ijs}(\nn_1\nn_2\nn_3)_{skl}=&\,\frac{1}{3}(\aaa_1\otimes\sss_3+\aaa_2\otimes\sss_4+\aaa_3\otimes\sss_5).\label{eps-n123-3}
\end{align}
Analogous to the derivation of (\ref{m1m2-m2m1+sym}), it holds
\begin{align}
&\,\langle(m_{1i}m_{3j}-m_{3i}m_{1j})(\mm_1\mm_3)_{kl}\rangle\nonumber\\
&\,\quad=z\epsilon^{jis}(\nn_1\nn_2\nn_3)_{ksl}
+\frac{1}{6}\Big(2\CB\big(Q^{(0)}_1\big)_{ijkl}-\CB\big(Q^{(0)}_2\big)_{ijkl}\Big)\nonumber\\
&\,\quad=z\epsilon^{jis}(\nn_1\nn_2\nn_3)_{ksl}
+\frac{1}{6}\Big((2s_1+s_2)\CB\big((\nn^2_1)_0\big)_{ijkl}+(2b_1+b_2)\CB\big(\nn^2_2-\nn^2_3\big)_{ijkl}\Big),\label{m1m3-m3m1+sym}\\
&\,\langle(m_{2i}m_{3j}-m_{3i}m_{2j})(\mm_2\mm_3)_{kl}\rangle\nonumber\\
&\,\quad=z\epsilon^{ijs}(\nn_1\nn_2\nn_3)_{skl}
+\frac{1}{6}\Big(\CB\big(Q^{(0)}_1\big)_{ijkl}-2\CB\big(Q^{(0)}_2\big)_{ijkl}\Big)\nonumber\\
&\,\quad=z\epsilon^{ijs}(\nn_1\nn_2\nn_3)_{skl}
+\frac{1}{6}\Big((s_1+2s_2)\CB\big((\nn^2_1)_0\big)_{ijkl}+(b_1+2b_2)\CB\big(\nn^2_2-\nn^2_3\big)_{ijkl}\Big).\label{m2m3-m3m2+sym}
\end{align}

Therefore, taking advantage of the definition of $\CN^{(0)}_{Q_1}$ and combining (\ref{RR0_3}) and (\ref{RR0_4}) with (\ref{m1m2-m2m1+sym})-(\ref{eps-n123-2}) and (\ref{m1m3-m3m1+sym}), and using $1-e_1-e_2=0$, we deduce that
\begin{align}\label{CN-1}
\CN^{(0)}_{Q_1}=&\,\frac{1}{2}\CR^{(0)}_4+\frac{1}{2}(e_1-e_2)\CR^{(0)}_3-\frac{z}{3}(\aaa_1\otimes\sss_3+\aaa_2\otimes\sss_4+\aaa_3\otimes\sss_5)\nonumber\\
&\,+(e_1+e_2)\frac{z}{3}(\aaa_1\otimes\sss_3+\aaa_2\otimes\sss_4+\aaa_3\otimes\sss_5)\nonumber\\
&\,+\frac{1}{6}\Big((2s_1+s_2)\CB\big((\nn^2_1)_0\big)_{ijkl}+(2b_1+b_2)\CB\big(\nn^2_2-\nn^2_3\big)_{ijkl}\Big)\nonumber\\
&\,+(e_1+e_2)\frac{1}{6}\Big((s_1-s_2)\CB\big((\nn^2_1)_0\big)_{ijkl}+(b_1-b_2)\CB\big(\nn^2_2-\nn^2_3\big)_{ijkl}\Big)\nonumber\\
=&\,\frac{1}{2}\CR^{(0)}_4+\frac{1}{2}(e_1-e_2)\CR^{(0)}_3+(s_1-b_1)\aaa_1\otimes\sss_3-(s_1+b_1)\aaa_2\otimes\sss_4+2b_1\aaa_3\otimes\sss_5.
\end{align}
Similarly, combining (\ref{RR0_3}) and (\ref{RR0_5}) with (\ref{m1m2-m2m1+sym})-(\ref{eps-n123-1}), (\ref{eps-n123-3}) and (\ref{m2m3-m3m2+sym}), then we have
\begin{align}\label{CN-2}
\CN^{(0)}_{Q_2}
=&\,\frac{1}{2}\CR^{(0)}_5+\frac{1}{2}(e_2-e_1)\CR^{(0)}_3+\frac{z}{3}(\aaa_1\otimes\sss_3+\aaa_2\otimes\sss_4+\aaa_3\otimes\sss_5)\nonumber\\
&\,
-(e_1+e_2)\frac{z}{3}(\aaa_1\otimes\sss_3+\aaa_2\otimes\sss_4+\aaa_3\otimes\sss_5)\nonumber\\
&\,+\frac{1}{6}\Big((s_1+2s_2)\CB\big((\nn^2_1)_0\big)_{ijkl}+(b_1+2b_2)\CB\big(\nn^2_2-\nn^2_3\big)_{ijkl}\Big)\nonumber\\
&\,-(e_1+e_2)\frac{1}{6}\Big((s_1-s_2)\CB\big((\nn^2_1)_0\big)_{ijkl}+(b_1-b_2)\CB\big(\nn^2_2-\nn^2_3\big)_{ijkl}\Big)\nonumber\\
=&\,\frac{1}{2}\CR^{(0)}_5+\frac{1}{2}(e_2-e_1)\CR^{(0)}_3+(s_2-b_2)\aaa_1\otimes\sss_3-(s_2+b_2)\aaa_2\otimes\sss_4+2b_2\aaa_3\otimes\sss_5.
\end{align}
The equations \eqref{N1-u}--\eqref{V-matrix} then come from \eqref{CN-1} and \eqref{CN-2}.

Finally, we deal with the third-order tensors. By a direct calculation, we get
\begin{align*}
(\nn_1\nn_2\nn_3)_{ijk}=&\,\frac{1}{6}\Big(n_{1i}(n_{2j}n_{3k}+n_{3j}n_{2k})+n_{2i}(n_{1j}n_{3k}+n_{3j}n_{1k})\\
&\,+n_{3i}(n_{1j}n_{2k}+n_{2j}n_{1k})\Big)\\
=&\,\frac{1}{3}\big(\nn_1\otimes\sss_5+\nn_2\otimes\sss_4+\nn_3\otimes\sss_3\big).
\end{align*}
Meanwhile, we also easily deduce that
\begin{align*}
\epsilon^{ijs}\big((\nn^2_1)_0\big)_{ks}+\epsilon^{iks}\big((\nn^2_1)_0\big)_{js}=&\,(n_{2i}n_{3j}-n_{3i}n_{2j})n_{1k}+(n_{2i}n_{3k}-n_{3i}n_{2k})n_{1j}\\
=&\,2(\nn_2\otimes\sss_4-\nn_3\otimes\sss_3),\\
\epsilon^{ijs}(\nn^2_2-\nn^2_3)_{ks}+\epsilon^{iks}(\nn^2_2-\nn^2_3)_{js}
=&\,n_{2k}(n_{3i}n_{1j}-n_{1i}n_{3j})-n_{3k}(n_{1i}n_{2j}-n_{2i}n_{1j})\\
&\,+n_{2j}(n_{3i}n_{1k}-n_{1i}n_{3k})-n_{3j}(n_{1i}n_{2k}-n_{2i}n_{1k})\\
=&\,2\big(\nn_3\otimes\sss_3-2\nn_1\otimes\sss_5+\nn_2\otimes\sss_4\big).
\end{align*}
Hence, by using (\ref{m1-m2m3-block}), we derive from Theorem \ref{Q-biaxial-theorem}
 that
\begin{align}\label{average-m1-m2m3-block}
\langle\mm_1\otimes\mm_2\mm_3\rangle_{ijk}=&\,
\langle\mm_1\mm_2\mm_3\rangle_{ijk}+\frac{1}{6}\Big(\epsilon^{ijs}\big(\langle(\mm^2_3)_0\rangle-\langle(\mm^2_2)_0\rangle\big)_{ks}\nonumber\\
&\,+\epsilon^{iks}\big(\langle(\mm^2_3)_0\rangle-\langle(\mm^2_2)_0\rangle\big)_{js}\Big)\nonumber\\
=&\,z(\nn_1\nn_2\nn_3)_{ijk}-\frac{1}{6}\Big(\epsilon^{ijs}\big(Q^{(0)}_1+2Q^{(0)}_2\big)_{ks}+\epsilon^{iks}\big(Q^{(0)}_1+2Q^{(0)}_2\big)_{js}\Big)\nonumber\\
=&\,z(\nn_1\nn_2\nn_3)_{ijk}-\frac{1}{6}(s_1+2s_2)\Big(\epsilon^{ijs}\big((\nn^2_1)_0\big)_{ks}+\epsilon^{iks}\big((\nn^2_1)_0\big)_{js}\Big)\nonumber\\
&\,-\frac{1}{6}(b_1+2b_2)\Big(\epsilon^{ijs}(\nn^2_2-\nn^2_3)_{ks}+\epsilon^{iks}(\nn^2_2-\nn^2_3)_{js}\Big)\nonumber\\
=&\,\frac{1}{3}(z+2b_1+4b_2)\nn_1\otimes\sss_5+\frac{1}{3}(z-s_1-2s_2-b_1-2b_2)\nn_2\otimes\sss_4\nonumber\\
&\,+\frac{1}{3}(z+s_1+2s_2-b_1-2b_2)\nn_3\otimes\sss_3,
\end{align}
for which the coefficient matrix $T_1$ under the basis $\nn_i\otimes\sss_j$ is given by
\begin{align}\label{T1a}
T_1=\left(
    \begin{array}{ccccc}
        0 &\, 0 &\, 0 &\, 0 &\, \frac{1}{3}(z+2b_1+4b_2)
        \vspace{3ex}\\
       0 &\, 0 &\, 0 &\, \frac{1}{3}(z-s_1-2s_2 &\,0
        \\
        &\, &\, &\, -b_1-2b_2) &\,\vspace{0.5ex}\\
        0&\,0 &\, \frac{1}{3}(z+s_1+2s_2 &\, 0 &\, 0\\
        &\, &\, -b_1-2b_2) &\, &\,
    \end{array}
    \right).
\end{align}
Following the same procedure, we obtain
\begin{align}
\langle\mm_2\otimes\mm_1\mm_3\rangle_{ijk}=&\,\langle\mm_1\mm_2\mm_3\rangle_{ijk}+\frac{1}{6}\Big(\epsilon^{ijs}\big(\langle(\mm^2_1)_0\rangle-\langle(\mm^2_3)_0\rangle\big)_{ks}\nonumber\\
&\,+\epsilon^{iks}\big(\langle(\mm^2_1)_0\rangle-\langle(\mm^2_3)_0\rangle\big)_{js}\Big)\nonumber\\
=&\,z(\nn_1\nn_2\nn_3)_{ijk}+\frac{1}{6}(2s_1+s_2)\Big(\epsilon^{ijs}\big((\nn^2_1)_0\big)_{ks}+\epsilon^{iks}\big((\nn^2_1)_0\big)_{js}\Big)\nonumber\\
&\,+\frac{1}{6}(2b_1+b_2)\Big(\epsilon^{ijs}(\nn^2_2-\nn^2_3)_{ks}+\epsilon^{iks}(\nn^2_2-\nn^2_3)_{js}\Big)\nonumber\\
=&\,\frac{1}{3}(z-4b_1-2b_2)\nn_1\otimes\sss_5+\frac{1}{3}(z+2s_1+s_2+2b_1+b_2)\nn_2\otimes\sss_4\nonumber\\
&\,+\frac{1}{3}(z-2s_1-s_2+2b_1+b_2)\nn_3\otimes\sss_3,\label{average-m2-m1m3-block}\\
\langle\mm_3\otimes\mm_1\mm_2\rangle_{ijk}=&\,\langle\mm_1\mm_2\mm_3\rangle_{ijk}+\frac{1}{6}\Big(\epsilon^{ijs}\big(\langle(\mm^2_2)_0\rangle-\langle(\mm^2_1)_0\rangle\big)_{ks}\nonumber\\
&\,+\epsilon^{iks}\big(\langle(\mm^2_2)_0\rangle-\langle(\mm^2_1)_0\rangle\big)_{js}\Big)\nonumber\\
=&\,z(\nn_1\nn_2\nn_3)_{ijk}-\frac{1}{6}(s_1-s_2)\Big(\epsilon^{ijs}\big((\nn^2_1)_0\big)_{ks}+\epsilon^{iks}\big((\nn^2_1)_0\big)_{js}\Big)\nonumber\\
&\,-\frac{1}{6}(b_1-b_2)\Big(\epsilon^{ijs}(\nn^2_2-\nn^2_3)_{ks}+\epsilon^{iks}(\nn^2_2-\nn^2_3)_{js}\Big)\nonumber\\
=&\,\frac{1}{3}(z+2b_1-2b_2)\nn_1\otimes\sss_5+\frac{1}{3}(z-s_1+s_2-b_1+b_2)\nn_2\otimes\sss_4\nonumber\\
&\,+\frac{1}{3}(z+s_1-s_2-b_1+b_2)\nn_3\otimes\sss_3,\label{average-m3-m1m2-block}
\end{align}
where the associated coefficient matrices $T_2, T_3$ in (\ref{average-m2-m1m3-block}) and (\ref{average-m3-m1m2-block}) can be written as
\begin{align}
T_2=&\,\left(
    \begin{array}{ccccc}
        0 &\, 0 &\, 0 &\, 0 &\, \frac{1}{3}(z-4b_1-2b_2)
        \vspace{3ex}\\
       0 &\, 0 &\, 0 &\, \frac{1}{3}(z+2s_1+s_2 &\,0
        \\
        &\, &\, &\, +2b_1+b_2) &\,\vspace{0.5ex}\\
        0&\,0 &\, \frac{1}{3}(z-2s_1-s_2 &\, 0 &\, 0\\
        &\, &\, +2b_1+b_2) &\, &\,
    \end{array}
    \right),\label{T2a}\\
T_3=&\,\left(
    \begin{array}{ccccc}
        0 &\, 0 &\, 0 &\, 0 &\, \frac{1}{3}(z+2b_1-2b_2)
        \vspace{3ex}\\
       0 &\, 0 &\, 0 &\, \frac{1}{3}(z-s_1+s_2 &\,0
        \\
        &\, &\, &\, -b_1+b_2) &\,\vspace{0.5ex}\\
        0&\,0 &\, \frac{1}{3}(z+s_1-s_2 &\, 0 &\, 0\\
        &\, &\, -b_1+b_2) &\, &\,
    \end{array}
    \right).\label{T3a}
\end{align}

Define
\begin{align}
    &\,w_i^T=(s_i,b_i,0,0,0),~i=1,2, \nonumber\\
    &\,W_1=\mathrm{diag}\Big(\frac{1}{3}(2s_1+1),\frac{1}{3}(1-s_1)+b_1,\frac{1}{3}(1-s_1)-b_1\Big), \nonumber\\
    &\,W_2=\mathrm{diag}\Big(\frac{1}{3}(2s_2+1),\frac{1}{3}(1-s_2)+b_2,\frac{1}{3}(1-s_2)-b_2\Big), \nonumber\\
    &\,W_3=\mathrm{diag}\Big(\frac{1}{3}(1-2s_1-2s_2),\frac{1}{3}(1+s_1+s_2)-b_1-b_2,\frac{1}{3}(1+s_1+s_2)+b_1+b_2\Big). \label{Q-var}
\end{align}
Then, the quasi-entropy $\Xi_4$ can be reduced to
\begin{align}
    \Xi_{4,\mathrm{Bi}}=&\,-\ln\det\left(
    \begin{array}{ccc}
        1 &\,  &\, \\
         &\, \Lambda &\,  \\
         &\, &\, \Lambda
    \end{array}
    \right)\left(
    \begin{array}{ccc}
        1 &\, w_1^T &\, (2w_2+w_1)^T\vspace{1ex}\\
        w_1 &\,
        R_1 &\, S \vspace{1ex}\\
        2w_2+w_1 &\,
        S^T &\,
        R_6
    \end{array}
    \right)\left(
    \begin{array}{ccc}
        1 &\,  &\, \\
         &\, \Lambda &\,  \\
         &\, &\, \Lambda
    \end{array}
    \right)\nonumber\\
    &\,-\ln\det\left(
    \begin{array}{cc}
        1 &\,  \\
         &\, \Lambda
    \end{array}
    \right)\left(
    \begin{array}{cc}
        W_1 &\, T_1 \vspace{1ex}\\
        T_1^T &\, R_3
    \end{array}
    \right)\left(
    \begin{array}{cc}
        1 &\,  \\
         &\, \Lambda
    \end{array}
    \right)\nonumber\\
    &\,-\ln\det\left(
    \begin{array}{cc}
        1 &\,  \\
         &\, \Lambda
    \end{array}
    \right)\left(
    \begin{array}{cc}
        W_2 &\, T_2 \vspace{1ex}\\
        T_2^T &\, R_4
    \end{array}
    \right)\left(
    \begin{array}{cc}
        1 &\,  \\
         &\, \Lambda
    \end{array}
    \right)\nonumber\\
    &\,-\ln\det\left(
    \begin{array}{cc}
        1 &\,  \\
         &\, \Lambda
    \end{array}
    \right)\left(
    \begin{array}{cc}
        W_3 &\, T_3 \vspace{1ex}\\
        T_3^T &\, R_5
    \end{array}
    \right)\left(
    \begin{array}{cc}
        1 &\,  \\
         &\, \Lambda
    \end{array}
    \right)\nonumber\\
    =&\,-\ln\det\left(
    \begin{array}{ccc}
        1 &\, w_1^T &\, (2w_2+w_1)^T\vspace{1ex}\\
        w_1 &\,
        R_1 &\, S \vspace{1ex}\\
        2w_2+w_1 &\,
        S^T &\,
        R_6
    \end{array}
    \right)
    -\ln\det\left(
    \begin{array}{cc}
        W_1 &\, T_1 \vspace{1ex}\\
        T_1^T &\, R_3
    \end{array}
    \right)\nonumber\\
    &\,-\ln\det\left(
    \begin{array}{cc}
        W_2 &\, T_2 \vspace{1ex}\\
        T_2^T &\, R_4
    \end{array}
    \right)
    -\ln\det\left(
    \begin{array}{cc}
        W_3 &\, T_3 \vspace{1ex}\\
        T_3^T &\, R_5
    \end{array}
    \right)-10\ln\det\Lambda.\label{qent-4th-coord}
\end{align}
The expressions of $R_i$, $S$, $T_i$ can all be found above.

\section{The uniaxial case: Theorem \ref{Q-unixial-theorem}}\label{Apprendix-D}

Assume that $Q_i$ are uniaxial, i.e. $b_i=0$ so that
\begin{align*}
  \,Q_i=s_i\Big(\nn_1^2-\frac{\Fi}{3}\Big),~i=1,2.
\end{align*}
By \eqref{range-2nd}, we require that the two scalars $s_i$ satisfy
\begin{equation}\label{range-uni-2nd}
    -\frac 12<s_1,~s_2,~-s_1-s_2<1.
\end{equation}

For the original entropy, the discussion is similar to the biaxial case.
\begin{lemma}
If $s_i$ satisfy \eqref{range-uni-2nd}, then there exists a unique density function
\begin{align*}
    \rho=\frac 1Z \exp(\sum_{i=1,2}\lambda_{i}(\mm_i\cdot\nn_1)^2)
\end{align*}
such that $\langle(\mm_i^2)_0\rangle=s_i(\nn_1^2)_0$.
\end{lemma}
We omit the rest of the derivation since it is the same as the biaxial case.

We turn to the quasi-entropy.
Here, we need to notice that
\begin{align}
    2X_6+X_4=\left(
    \begin{array}{ccccc}
        0 &\, \frac{3}{14} &\,  &\,  &\,
        \vspace{0.5ex}\\
        \frac{3}{14} &\, 0 &\,  &\,  &\,
        \vspace{1ex}\\
        &\, &\, \frac{4}{7} &\, &\,
        \vspace{0.5ex}\\
        &\, &\, &\, -\frac{4}{7} &\,
        \vspace{0.5ex}\\
        &\, &\, &\, &\, 0
    \end{array}
    \right)\eqdefa X_6', \\
    8X_6+8X_5+X_4=\left(
    \begin{array}{ccccc}
        0 &\, 0 &\,  &\,  &\,
        \vspace{0.5ex}\\
        0 &\, 1 &\,  &\,  &\,
        \vspace{1ex}\\
        &\, &\, 0 &\, &\,
        \vspace{0.5ex}\\
        &\, &\, &\, 0 &\,
        \vspace{0.5ex}\\
        &\, &\, &\, &\, -4
    \end{array}
    \right)\eqdefa X_5'.
\end{align}
Let us define
\begin{align}
    &\,a_1'=a_1-\frac 12 a_3+\frac 38 a_2,\quad a_2'=\frac 18 a_2,\quad a_3=\frac 12(a_3-a_2),\nonumber\\
    &\,\tilde{a}_1'=\tilde{a}_1-\frac 12 \tilde{a}_3+\frac 38 \tilde{a}_2,\quad \tilde{a}_2'=\frac 18 \tilde{a}_2,\quad \tilde{a}_3=\frac 12(\tilde{a}_3-\tilde{a}_2), \nonumber\\
    &\,\bar{a}_1'=\bar{a}_1-\frac 12 \bar{a}_3+\frac 38 \bar{a}_2,\quad \bar{a}_2'=\frac 18 \bar{a}_2,\quad \bar{a}_3=\frac 12(\bar{a}_3-\bar{a}_2). \nonumber
\end{align}
It can be verified that
\begin{align*}
    a_1X_4+a_2X_5+a_3X_6=a_1'X_4+a_2'X_5'+a_3'X_6'.
\end{align*}

In what follows, we show that when $Q_i$ are uniaxial, $\Xi_{4,\mathrm{Bi}}$ reaches its minimum only when $a_2'=a_3'=\tilde{a}_2'=\tilde{a}_3'=\bar{a}_2'=\bar{a}_3'=z=0$.

Let us discuss each of the log-determinant in \eqref{qent-4th-coord}.
We could rearrange the indices to arriave at
\begin{align}
    &\,\hspace{-36pt}-\ln\det\left(
    \begin{array}{ccc}
        1 &\, w_1^T &\, (2w_2+w_1)^T\vspace{1ex}\\
        w_1 &\,
        R_1 &\, S \vspace{1ex}\\
        2w_2+w_1 &\,
        S^T &\,
        R_6
    \end{array}
    \right)\nonumber\\
    =&\,-\ln\det\left(\begin{array}{ccc}
         1 &\, (s_1,2s_1+s_2) &\, 0_{1\times 2} \vspace{1ex}\\
         (s_1,2s_1+s_2)^T &\, \Upsilon_1  &\, \frac{3}{14}\Theta_3- z\Pi_1   \vspace{1ex}\\
         0_{2\times 1} &\, \frac{3}{14}\Theta_3+ z\Pi_1 &\,  \Upsilon_2 +\Theta_2
    \end{array}
    \right)\nonumber\\
    &\,-\ln\det(\Upsilon_3+\frac{4}{7}\Theta_3)-\ln\det(\Upsilon_3-\frac{4}{7}\Theta_3)
    -\ln\det(4\Upsilon_2-4\Theta_2),
\end{align}
where the blocks $\Upsilon_i$, $\Theta_i$ and $\Pi_1$ are given by
\begin{align*}
    &\,\Upsilon_1=\left(\begin{array}{cc}
        \frac 15+\frac 27 s_1+\frac{18}{35}a_1' & -\frac 27(s_1+2s_2) +\frac{18}{35}(a_1'+2\bar{a}_1')
        \vspace{1ex}\\
        -\frac 27(s_1+2s_2)+\frac{18}{35}(a_1'+2\bar{a}_1') & \frac 35-\frac 67 s_1+\frac{18}{35}(a_1'+4\tilde{a}_1'+4\bar{a}_1')
    \end{array}\right),\\
    &\,\Upsilon_2=\left(\begin{array}{cc}
        \frac {1}{15}-\frac {2}{21} s_1+\frac{1}{35}a_1' & \frac{2}{21}(s_1+2s_2) +\frac{1}{35}(a_1'+2\bar{a}_1')
        \vspace{1ex}\\
        \frac{2}{21}(s_1+2s_2) +\frac{1}{35}(a_1'+2\bar{a}_1') & \frac 15+\frac 27 s_1 +\frac{1}{35}(a_1'+4\tilde{a}_1'+4\bar{a}_1')
    \end{array}\right),\\
    &\,\Upsilon_3=\left(\begin{array}{cc}
        \frac {4}{15}+\frac {4}{21} s_1-\frac{16}{35}a_1' & -\frac{4}{21}(s_1+2s_2) -\frac{16}{35}(a_1'+2\bar{a}_1')
        \vspace{1ex}\\
        -\frac{4}{21}(s_1+2s_2) -\frac{16}{35}(a_1'+2\bar{a}_1') & \frac 45-\frac 47 s_1 -\frac{16}{35}(a_1'+4\tilde{a}_1'+4\bar{a}_1')
    \end{array}\right),\\
    &\,\Theta_2=\left(\begin{array}{cc}
        a_2' & a_2'+2\bar{a}_2'
        \vspace{1ex}\\
        a_2'+2\bar{a}_2' & a_2'+4\tilde{a}_2'+4\bar{a}_2'
    \end{array}\right),\nonumber\\
    &\,\Theta_3=\left(\begin{array}{cc}
        a_3' & a_3'+2\bar{a}_3'
        \vspace{1ex}\\
        a_3'+2\bar{a}_3' & a_3'+4\tilde{a}_3'+4\bar{a}_3'
    \end{array}\right),\nonumber\\
    &\,\Pi_1=\left(
    \begin{array}{cc}
        0 &\, \frac 32
        \vspace{1ex}\\
        -\frac 32 &\, 0
    \end{array}
    \right).
\end{align*}
Notice that $\Upsilon_i$ does not depend on $a_i',\tilde{a}_i',\bar{a}_i'$ for $i=2,3$, and $\Theta_2$, $\Theta_3$ only depend on them.

By Lemma \ref{det-est1}, we deduce that
\begin{align}
    &\,-\ln\det\left(\begin{array}{ccc}
         1 &\, (s_1,2s_1+s_2) &\, 0_{1\times 2} \vspace{1ex}\\
         (s_1,2s_1+s_2)^T &\, \Upsilon_1  &\, \frac{3}{14}\Theta_3- z\Pi_1   \vspace{1ex}\\
         0_{2\times 1} &\, \frac{3}{14}\Theta_3+ z\Pi_1 &\,  \Upsilon_2 +\Theta_2
    \end{array}
    \right)\nonumber\\
    &\,\quad\ge -\ln\det\left(\begin{array}{cc}
         1 &\, (s_1,2s_1+s_2) \vspace{1ex}\\
         (s_1,2s_1+s_2)^T &\, \Upsilon_1
    \end{array}
    \right)
    -\ln\det(\Upsilon_2 +\Theta_2). \label{est1}
\end{align}
The equality holds if and only if $z=0$ and $\Theta_3=0$.
In addition, it shall be noticed that $-\ln\det A$ is strictly convex about $A$ (see, for example, Lemma 4.5 in \cite{Xu3} for a proof).
Therefore, we obtain
\begin{align}
    &\,-\ln\det(\Upsilon_2 +\Theta_2)-\ln\det(4\Upsilon_2-4\Theta_2)\ge -2\ln\det\Upsilon_2-2\ln 4, \nonumber\\
    &\,-\ln\det(\Upsilon_3+\frac{4}{7}\Theta_3)-\ln\det(\Upsilon_3-\frac{4}{7}\Theta_3)\ge -2\ln\det\Upsilon_3. \label{est2}
\end{align}
The equalities hold if and only if $\Theta_2=\Theta_3=0$.

Let us look into another log-determinant in \eqref{qent-4th-coord}.  It follows that
\begin{align}
    &\,-\ln\det\left(
    \begin{array}{cc}
        W_1 &\, T_1 \vspace{1ex}\\
        T_1^T &\, R_3
    \end{array}
    \right)=\nonumber\\
    &\,\quad-\ln\det\left(\begin{array}{cccccccc}
        \xi_1 &\,  &\,  &\,  &\,  &\,  &\,  &\, \frac 13 z\\
         &\, \xi_2 &\,  &\,  &\,  &\,  &\, -\xi_3+\frac 13 z &\, \\
         &\,  &\, \xi_2 &\,  &\,  &\, \xi_3+\frac 13z &\,  &\, \\
         &\,  &\,  &\, \xi_4 &\, \frac{3}{14}\bar{a}_2' &\,  &\,  &\, \vspace{1ex}\\
         &\,  &\,  &\, \frac{3}{14}\bar{a}_2' &\, \xi_5+\bar{a}'_3 &\,  &\,  &\, \\
         &\,  &\, \xi_3+\frac 13 z &\,  &\,  &\, \xi_6+\frac{4}{7}\bar{a}'_2 &\,  &\, \\
         &\, -\xi_3+\frac 13 z &\,  &\,  &\,  &\,  &\, \xi_6-\frac{4}{7}\bar{a}'_2 &\, \\
        \frac 13 z &\,  &\,  &\,  &\,  &\,  &\,  &\, 4\xi_5-4\bar{a}'_3
    \end{array}\right).\nonumber
\end{align}
In the above, those $\xi_i$ are given by
\begin{align*}
    &\,\xi_1=\frac{1}{3}(2s_1+1),\quad \xi_2=\frac{1}{3}(1-s_1),\quad
    \xi_3=\frac{1}{3}(s_1+2s_2), \\
    &\,\xi_4=\frac{3}{5}+\frac{6}{7}(s_1+s_2)+\frac{72}{35}\bar{a}_1', \\
    &\,\xi_5=\frac{1}{5}-\frac{2}{7}(s_1+s_2)+\frac{4}{35}\bar{a}_1', \\
    &\,\xi_6=\frac{4}{5}+\frac{4}{7}(s_1+s_2)-\frac{64}{35}\bar{a}_1'.
\end{align*}
Since the function $-\ln x$ is monotonely decreasing and strictly convex, we have the inequality
\begin{align}
    &\,-\ln\Big(\xi_1(4\xi_5-4\bar{a}_3')-\frac{1}{9}z^2\Big)
    -\ln\Big(\xi_2(\xi_6-\frac{4}{7}\bar{a}_2')-(\xi_3-\frac{1}{3}z)^2\Big)\nonumber\\
    &\,-\ln\Big(\xi_2(\xi_6+\frac{4}{7}\bar{a}_2')-(\xi_3+\frac{1}{3}z)^2\Big)
    -\ln\Big(\xi_4(\xi_5+\bar{a}_3')-(\frac{3}{14}\bar{a}_2')^2\Big)\nonumber\\
    \ge &\,-\ln\Big(\xi_1(4\xi_5-4\bar{a}_3')\Big)
    -\ln\Big(\xi_2(\xi_6-\frac{4}{7}\bar{a}_2')-\xi_3^2+\frac{2}{3}\xi_3z\Big)\nonumber\\
    &\,-\ln\Big(\xi_2(\xi_6+\frac{4}{7}\bar{a}_2')-\xi_3^2-\frac{2}{3}\xi_3z\Big)
    -\ln\Big(\xi_1(\xi_5+\bar{a}_3')\Big)\nonumber\\
    =&\, -\ln\xi_1-\ln\xi_4-\ln 4\nonumber\\
    &\, -\ln(\xi_5-\bar{a}_3')-\ln(\xi_5+\bar{a}_3')\nonumber\\
    &\, -\ln\Big(\xi_2(\xi_6-\frac{4}{7}\bar{a}_2')-\xi_3^2+\frac{2}{3}\xi_3z\Big)-\ln\Big(\xi_2(\xi_6+\frac{4}{7}\bar{a}_2')\Big)\nonumber\\
    \ge &\, -\ln\xi_1-\ln\xi_4-\ln 4-2\ln\xi_5-2\ln(\xi_2\xi_6-\xi_3^2).\label{est3}
\end{align}
The equalities hold if and only if $\bar{a}_2'=\bar{a}_3'=z=0$.

Similarly, we could deal with the other two log-determinants in \eqref{qent-4th-coord}.
Summarizing \eqref{est1}, \eqref{est2} and \eqref{est3}, we conclude that when $Q_i$ are uniaxial, at the minimizer we must have $a_2'=a_3'=\tilde{a}_2'=\tilde{a}_3'=\bar{a}_2'=\bar{a}_3'=z=0$.

\section{The orientational elasticity}\label{Apprendix-E}

For the readers' convenience, we present the orientational elasticity for the biaxial nematic phases that can be found in \cite{Xu2}, where the elasitic constants expressed the coefficients in the molecular-theory-based static $Q$-tensor model.
In addition, the variational derivatives with respect to the orthonomal frame $\Fp=(\nn_1,\nn_2,\nn_3)$ are derived.

We first write down an equivalent formulation of \eqref{elas_Bi}.
Using the following relations
\begin{align*}
  &\,\nabla\cdot\nn_2=-D_{31}+D_{13},\qquad
  \nn_2\cdot\nabla\times\nn_2=D_{33}+D_{11}, \\
  &\,\nn_3\cdot\nabla\times\nn_2=-D_{23},\qquad
  \nn_1\cdot\nabla\times\nn_2=-D_{21},\\
  &\,|\nn_2\times\nabla\times\nn_2|^2=(\nn_1\cdot\nabla\times\nn_2)^2+(\nn_3\cdot\nabla\times\nn_2)^2,
\end{align*}
 together with (\ref{elas_Bi}) yields that the equivalent expression analogous to the Oseen-Frank energy can be given by
\begin{align}
\frac{\CF_{Bi}(\Fp)}{ck_BT}=&\,\int\ud\bm{x}~\frac{1}{2}\Big(K_1(\nabla\cdot\nn_1)^2+K_2(\nn_1\cdot\nabla\times\nn_1)^2
+K_3(\nn_1\times\nabla\times\nn_1)^2 \nonumber\\
&\,+K_4(\nabla\cdot\nn_2)^2+K_5(\nn_2\cdot\nabla\times\nn_2)^2
+K_6(\nn_2\times\nabla\times\nn_2)^2\nonumber\\
&\,+K_7(\nabla\cdot\nn_3)^2+K_8(\nn_3\cdot\nabla\times\nn_3)^2
+K_9(\nn_3\times\nabla\times\nn_3)^2\nonumber\\
&\,+K_{10}(\nn_1\cdot\nabla\times\nn_3)^2+K_{11}(\nn_2\cdot\nabla\times\nn_1)^2+K_{12}(\nn_3\cdot\nabla\times\nn_2)^2\Big),\label{elas_Bi2}
\end{align}
where the elastic coefficients $K_i(i=1,\cdots,12)$ can be expressed by $K_{ijkl}(i,j,k,l=1,2,3)$ (see \cite{GV1} for details).
In the above, we also neglect the surface terms \eqref{surfterm}.

The next task is to provide the biaxial elastic energy with the form (\ref{elas_Bi}) derived from the molecular-theory-based static tensor model (\ref{free-energy}), where the elastic coefficients $K_{ijkl}$ are expressed by molecular parameters. We refer to \cite{Xu2} for more detailed discussion.

Assume that the minimizers of the bulk energy in (\ref{free-energy}) has the following biaxial form:
\begin{align*}
Q_{\alpha}=(s_{\alpha}+b_{\alpha})\nn^2_1+2b_{\alpha}\nn^2_2-\Big(\frac13s_{\alpha}+b_{\alpha}\Big)\Fi,\quad \alpha=1,2.
\end{align*}
Then the corresponding derivative terms are calculated as
\begin{align*}
|\nabla Q_{\alpha}|^2=&\,2(s_{\alpha}+b_{\alpha})^2(\partial_kn_{1i})^2+8b^2_{\alpha}(\partial_kn_{2i})^2+8b_{\alpha}(s_{\alpha}+b_{\alpha})n_{1i}n_{2j}\partial_kn_{1j}\partial_kn_{2i},\\
\partial_iQ_{1jk}\partial_iQ_{2jk}=&\,2(s_1+b_1)(s_2+b_2)(\partial_in_{1j})^2+8b_1b_2(\partial_in_{2j})^2\\
&\,+4[b_1(s_2+b_2)+b_2(s_1+b_1)]n_{1j}n_{2k}\partial_in_{1k}\partial_in_{2j},\\
\partial_iQ_{\alpha ik}\partial_jQ_{\beta jk}=&\,(s_{\alpha}+b_{\alpha})(s_{\beta}+b_{\beta})\big(|\nabla\cdot\nn_1|^2+n_{1i}n_{1j}\partial_in_{1k}\partial_jn_{1k}\big)\\
&\,+2\big[b_{\alpha}(s_{\beta}+b_{\beta})+b_{\beta}(s_{\alpha}+b_{\alpha})\big]\Big((\nabla\cdot\nn_1)n_{1k}n_{2j}\partial_jn_{2k}+(\nabla\cdot\nn_2)n_{1i}n_{2k}\partial_in_{1k}\\
&\,+n_{1i}n_{2j}\partial_in_{1k}\partial_jn_{2k}\Big)
+4b_{\alpha}b_{\beta}\big(|\nabla\cdot\nn_2|^2+n_{2i}n_{2j}\partial_in_{2k}\partial_jn_{2k}\big).
\end{align*}
From which and the elastic energy in (\ref{free-energy}) implies that
\begin{align}\label{elest-bi-11}
\frac{\CF_{Bi}(\Fp)}{ck_BT}=&\,\int\ud\xx\frac{1}{2}\Big[J_1(\partial_in_{1j})^2+J_2(\partial_in_{2j})^2+J_3n_{1i}n_{2j}\partial_kn_{1j}\partial_kn_{2i}\nonumber\\
&\,+J_4\big(|\nabla\cdot\nn_1|^2+n_{1i}n_{1j}\partial_in_{1k}\partial_jn_{1k}\big)
+J_5\big(|\nabla\cdot\nn_2|^2+n_{2i}n_{2j}\partial_in_{2k}\partial_jn_{2k}\big)\nonumber\\
&\,+J_6\big((\nabla\cdot\nn_1)n_{1k}n_{2j}\partial_jn_{2k}+(\nabla\cdot\nn_2)n_{1i}n_{2k}\partial_in_{1k}
+n_{1i}n_{2j}\partial_in_{1k}\partial_jn_{2k}\big)
\Big],
\end{align}
where the coefficients $J_i(i=1,\cdots,6)$ are given by \eqref{J-coefficients-6}.

We need to express the derivative terms in (\ref{elest-bi-11}) by the nine invariant $D_{\lambda\delta}(\lambda,\delta=1,2,3)$. For example, the following four terms can be respectively expressed as
\begin{align*}
(\partial_in_{1j})^2=&\,\delta_{jl}\delta_{ik}\partial_kn_{1l}\partial_in_{1j}\\
=&\,(n_{2j}n_{2l}+n_{3j}n_{3l})(n_{1i}n_{1k}+n_{2i}n_{2k}+n_{3i}n_{3k})\partial_kn_{1l}\partial_in_{1j}\\
=&\,\big(n_{1i}n_{2j}n_{1k}n_{2l}+n_{2i}n_{2j}n_{2k}n_{2l}+n_{3i}n_{2j}n_{3k}n_{2l}+n_{1i}n_{3j}n_{1k}n_{3l}\\
&\,+n_{2i}n_{3j}n_{2k}n_{3l}+n_{3i}n_{3j}n_{3k}n_{3l}\big)\partial_kn_{1l}\partial_in_{1j}\\
=&\,D^2_{13}+D^2_{23}+D^2_{33}+D^2_{12}+D^2_{22}+D^2_{32},\\
\partial_jn_{1j}=&\,\delta_{ij}\partial_in_{1j}=(n_{2i}n_{2j}+n_{3i}n_{3j})\partial_in_{1j}\\
=&\,D_{32}-D_{23},\\
n_{1i}n_{2j}\partial_kn_{1j}\partial_kn_{2i}=&\,\delta_{jl}\delta_{ks}n_{1i}n_{2l}\partial_kn_{1j}\partial_sn_{2i}\\
=&\,n_{2j}n_{2l}(n_{1k}n_{1s}+n_{2k}n_{2s}+n_{3k}n_{3s})n_{1i}n_{2l}\partial_kn_{1j}\partial_sn_{2i}\\
=&\,(n_{1k}n_{2j}n_{1s}n_{1i}+n_{2k}n_{2j}n_{2s}n_{1i}+n_{3k}n_{2j}n_{3s}n_{1i})\partial_kn_{1j}\partial_sn_{2i}\\
=&\,-(D^2_{13}+D^2_{23}+D^2_{33}),\\
n_{1i}n_{1j}\partial_in_{1k}\partial_jn_{1k}=&\,\delta_{kl}n_{1i}n_{1j}\partial_in_{1l}\partial_jn_{1k}\\
=&\,(n_{2k}n_{2l}+n_{3k}n_{3l})n_{1i}n_{1j}\partial_in_{1l}\partial_jn_{1k}\\
=&\,D^2_{12}+D^2_{13}.
\end{align*}
While the remaining four terms
can be similarly expressed as follows:
\begin{align*}
(\partial_in_{2j})^2
=&\,D^2_{13}+D^2_{23}+D^2_{33}+D^2_{11}+D^2_{21}+D^2_{31},\\
\partial_jn_{2j}
=&\,D_{13}-D_{31},\\
n_{2i}n_{2j}\partial_in_{2k}\partial_jn_{2k}
=&\,D^2_{21}+D^2_{23},\\
n_{1i}n_{2j}\partial_in_{1k}\partial_jn_{2k}
=&\,-D_{12}D_{21}.
\end{align*}
Plugging the above eight relations into (\ref{elest-bi-11}), we immediately obtain the biaxial elastic energy (\ref{elas_Bi}), where the elastic coefficients $K_{ijkl}(i,j,k,l=1,2,3)$, completely determined by the molecular parameters, are given by \eqref{K-coeffi}.

Then, we calculate the variational derivative about the frame $\mathfrak{p}$, and derive the variational derivative along the infinitesimal rotation round $\nn_i(i=1,2,3)$. For instance, the variational derivative along the infinitesimal rotation round $\nn_1$ is given by
\begin{align*}
    n_{2\alpha}\frac{\delta}{\delta n_{3\alpha}}-n_{3\alpha}\frac{\delta}{\delta n_{2\alpha}},
\end{align*}
where the operator $\frac{\delta}{\delta n_{3\alpha}}$ represents the variational derivative about $\nn_3$ assuming that $\nn_3$ is an independent vector (ignoring the constraints that $\nn_3\cdot\nn_3=1$ and $\nn_3\cdot\nn_1=\nn_3\cdot\nn_2=0$).

Therefore, the variational derivatives of the elastic energy (\ref{elas_Bi}) with respect to the frame $\Fp$ can be respectively calculated as follows:
\begin{align}
\frac{\delta \CF_{Bi}}{\delta n_{1\alpha}}=&\,K_{1111}D_{11}n_{2k}\partial_{\alpha}n_{3k}-K_{2222}\partial_k(D_{22}n_{2k}n_{3\alpha})+K_{3333}D_{33}n_{3k}\partial_kn_{2\alpha}\nonumber\\
&\,+K_{1212}\big(D_{12}n_{3k}\partial_{\alpha}n_{1k}-\partial_k(D_{12}n_{1k}n_{3\alpha})\big)+K_{2323}D_{23}n_{2k}\partial_kn_{2\alpha}\nonumber\\
&\,-K_{3232}\partial_k(D_{32}n_{3k}n_{3\alpha})+K_{1313}D_{13}n_{1k}(\partial_kn_{2\alpha}+\partial_{\alpha}n_{2k})\nonumber\\
&\,+\frac{1}{2}K_{1221}\big(D_{21}n_{3k}\partial_{\alpha}n_{1k}-\partial_k(D_{21}n_{1k}n_{3\alpha})\big)\nonumber\\
&\,+\frac{1}{2}K_{2332}\big(D_{32}n_{2k}\partial_kn_{2\alpha}-\partial_k(D_{23}n_{3k}n_{3\alpha})\big)\nonumber\nonumber\\
&\,+\frac{1}{2}K_{1331}D_{31}n_{1k}(\partial_kn_{2\alpha}+\partial_{\alpha}n_{2k}),\label{variational-derivative-n1}\\
\frac{\delta \CF_{Bi}}{\delta n_{2\alpha}}=&\,K_{1111}D_{11}n_{1k}\partial_kn_{3\alpha}+K_{2222}D_{22}n_{3k}\partial_{\alpha}n_{1k}
-K_{3333}\partial_k(D_{33}n_{3k}n_{1\alpha})\nonumber\\
&\,+K_{2121}D_{21}n_{2k}(\partial_kn_{3\alpha}+\partial_{\alpha}n_{3k})+K_{2323}\big(D_{23}n_{1k}\partial_{\alpha}n_{2k}-\partial_k(D_{23}n_{2k}n_{1\alpha})\big)\nonumber\\
&\,+K_{3131}D_{31}n_{3k}\partial_kn_{3\alpha}-K_{1313}\partial_k(D_{13}n_{1k}n_{1\alpha})\nonumber\\
&\,+\frac{1}{2}K_{1221}D_{12}n_{2k}(\partial_kn_{3\alpha}+\partial_{\alpha}n_{3k})\nonumber\\
&\,+\frac{1}{2}K_{2332}\big(D_{32}n_{1k}\partial_{\alpha}n_{2k}-\partial_k(D_{32}n_{2k}n_{1\alpha})\big)\nonumber\\
&\,+\frac{1}{2}K_{1331}\big(D_{13}n_{3k}\partial_kn_{3\alpha}-\partial_k(D_{31}n_{1k}n_{1\alpha})\big),\label{variational-derivative-n2}\\
\frac{\delta \CF_{Bi}}{\delta n_{3\alpha}}=&\,-K_{1111}\partial_k(D_{11}n_{1k}n_{2\alpha})+K_{2222}D_{22}n_{2k}\partial_kn_{1\alpha}+K_{3333}D_{33}n_{1k}\partial_{\alpha}n_{2k}\nonumber\\
&\,+K_{1212}D_{12}n_{1k}\partial_kn_{1\alpha}-K_{2121}\partial_k(D_{21}n_{2k}n_{2\alpha})+K_{3232}D_{32}n_{3k}(\partial_kn_{1\alpha}+\partial_{\alpha}n_{1k})\nonumber\\
&\,+K_{3131}\big(D_{31}n_{2k}\partial_{\alpha}n_{3k}-\partial_k(D_{31}n_{3k}n_{2\alpha})\big)\nonumber\\
&\,+\frac{1}{2}K_{1221}\big(D_{21}n_{1k}\partial_kn_{1\alpha}-\partial_k(D_{12}n_{2k}n_{2\alpha})\big)\nonumber\\
&\,+\frac{1}{2}K_{2332}D_{23}n_{3k}(\partial_kn_{1\alpha}+\partial_{\alpha}n_{1k})\nonumber\\
&\,+\frac{1}{2}K_{1331}\big(D_{13}n_{2k}\partial_{\alpha}n_{3k}-\partial_k(D_{13}n_{3k}n_{2\alpha})\big).\label{variational-derivative-n3}
\end{align}
Using the variational derivatives (\ref{variational-derivative-n2}) and (\ref{variational-derivative-n3}), we have
\begin{align}\label{var-direction-n1}
&\,n_{2\alpha}\frac{\delta \CF_{Bi}}{\delta n_{3\alpha}}-n_{3\alpha}\frac{\delta \CF_{Bi}}{\delta n_{2\alpha}}\nonumber\\
&\,\quad=-K_{1111}n_{2\alpha}\partial_k(D_{11}n_{1k}n_{2\alpha})-K_{2222}D_{22}(D_{23}+D_{32})\nonumber\\
&\,\qquad+K_{3333}\big(D_{33}D_{23}+n_{3\alpha}\partial_k(D_{33}n_{3k}n_{1\alpha})\big)-K_{1212}D_{12}D_{13}\nonumber\\
&\,\qquad-K_{2121}\big(D_{21}D_{31}+n_{2\alpha}\partial_k(D_{21}n_{2k}n_{2\alpha})\big)+K_{3232}D_{32}(D_{22}-D_{33})\nonumber\\
&\,\qquad-K_{2323}\big(D_{33}D_{23}-n_{3\alpha}\partial_k(D_{23}n_{2k}n_{1\alpha})\big)\nonumber\\
&\,\qquad+K_{3131}\big(D_{31}D_{21}-n_{2\alpha}\partial_k(D_{31}n_{3k}n_{2\alpha})\big)+K_{1313}n_{3\alpha}\partial_k(D_{13}n_{1k}n_{1\alpha})\nonumber\\
&\,\qquad-\frac{1}{2}K_{1221}\big(D_{21}D_{13}+D_{12}D_{31}+n_{2\alpha}\partial_k(D_{12}n_{2k}n_{2\alpha})\big)\nonumber\\
&\,\qquad+\frac{1}{2}K_{2332}\big(D_{23}(D_{22}-D_{33})-D_{33}D_{32}+n_{3\alpha}\partial_k(D_{32}n_{2k}n_{1\alpha})\big)\nonumber\\
&\,\qquad+\frac{1}{2}K_{1331}\big(D_{13}D_{21}-n_{2\alpha}\partial_k(D_{13}n_{3k}n_{2\alpha})+n_{3\alpha}\partial_k(D_{31}n_{1k}n_{1\alpha})\big).
\end{align}
Similarly, we have
\begin{align}
&\,n_{3\alpha}\frac{\delta \CF_{Bi}}{\delta n_{1\alpha}}-n_{1\alpha}\frac{\delta \CF_{Bi}}{\delta n_{3\alpha}}\nonumber\\
&\,\quad=K_{1111}D_{11}(D_{13}+D_{31})-K_{2222}\partial_k(n_{2k}D_{22})-K_{3333}D_{33}(D_{13}+D_{31})\nonumber\\
&\,\qquad+K_{1212}\big(D_{12}D_{32}-\partial_k(n_{1k}D_{12})\big)+(K_{2121}-K_{2323})D_{23}D_{21}\nonumber\\
&\,\qquad-K_{3232}\big(D_{12}D_{32}+\partial_k(n_{3k}D_{32})\big)+K_{1313}D_{13}(D_{33}-D_{11})\nonumber\\
&\,\qquad-K_{3131}D_{31}(D_{11}-D_{33})+\frac{1}{2}K_{1221}\big(D_{21}D_{32}+D_{12}D_{23}-\partial_k(n_{1k}D_{21})\big)\nonumber\\
&\,\qquad-\frac{1}{2}K_{2332}\big(D_{32}D_{21}+D_{23}D_{12}+\partial_k(n_{3k}D_{23})\big)\nonumber\\
&\,\qquad+\frac{1}{2}K_{1331}(D_{31}+D_{13})(D_{33}-D_{11}),\label{var-direction-n2}
\end{align}
and
\begin{align}
&\,n_{1\alpha}\frac{\delta \CF_{Bi}}{\delta n_{2\alpha}}-n_{2\alpha}\frac{\delta \CF_{Bi}}{\delta n_{1\alpha}}\nonumber\\
&\,\quad=-K_{1111}D_{11}(D_{12}+D_{21})+K_{2222}D_{22}(D_{12}+D_{21})-K_{3333}\partial_k(n_{3k}D_{33})\nonumber\\
&\,\qquad+K_{2121}D_{21}(D_{11}-D_{22})+K_{2323}\big(D_{23}D_{13}-\partial_k(n_{2k}D_{23})\big)\nonumber\\
&\,\qquad+(K_{3232}-K_{3131})D_{31}D_{32}-K_{1313}\big(D_{13}D_{23}+\partial_k(n_{1k}D_{13})\big)\nonumber\\
&\,\qquad-K_{1212}D_{12}(D_{22}-D_{11})+\frac{1}{2}K_{1221}(D_{12}+D_{21})(D_{11}-D_{22})\nonumber\\
&\,\qquad+\frac{1}{2}K_{2332}\big(D_{32}D_{13}+D_{23}D_{31}-\partial_k(n_{2k}D_{32})\big)\nonumber\\
&\,\qquad-\frac{1}{2}K_{1331}\big(D_{32}D_{13}+D_{23}D_{31}+\partial_k(n_{1k}D_{31})\big).\label{var-direction-n3}
\end{align}

\bigskip
\noindent{\bf Acknowledgments.}
Sirui Li is partially supported by the
NSF of China under grant No. 12061019 and by the Growth
Foundation for Youth Science and Technology Talent of
Educational Commission of Guizhou Province of China under grant No. [2021]087. Jie Xu is partially supported by the NSF of China under grant No. 12001524.

\end{document}